\documentclass[structabstract]{aa}  
\usepackage{graphicx}
\usepackage{amsmath,amssymb}
\usepackage{txfonts}
\begin{document}

\def\Arrow{\mathop{\longrightarrow}\limits}
\def\Harpoons{\mathop{\rightleftharpoons}\limits}

   \title{A molecular line study of the filamentary infrared dark cloud G304.74+01.32\thanks{This publication is based on data acquired with the Atacama Pathfinder EXperiment (APEX) under programmes 083.F-9302A and 087.F-9318A. APEX is a collaboration between the Max-Planck-Institut f\"{u}r Radioastronomie, the European Southern Observatory, and the Onsala Space Observatory.}}

   \author{Oskari Miettinen}

 \offprints{O. Miettinen}

   \institute{Department of Physics, P.O. Box 64, FI-00014 University of Helsinki, Finland\\ \email{oskari.miettinen@helsinki.fi}}

   \date{Received ; accepted}

\authorrunning{Miettinen}
\titlerunning{IRDC G304.74+01.32}

  \abstract
   {Infrared dark clouds (IRDCs) are promising sites to study the earliest 
formation stages of stellar clusters and high-mass stars, and the physics of 
molecular-cloud formation and fragmentation.}
   {The aim of this study is to better understand the physical and chemical 
properties of the filamentary IRDC G304.74+01.32 (hereafter, 
G304.74). In particular, we aim to investigate the kinematics and dynamical 
state of the cloud and clumps within it, and the amount of CO depletion.}
   {All the submillimetre peak positions in the cloud identified from our 
previous LABOCA 870-$\mu$m map were observed in C$^{17}$O$(2-1)$ with APEX. 
These are the first line observations along the whole filament made so far. 
Selected positions were also observed in the $^{13}$CO$(2-1)$, SiO$(5-4)$, and 
CH$_3$OH$(5_k-4_k)$ transitions at $\sim1$ mm wavelength.} 
   {The C$^{17}$O lines were detected towards all target positions at 
similar radial velocities. CO does not appear to be significantly 
depleted in the clumps; the largest depletion factors are only $\sim2$.
Two- to three methanol $5_k-4_k$ lines near $\sim241.8$ GHz were detected 
towards all selected target positions, whereas SiO$(5-4)$ was seen in only one 
of these positions, namely SMM 3. In the band covering SiO$(5-4)$, we 
also detected DCN$(3-2)$ line towards SMM 3. The $^{13}$CO$(2-1)$ lines show 
blue asymmetric profiles, indicating large-scale infall motions. The clumps 
show trans- to supersonic non-thermal motions, and virial-parameter analysis 
suggests that most of them are gravitationally bound. The external pressure 
may also play a non-negligible role in the dynamics. The analysis 
suggests that the fragmentation of the filament into clumps is caused by 
``sausage''-type instability, in agreement with results from other IRDCs.}
   {The uniform C$^{17}$O radial velocities along the G304.74 cloud shows that 
it is a coherent filamentary structure. Although the clumps appear to be 
gravitationally bound, the ambient turbulent ram pressure may be an important 
factor in the cloud dynamics. This is qualitatively consistent with our 
earlier suggestion that the filament was formed by converging supersonic 
turbulent flows. The poloidal magnetic field could resist the radial cloud 
collapse, which conforms to the low infall velocites derived.
The cloud may not be able to form high-mass stars on the 
basis of mass-size threshold, but ``only'' stellar clusters and/or 
intermediate-mass stars. The star-formation activity in the cloud, such as 
outflows, is likely responsible in releasing CO from the icy grain mantles 
back into the gas phase. Shocks related to outflows may have also injected 
CH$_3$OH, SiO, and DCN into the gas-phase in SMM 3.}


   \maketitle
%

\section{Introduction}

Infrared dark clouds (IRDCs) are seen as dark absorption features against 
the Galactic background radiation, particularly at mid-infrared wavelengths 
(\cite{perault1996}; \cite{egan1998}; \cite{simon2006}; \cite{peretto2009}). 
Since their discovery in the mid to late nineties, and after the first 
detailed studies (\cite{carey1998}, 2000), IRDCs have been the subject of 
great research interest (e.g., \cite{teyssier2002}; \cite{johnstone2003}; 
\cite{rathborne2005}, 2006; \cite{pillai2006}; \cite{beuther2007}; 
\cite{sakai2008}; \cite{ragan2009}; \cite{jimenez2010}; \cite{leurini2011}; 
\cite{kainulainen2011b}, and many more). One of the reasons for such an active 
study of IRDCs is that they are found to be promising targets to study the 
initial conditions and early stages of high-mass star ($>8$ M$_{\sun}$) and 
stellar cluster formation, both of which are quite poorly known compared to 
the formation of solar-type stars. Moreover, IRDCs are interesting objects for 
the studies of molecular-cloud formation and their fragmentation into clumps 
and cores [e.g., \cite{miettinenharju2010} (hereafter, Paper I); 
\cite{jackson2010}; \cite{wang2011}]. Infrared dark clouds are often found to 
be filamentary in shape, resembling the morphologies seen in numerical 
studies of cloud formation (e.g., \cite{nakajima1996}; 
\cite{ballesteros1999}; \cite{klessen2000}; \cite{klessen2001}; 
\cite{heitsch2008}, 2011). On the other hand, theories and models of 
instability and fragmentation of filamentary, or cylindrical, gas structures 
have been extensively studied in the past (e.g., \cite{chandrasekhar1953}; 
\cite{stodolkiewicz1963}; \cite{ostriker1964}; \cite{bastien1983}; 
\cite{inutsuka1992}; \cite{curry2000}). The filamentary IRDCs are well-suited 
for testing these theories/models, such as their predictions about the 
preferred fragmentation length-scales.

\begin{figure*}
\begin{center}
\includegraphics[width=0.8\textwidth]{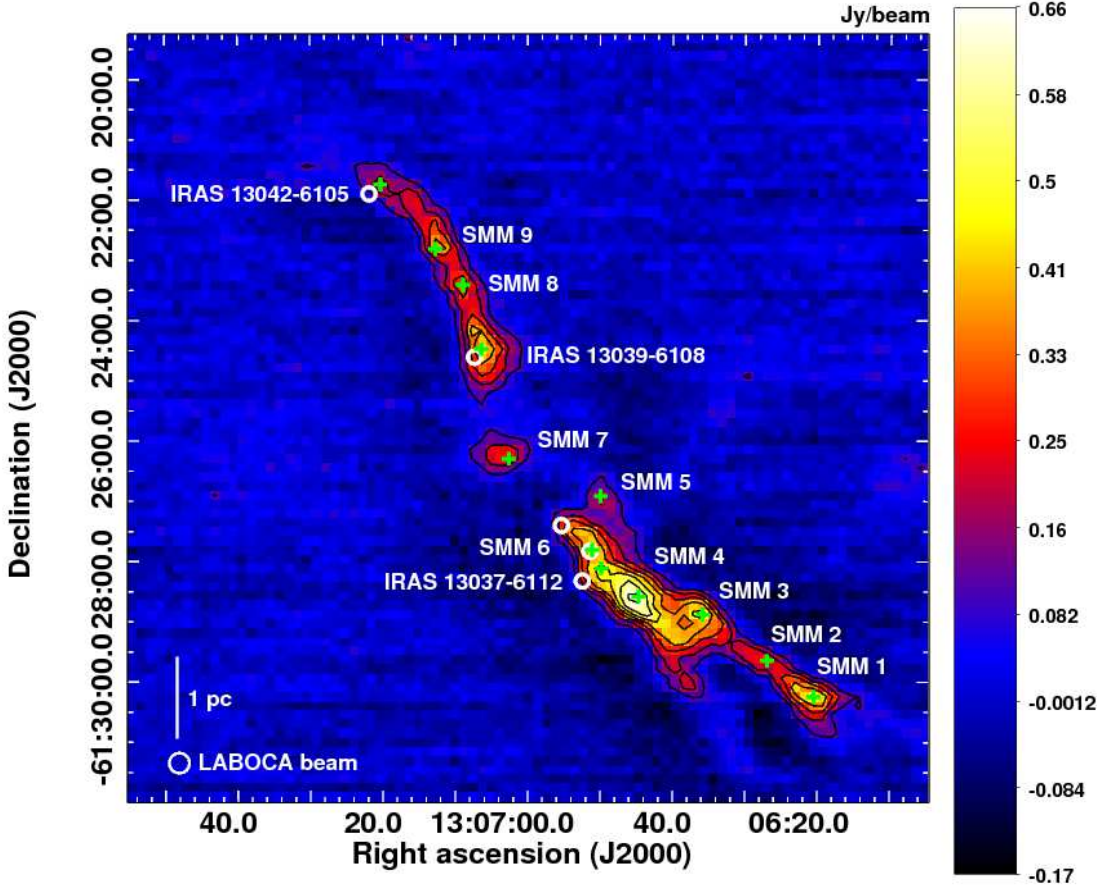}
\caption{LABOCA 870-$\mu$m image of the IRDC G304.74. The image is shown 
with linear scaling, and the colour bar indicates the flux density in 
units of Jy~beam$^{-1}$. The rms level is 0.03 Jy~beam$^{-1}$ ($1\sigma$). 
The first contour and the separation between contours is $3\sigma$. The green 
plus signs show the submm peak positions, which were used as the 
line-observation target positions in the present study. The white circles 
indicate the nominal catalogue positions of the \textit{MSX} 8-$\mu$m sources. 
The 1-pc scale bar and the effective LABOCA beam HPBW ($\sim20\arcsec$) are 
shown in the bottom left (modified from Paper I).}
\label{figure:laboca}
\end{center}
\end{figure*}

The target cloud of the present study is the IRDC G304.74+01.32 (hereafter, 
G304.74). We have recently mapped this IRDC in the 870-$\mu$m dust continuum 
emission with the LABOCA bolometer on APEX (Paper I). The cloud was found to 
have a integral-shaped/``hub-filament''-kind of structure 
(\cite{myers2009}), and it was resolved into twelve clumps with the 2D 
{\tt clumpfind} algorithm (\cite{williams1994}). Four of the clumps were found 
to be associated with \textit{MSX} 8-$\mu$m point-like sources, indicating 
embedded star formation. Moreover, three of these \textit{MSX}-bright clumps 
are associated with \textit{IRAS} point sources (the \textit{IRAS} sources 
13037-6112, 13039-6108, and 13042-6105; hereafter, IRAS13037 etc.). Eight 
clumps in G304.74 were found to be dark in the \textit{MSX} 8-$\mu$m image. 
The clump masses were estimated to be in the range $\sim40-220$ M$_{\sun}$ 
(assuming a dust temperature of 15 or 22 K), within the effective radii of 
$\sim0.3-0.5$ pc. Unfortunately, G304.74 is not covered by the \textit{Spitzer} 
infrared maps, which ``only'' cover the Galactic latitudes 
$\left| b \right|\leq1\degr$. Our 870-$\mu$m LABOCA map of G304.74 is 
shown in Fig.~\ref{figure:laboca} (modified from Fig.~1 of Paper I). See also 
Fig.~2 of Paper I for the \textit{MSX} 8-$\mu$m image of the region overlaid 
with LABOCA contours.

To our knowledge, besides the IRDC identification work by Simon et al. (2006) 
and our Paper I, the only studies of G304.74, or a clump within it, are those 
by Fontani et al. (2005) and Beltr{\'a}n et al. (2006). The latter authors 
performed SIMBA 1.2-mm dust continuum mapping of the cloud, and identified 
eight clumps, including the \textit{MSX}-bright sources (see our Paper I for 
more details). Fontani et al. (2005) carried out pointed CS and C$^{17}$O 
observations towards one source in G304.74, namely IRAS13039. By employing the 
Galactic rotation curve of Brand \& Blitz (1993) and the radial velocity 
$-26.2$ km~s$^{-1}$ derived from CS observations, Fontani et al. (2005) 
determined the source distance of 2.44 kpc. As was recently discussed by 
Beuther et al. (2011), G304.74 constitutes a small part of the kpc-distance 
cloud complexes, which form a portion of the Coalsack dark cloud in the plane 
of the sky. 

In the present paper we will present the first molecular-line 
observations along the whole filamentary structure of G304.74. These data will 
be used to examine the basic phy\-sical properties of the cloud, such as gas 
kinematics and dynamical state of the clumps. The cloud's kinematic distance, 
and the distance-dependent parameters presented in Paper I, will also be 
revised. This study will also address some che\-mical properties 
of the clumps. For example, we investigate the amount of CO depletion, which 
has only recently been studied in IRDCs (\cite{zhang2009}; 
\cite{hernandez2011}; \cite{miettinen2011}; \cite{chen2011}).

The rest of this paper is structured as follows. The observations and 
data-reduction procedures are described in Sect.~2. The direct observational 
results are presented in Sect.~3. In Sect.~4/Appendix A we describe 
the analysis and present the results of the physical and chemical properties 
of the cloud/clumps. Discussion of our results is presented in Sect.~5, and in 
Sect.~6 we summarise the main conclusions of this study.

\section{Observations and data reduction}

The observations presented in this paper were made using the APEX 12-m 
telescope at Llano de Chajnantor (Chilean Andes). 
The telescope and its performance are described in the paper by G{\"u}sten 
et al. (2006). All the submm peak positions in G304.74 identified from our 
previous LABOCA 870-$\mu$m map were observed in C$^{17}$O$(2-1)$, and selected 
positions in the southern part of the filament were observed in 
$^{13}$CO$(2-1)$, SiO$(5-4)$, and CH$_3$OH$(5_k-4_k)$ with the 
Swedish Heterodyne Facility Instrument (SHeFI; \cite{belitsky2007}; 
\cite{vassilev2008a}). The target positions with their physical properties are 
listed in Table~\ref{table:sources}. The observed transitions and 
observational parameters are listed in Table~\ref{table:lines}. 
The observations took place on 18--19 and 26--28 
May 2011. As frontend for the observations, we used the APEX-1 receiver of the 
SHeFI (\cite{vassilev2008b}). The backend for all observations was the Fast 
Fourier Transfrom Spectrometer (FFTS; \cite{klein2006}) with a 1 GHz bandwidth 
divided into 8\,192 channels.

The observations were performed in the wobbler-switching mode with a 
$150\arcsec$ azimuthal throw (symmetric offsets) and a chopping rate of 0.5 Hz.
The telescope pointing and focus corrections were checked by 
continuum scans on the planet Saturn, and the pointing was found to be better 
than $\sim5\arcsec$. Calibration was done by the chopper-wheel technique, and 
the output intensity scale given by the system is $T_{\rm A}^*$, which 
represents the antenna temperature corrected for atmospheric attenuation. 
The observed intensities were converted to the main-beam brightness 
temperature scale by $T_{\rm MB}=T_{\rm A}^*/\eta_{\rm MB}$, where 
$\eta_{\rm MB}$ is the main beam efficiency. The absolute calibration 
uncertainty is estimated to be around 10\%.

The spectra were reduced using the CLASS90 programme of the IRAM's GILDAS 
software package\footnote{{\tt http://www.iram.fr/IRAMFR/GILDAS}}. 
The individual spectra were averaged and the resulting spectra were 
Hanning-smoothed in order to improve the signal-to-noise ratio of the 
data. A first- or third-order polynomial was applied to correct the baseline 
in the final spectra. The resulting $1\sigma$ rms noise levels are 
$\sim40-100$ mK at the smoothed resolutions. 

The $^{13}$CO$(2-1)$ and C$^{17}$O$(2-1)$ rotational lines
contain three and nine hyperfine (hf) components, respectively. 
We fitted these hf structures using ``method hfs'' of CLASS90 to derive the LSR 
velocity (${\rm v}_{\rm LSR}$) of the emission, and FWHM linewidth 
($\Delta {\rm v}$). The hf-line fitting can also be used to derive the line 
optical thickness, $\tau$. However, in all spectra the hf components are 
blended together, and thus the optical thickness could not be reliably 
determined. For the rest frequencies of the $^{13}$CO$(2-1)$ hf components 
we used the values from Cazzoli et al. (2004; Table 2 therein), whereas 
those for C$^{17}$O$(2-1)$ were taken from Ladd et al. (1998; Table 6 therein).

\begin{table*}
\caption{Source list.}
\begin{minipage}{2\columnwidth}
\centering
\renewcommand{\footnoterule}{}
\label{table:sources}
\begin{tabular}{c c c c c c c c}
\hline\hline 
Source & $\alpha_{2000.0}$ & $\delta_{2000.0}$ & $R$ & $M$ & $N({\rm H_2})$ & $\langle n({\rm H_2}) \rangle$ & \textit{MSX} 8 $\mu$m \\
       & [h:m:s] & [$\degr$:$\arcmin$:$\arcsec$] & [pc] & [M$_{\sun}$] & [$10^{22}$ cm$^{-2}$] & [$10^4$ cm$^{-3}$] & \\
\hline
SMM 1 & 13 06 20.5 & -61 30 15 & $0.44\pm0.12$ & $121\pm92$ & $2.1\pm1.2$ & $0.7\pm0.5$ & dark\\
SMM 2 & 13 06 26.9 & -61 29 39 & $0.34\pm0.09$ & $61\pm46$ & $1.4\pm0.8$ & $0.7\pm0.5$ & dark\\
SMM 3 & 13 06 35.8 & -61 28 53 & $0.52\pm0.13$ & $206\pm156$ & $2.3\pm1.3$ & $0.7\pm0.5$ & dark\\
SMM 4 & 13 06 44.7 & -61 28 35 & $0.47\pm0.12$ & $248\pm188$ & $3.0\pm1.7$ & $1.1\pm0.8$ & dark\\
IRAS 13037-6112 & 13 06 49.8 & -61 28 07 & $0.32\pm0.08$ & $54\pm28$ & $1.2\pm0.1$ & $0.8\pm0.4$ & point\\
SMM 5 & 13 06 49.9 & -61 26 55 & $0.32\pm0.08$ & $42\pm32$ & $0.9\pm0.5$ & $0.6\pm0.4$ & dark\\
SMM 6 & 13 06 51.1 & -61 27 49 & $0.37\pm0.10$ & $109\pm83$ & $2.1\pm1.2$ & $1.0\pm0.8$ & point\\
SMM 7 & 13 07 02.6 & -61 26 18 & $0.37\pm0.10$ & $55\pm41$ & $1.2\pm0.7$ & $0.5\pm0.4$ & dark\\
IRAS 13039-6108 & 13 07 06.4 & -61 24 29 & $0.49\pm0.13$ & $88\pm46$ & $1.2\pm0.1$ & $0.3\pm0.2$ & point\\
SMM 8 & 13 07 08.9 & -61 23 25 & $0.36\pm0.09$ & $67\pm50$ & $1.4\pm0.8$ & $0.7\pm0.5$ & dark\\
SMM 9 & 13 07 12.7 & -61 22 49 & $0.39\pm0.10$ & $97\pm73$ & $1.7\pm1.0$ & $0.8\pm0.6$ & dark\\
IRAS 13042-6105 & 13 07 20.3 & -61 21 45 & $0.33\pm0.09$ & $48\pm37$ & $1.0\pm0.5$ & $0.6\pm0.5$ & point\\
\hline 
\end{tabular} 
\tablefoot{The clump physical properties are revised from those presented in 
Paper I (see Sect.~4). Columns (2) and (3) give the equatorial coordinates 
[$(\alpha, \,\delta)_{2000.0}$] of the LABOCA peak position. Columns (4)--(7) 
list, respectively, the clump effective radius, mass, beam-averaged peak H$_2$ 
column density, and the volume-averaged H$_2$ number density. In the 
last column we indicate whether the clump appears dark or bright in the 
\textit{MSX} 8-$\mu$m image.}
\end{minipage}
\end{table*}

\begin{table*}
\caption{Observed spectral-line transitions and observational parameters.}
\begin{minipage}{2\columnwidth}
\centering
\renewcommand{\footnoterule}{}
\label{table:lines}
\begin{tabular}{c c c c c c c c c c}
\hline\hline 
Transition & $\nu$\tablefootmark{a} & HPBW & $\eta_{\rm MB}$ & $T_{\rm sys}$ & PWV & \multicolumn{2}{c}{Channel spacing\tablefootmark{b}} & $t_{\rm int}$ & rms\\
      & [MHz] & [\arcsec] & & [K] & [mm] & [kHz] & [km~s$^{-1}$] & [min] & [mK]\\
\hline        
SiO$(5-4)$ & 217\,104.980 & 28.7 & 0.75 & 324--360 & 0.7--1.2 & 122.07 & 0.17 & 9.5 & 38--43\\
DCN$(3-2)$\tablefootmark{c} & 217\,238.61\tablefootmark{c} & 28.7 & 0.75 & 337 & 0.8--0.9 & 122.07 & 0.17 & 9.5 & 39\\
$^{13}$CO$(2-1)$ & 220\,398.70056\tablefootmark{d} & 28.3 & 0.75 & 282--371 & 0.8--1.0 & 122.07 & 0.17 & 9.5 & 43--46\\
C$^{17}$O$(2-1)$ & 224\,714.199\tablefootmark{e} & 27.8 & 0.75 & 392--418 & 0.7--1.7 & 122.07 & 0.16 & 2.6 & 81--102\\
CH$_3$OH$(5_{-1,5}-4_{-1,4})$-E & 241\,767.224 & 25.8 & 0.75 & 430--445 & 0.4--0.8 & 122.07 & 0.15 & 10.5--15.5 & 39--49 \\
CH$_3$OH$(5_{0,5}-4_{0,4})$-A$^+$ & 241\,791.431 & \ldots & \ldots & \ldots & \ldots & \ldots & \ldots & \ldots & \ldots\\ 
CH$_3$OH$(5_{4,*}-4_{4,*})$-A$^{+/-}$\tablefootmark{f} & 241\,806.508 & \ldots & \ldots & \ldots & \ldots & \ldots & \ldots & \ldots & \ldots\\
\hline 
\end{tabular} 
\tablefoot{Columns (2)--(10) give the rest frequencies of the observed 
transitions ($\nu$), the APEX beamsize (HPBW) and the main beam efficiency 
($\eta_{\rm MB}$) at the observed frequencies, the single-sideband system 
temperatures during the observations ($T_{\rm sys}$ in $T_{\rm MB}$ scale, see 
text), the amount of precipitable water vapour (PWV), channel widths (both in 
kHz and km~s$^{-1}$) of the original data, the on-source integration times per 
position ($t_{\rm int}$), and the $1\sigma$ rms noise at the smoothed 
resolution.\\
\tablefoottext{a}{The rest frequencies were taken from the Cologne Database 
for Molecular Spectroscopy (CDMS; 
{\tt http://www.astro.uni-koeln.de/cdms/catalog}) (\cite{muller2005}) unless 
otherwise stated.}\tablefoottext{b}{The original channel spacings. The final 
spectra were Hanning-smoothed which divides the number of channels by 
two.}\tablefoottext{c}{The DCN$(3-2)$ transition was additionally detected in 
the band covering the SiO$(5-4)$ line. The quoted frequency refers to the 
strongest hyperfine component $F=4-3$.}\tablefoottext{d}{The frequency was 
taken from Cazzoli et al. (2004), and it refers to the strongest hyperfine 
component $F=5/2-3/2$.}\tablefoottext{e}{The frequency was taken from Ladd et 
al. (1998), and it refers to the strongest hyperfine component $F=9/2-7/2$.} 
\tablefoottext{f}{A blend of CH$_3$OH$(5_{4,1}-4_{4,0})$-A$^+$ and 
CH$_3$OH$(5_{4,2}-4_{4,1})$-A$^-$, which have the same frequency.}}
\end{minipage}
\end{table*}

\section{Observational results}

\begin{figure*}
\begin{center}
\includegraphics[width=3.1cm, angle=-90]{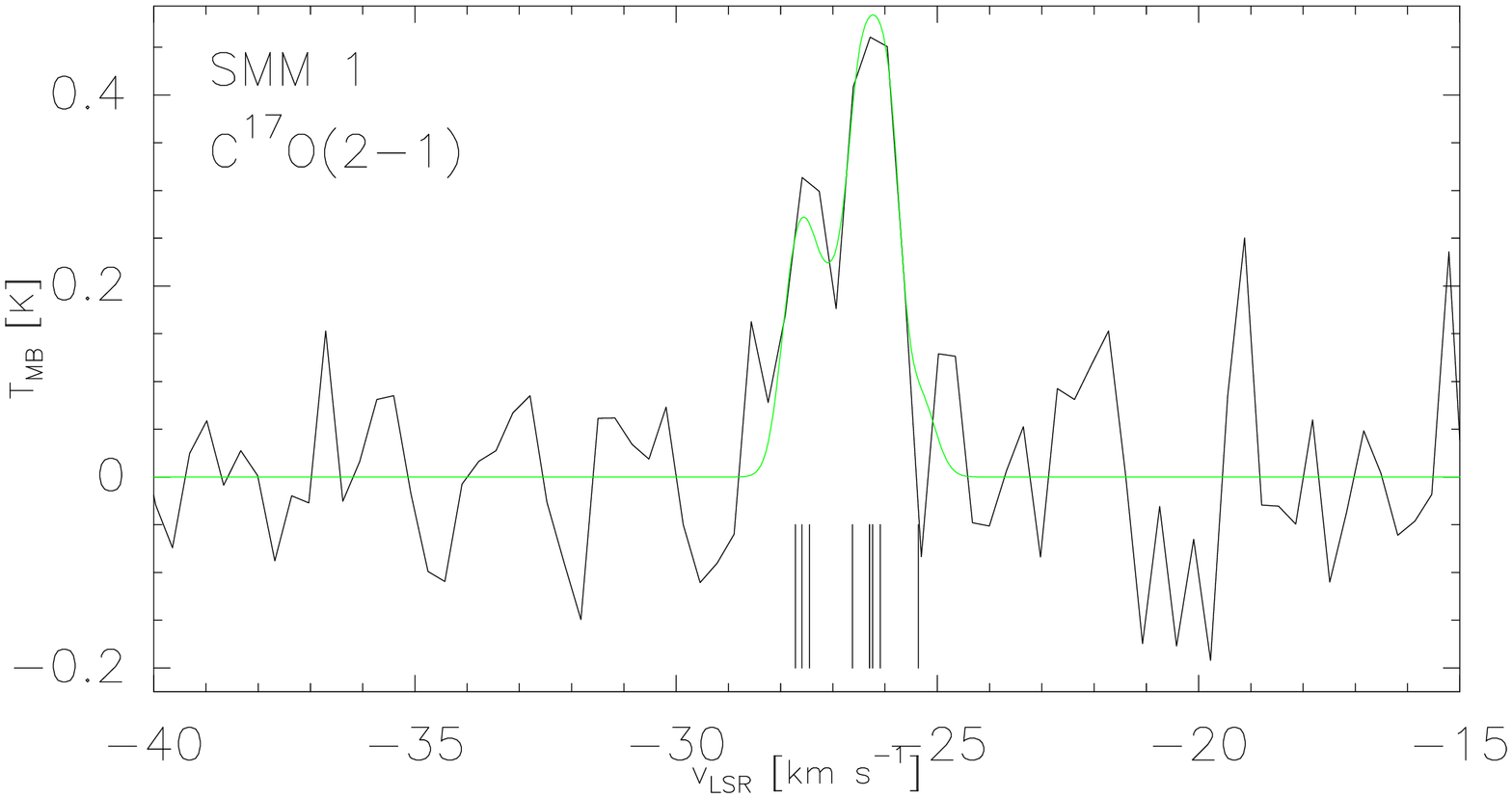}
\includegraphics[width=3.1cm, angle=-90]{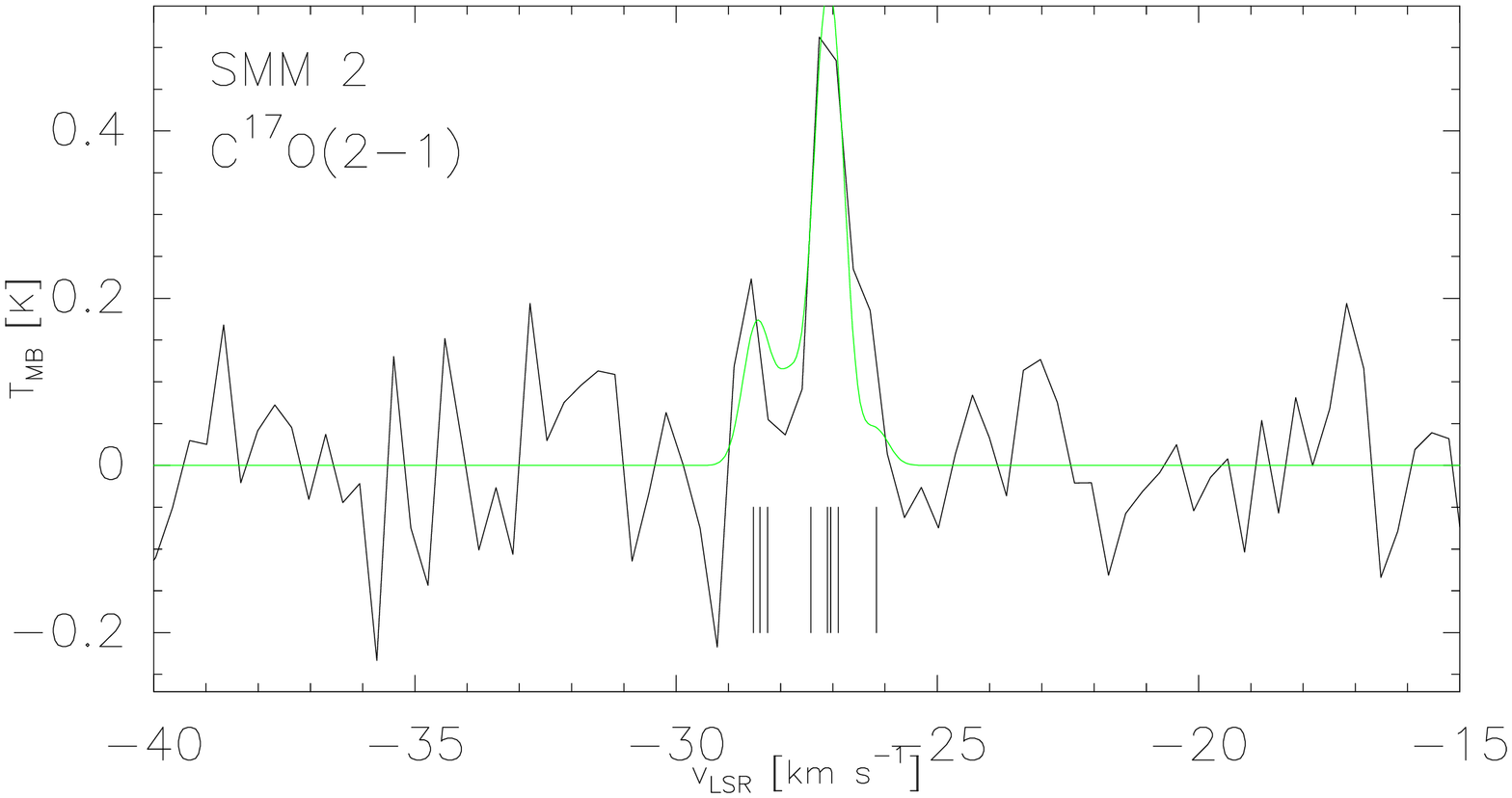}
\includegraphics[width=3.1cm, angle=-90]{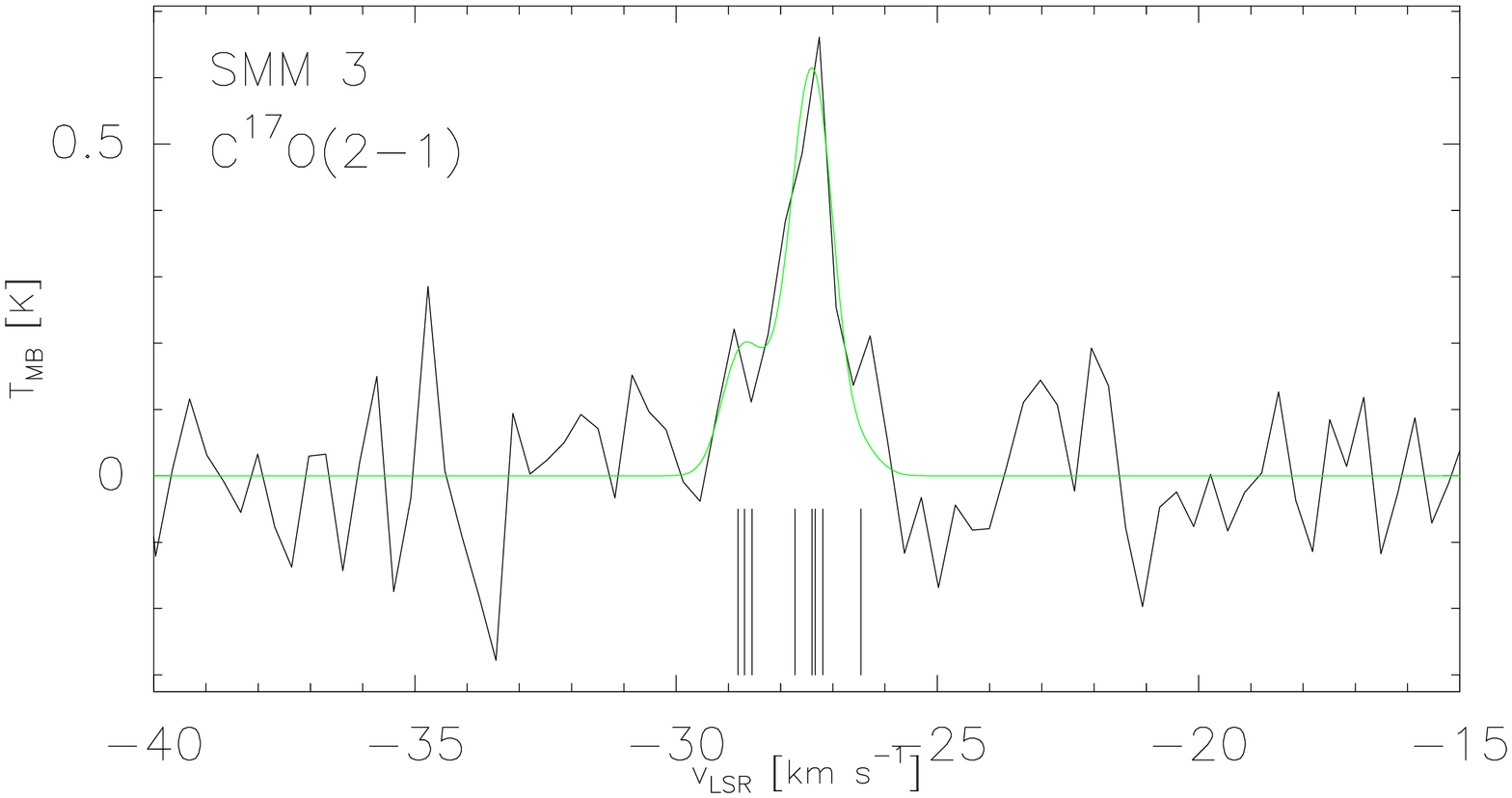}
\includegraphics[width=3.1cm, angle=-90]{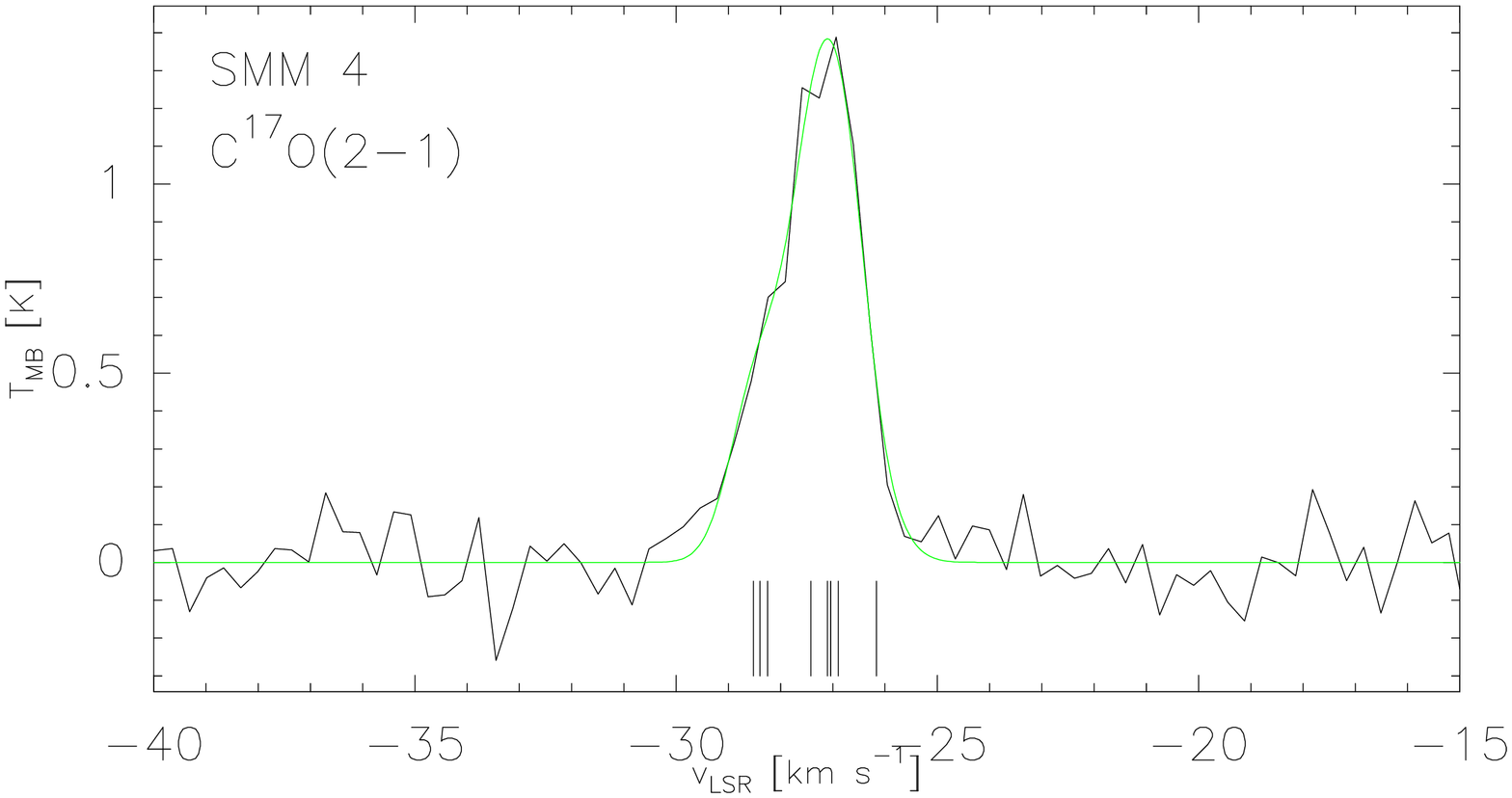}
\includegraphics[width=3.1cm, angle=-90]{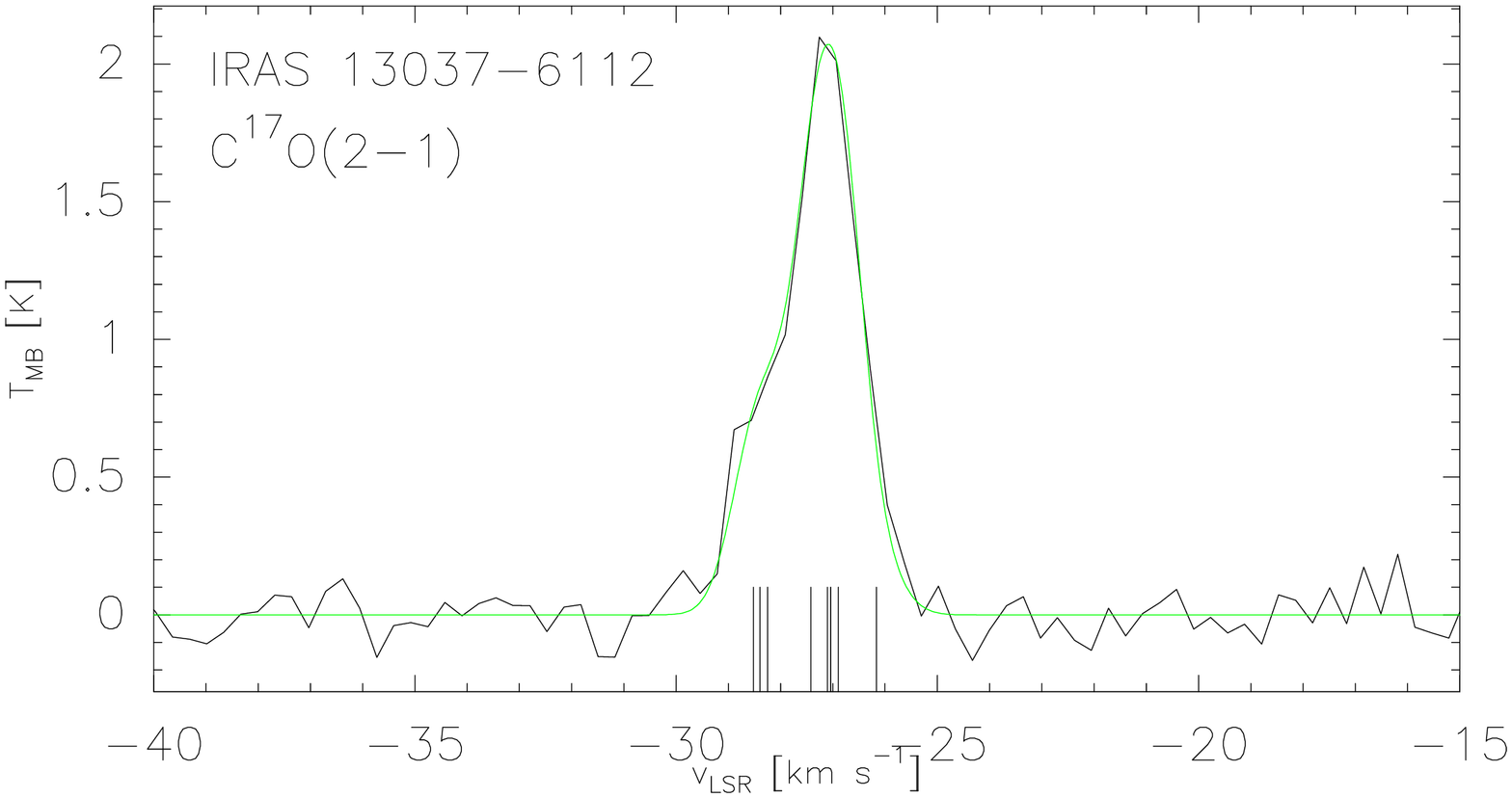}
\includegraphics[width=3.1cm, angle=-90]{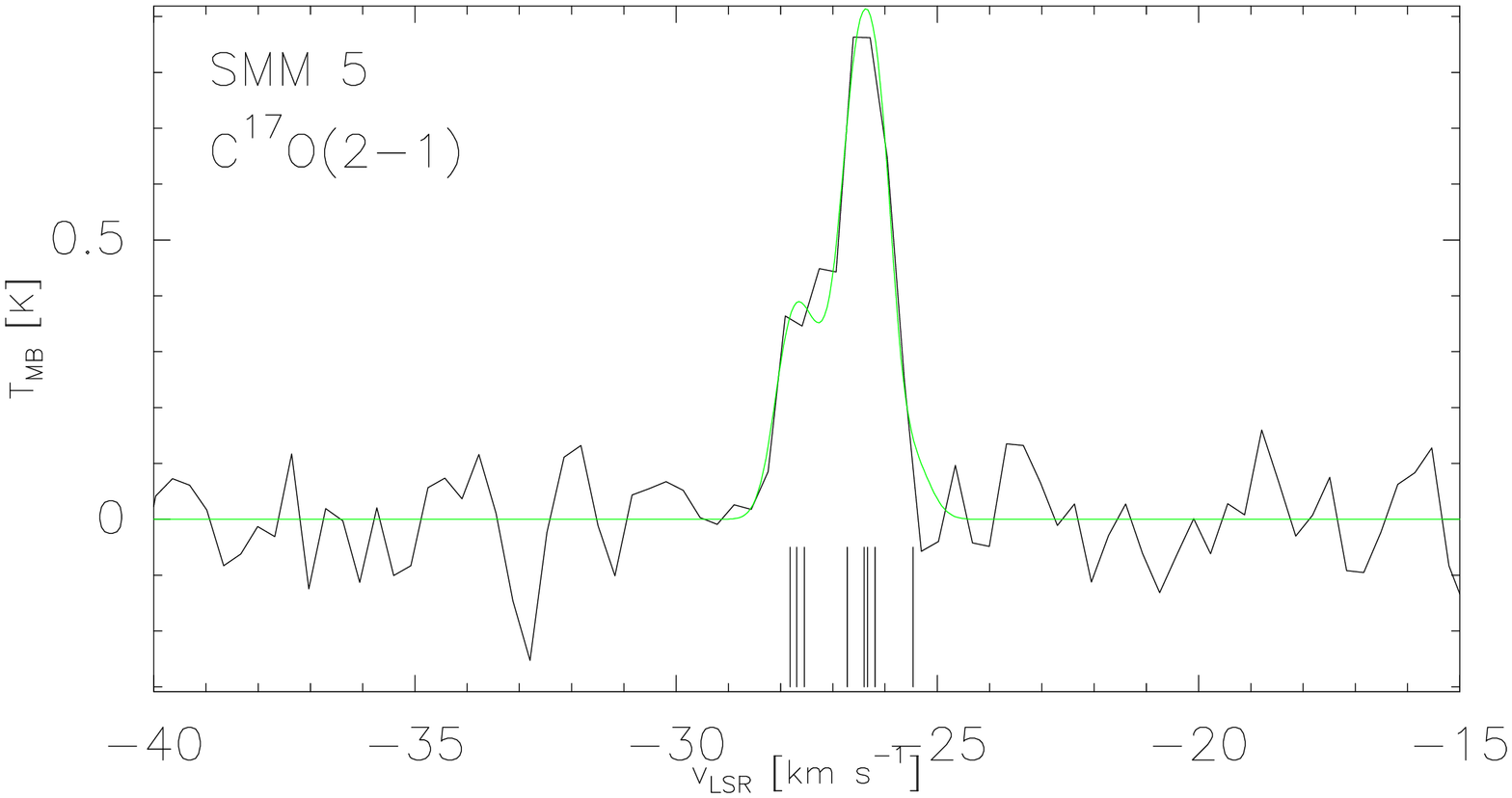}
\includegraphics[width=3.1cm, angle=-90]{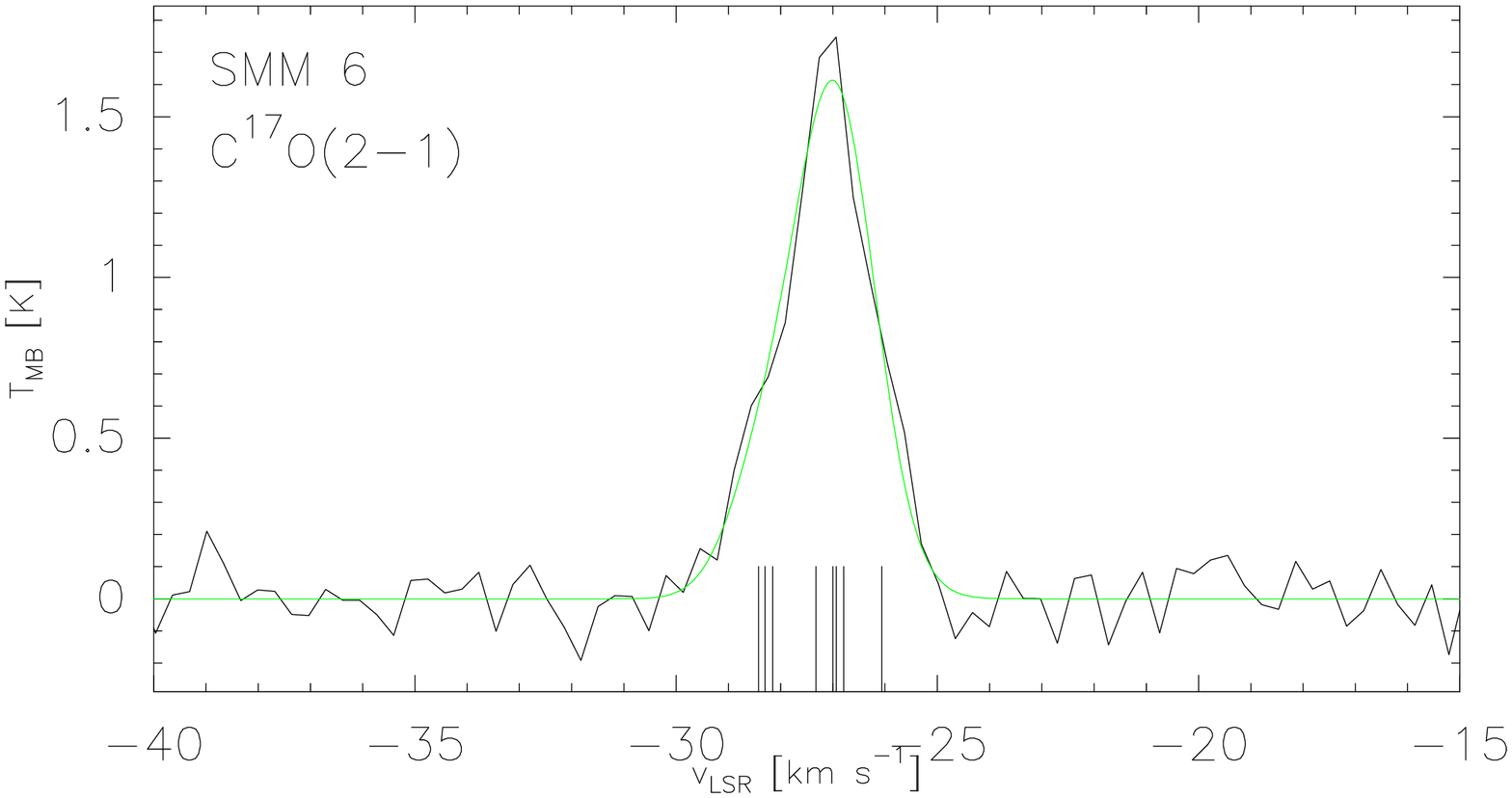}
\includegraphics[width=3.1cm, angle=-90]{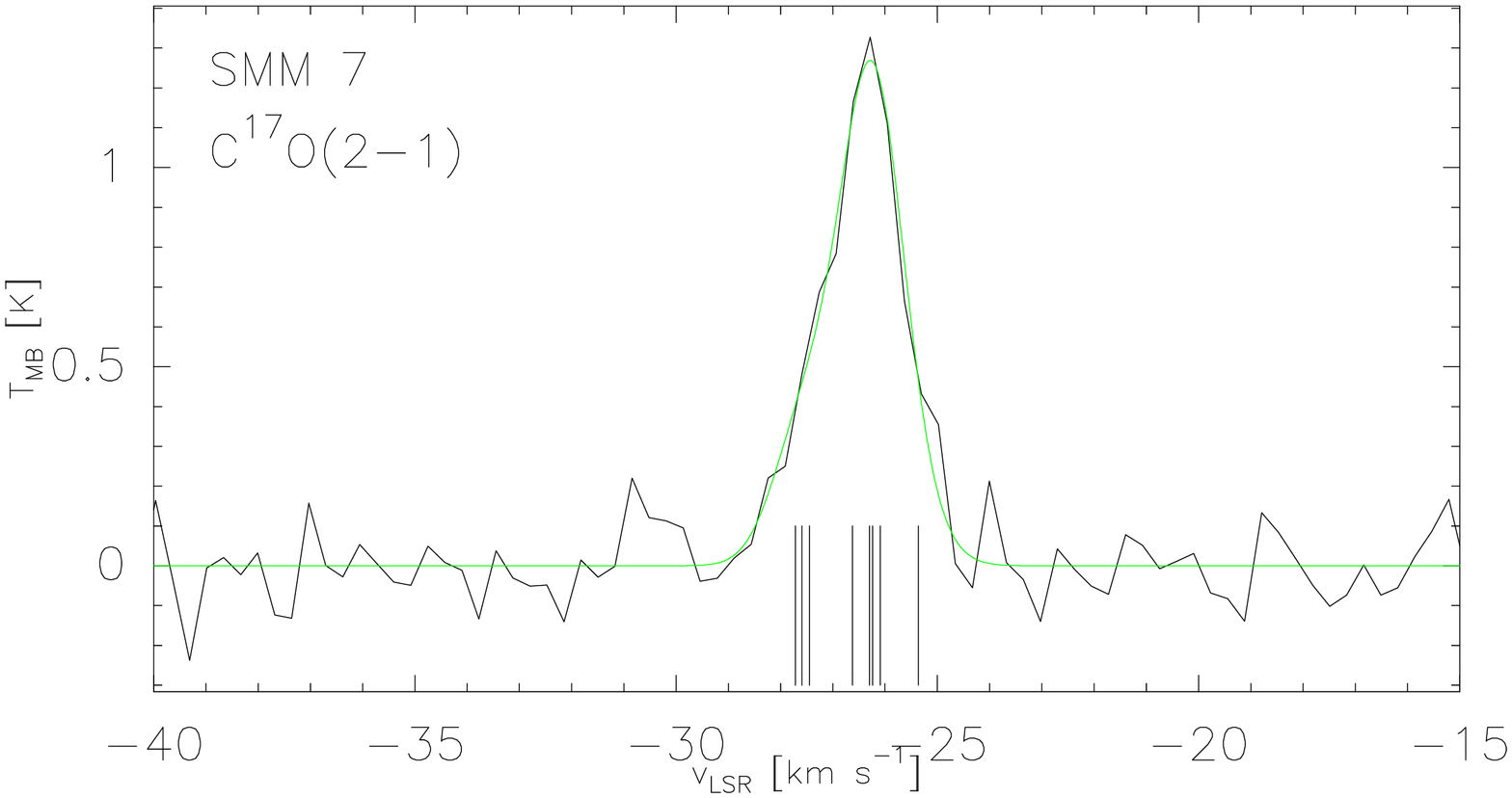}
\includegraphics[width=3.1cm, angle=-90]{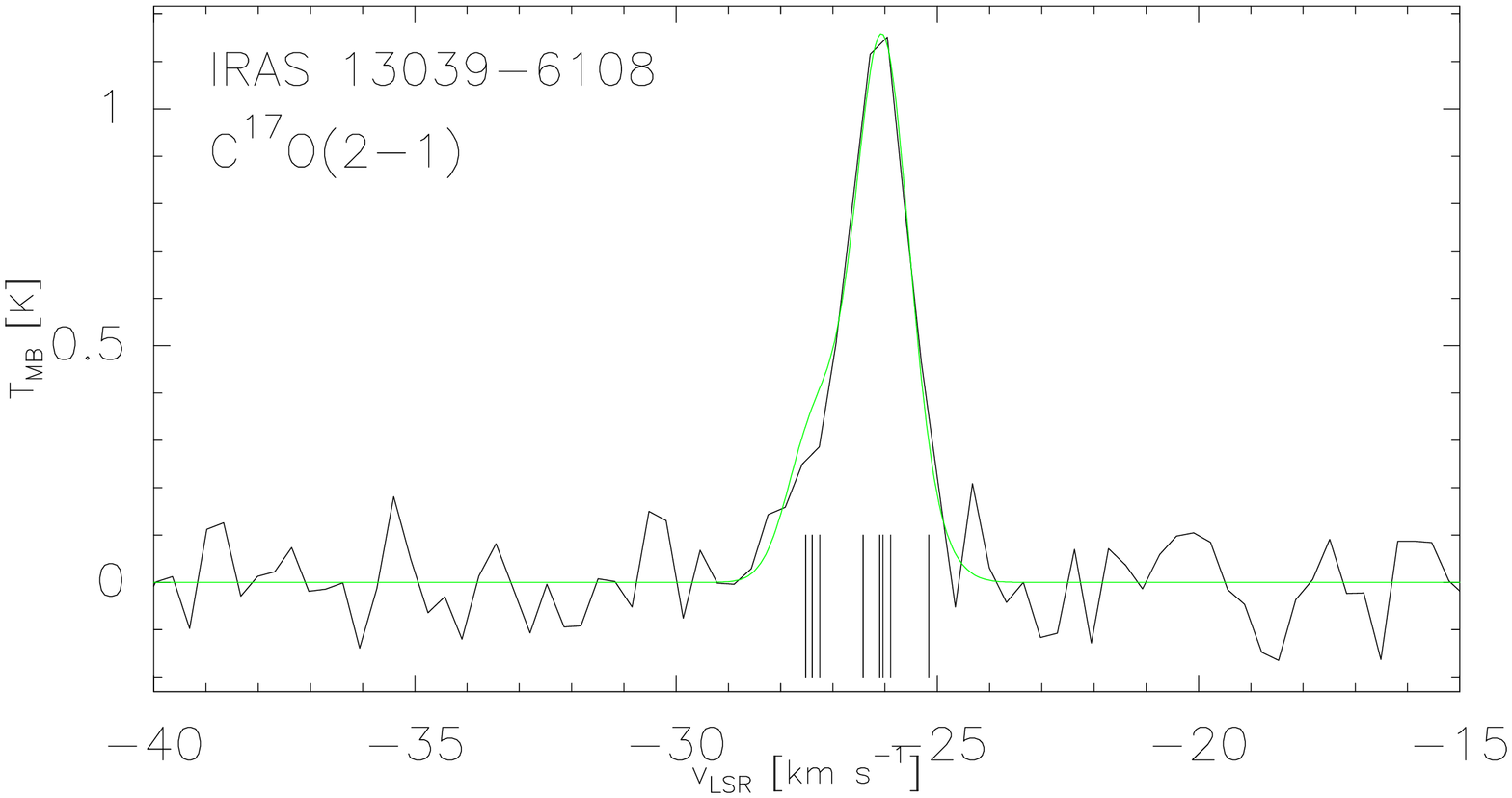}
\includegraphics[width=3.1cm, angle=-90]{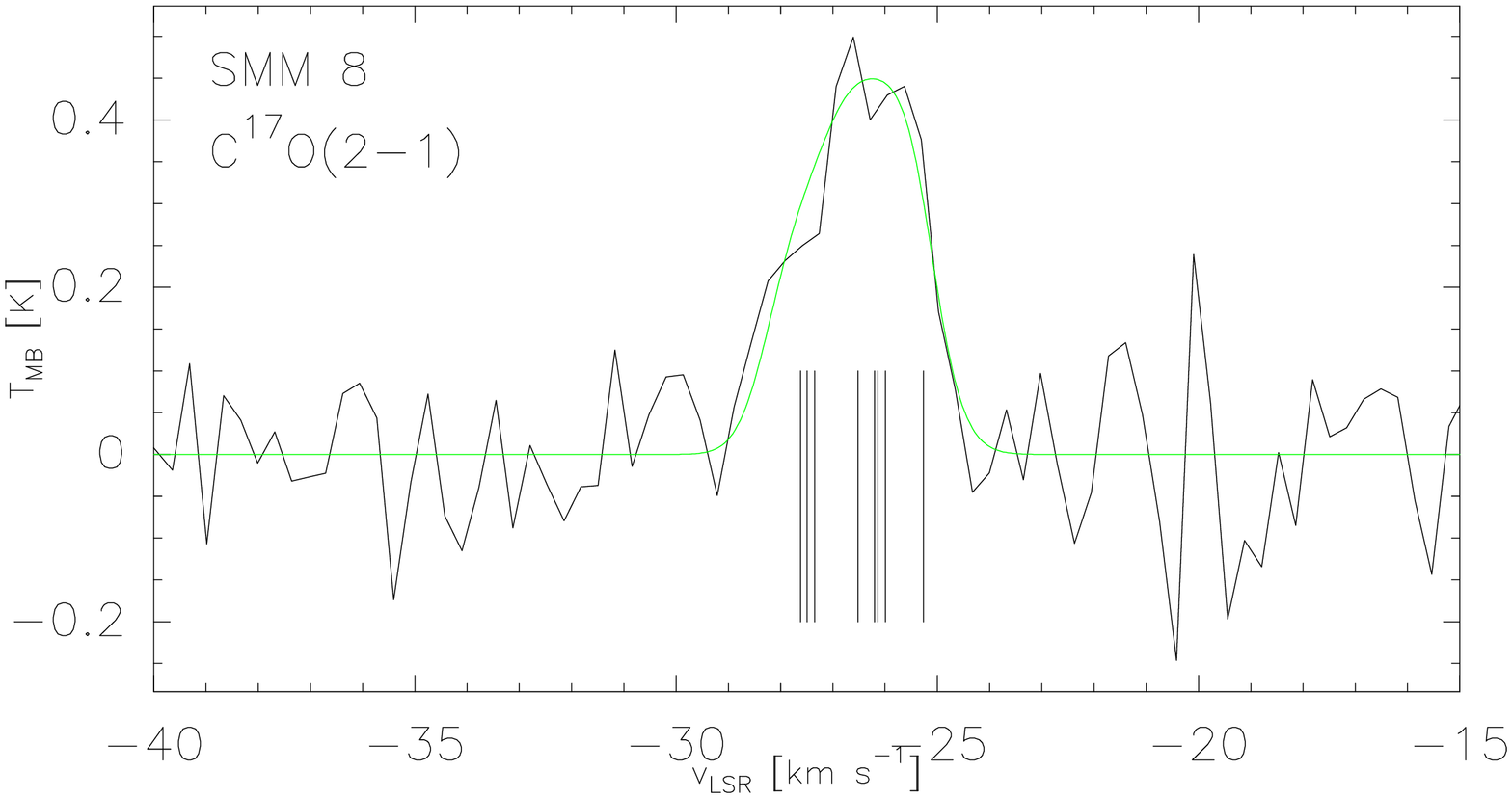}
\includegraphics[width=3.1cm, angle=-90]{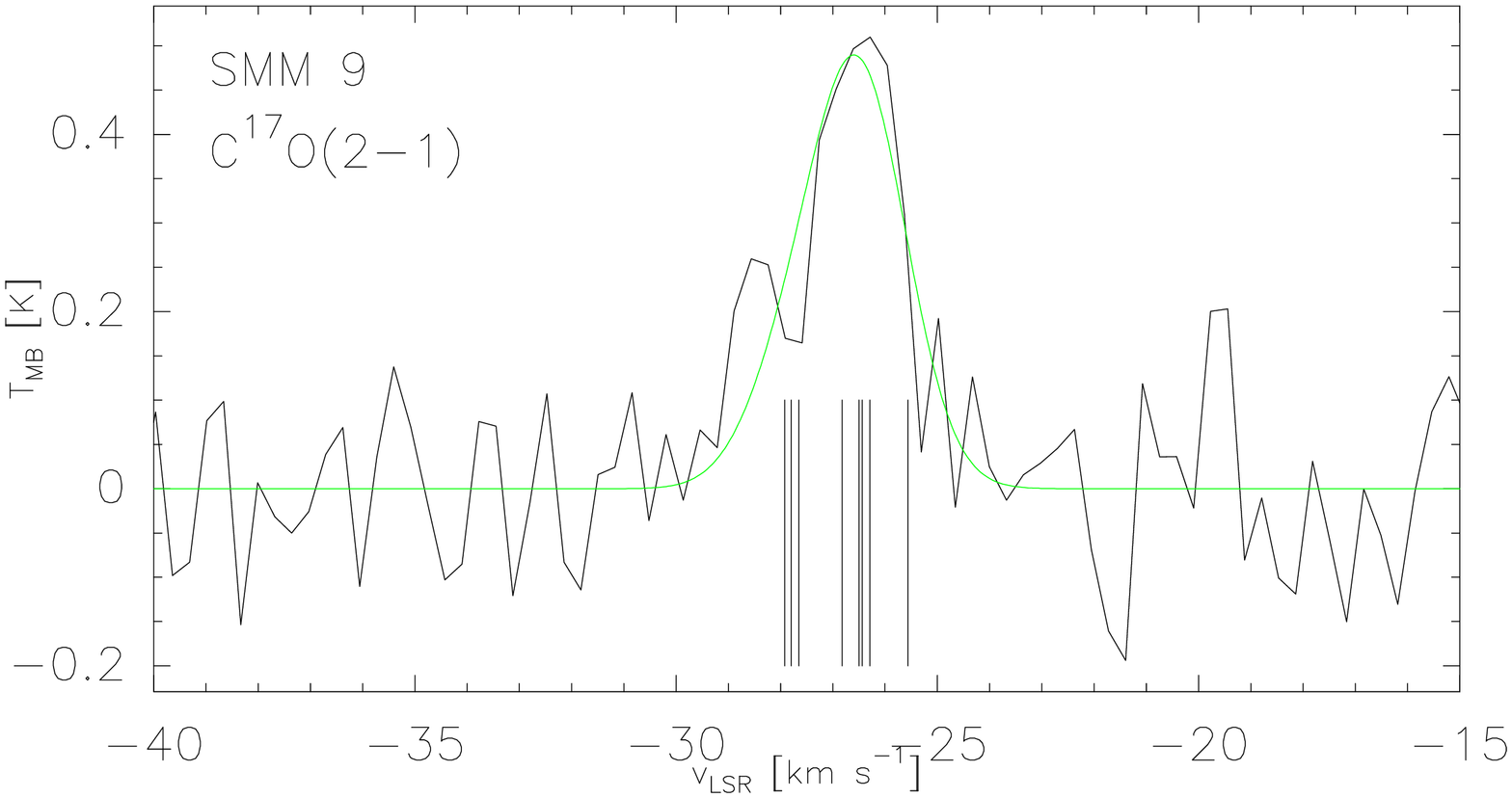}
\includegraphics[width=3.1cm, angle=-90]{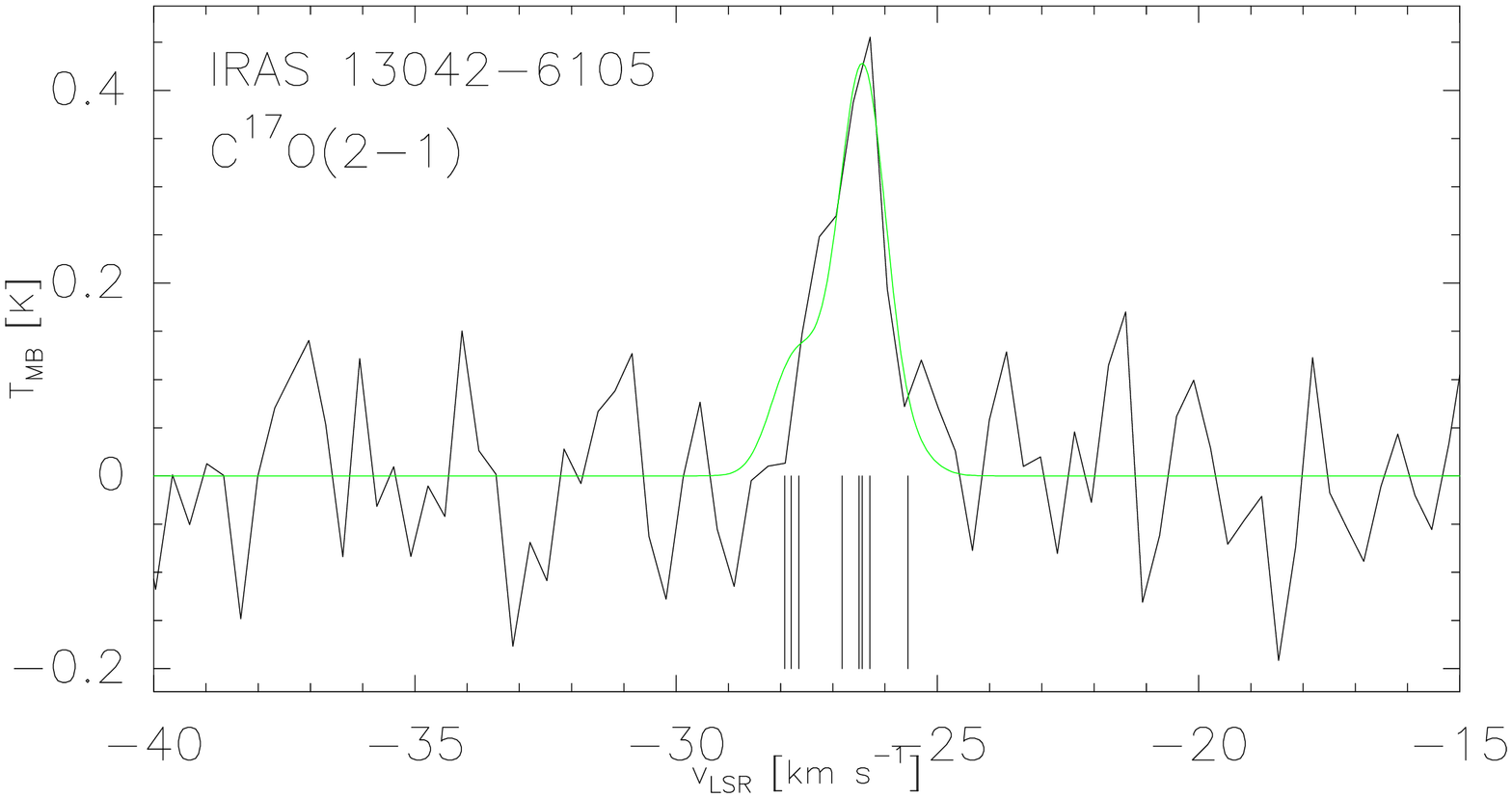}
\caption{Smoothed C$^{17}$O$(2-1)$ spectra. Overlaid on the spectra are the 
hf-structure fits. The relative velocities of individual hf components 
are indicated with vertical lines under the spectra.}
\label{figure:spectra}
\end{center}
\end{figure*}

\begin{figure*}
\begin{center}
\includegraphics[width=3.1cm, angle=-90]{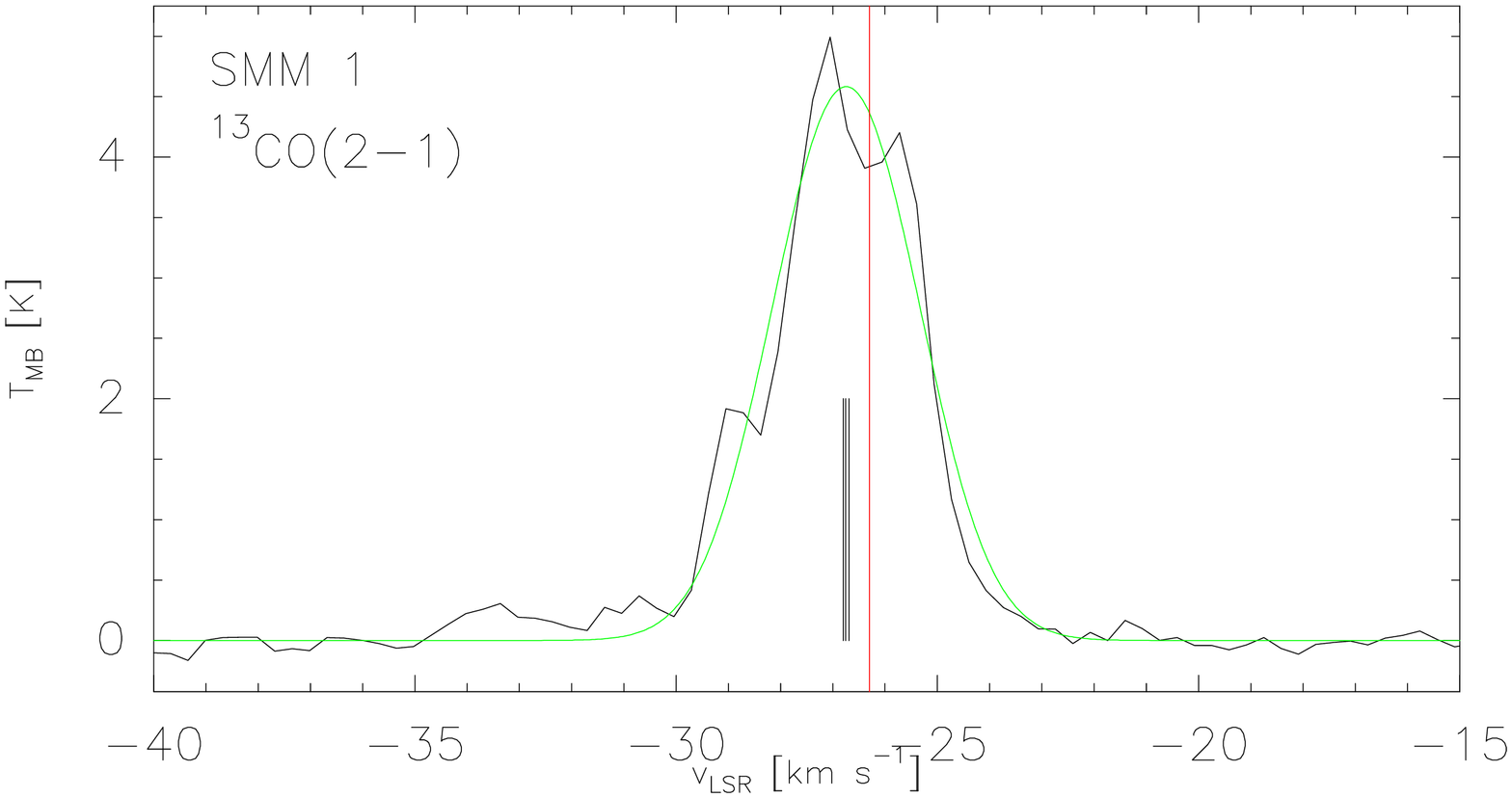}
\includegraphics[width=3.1cm, angle=-90]{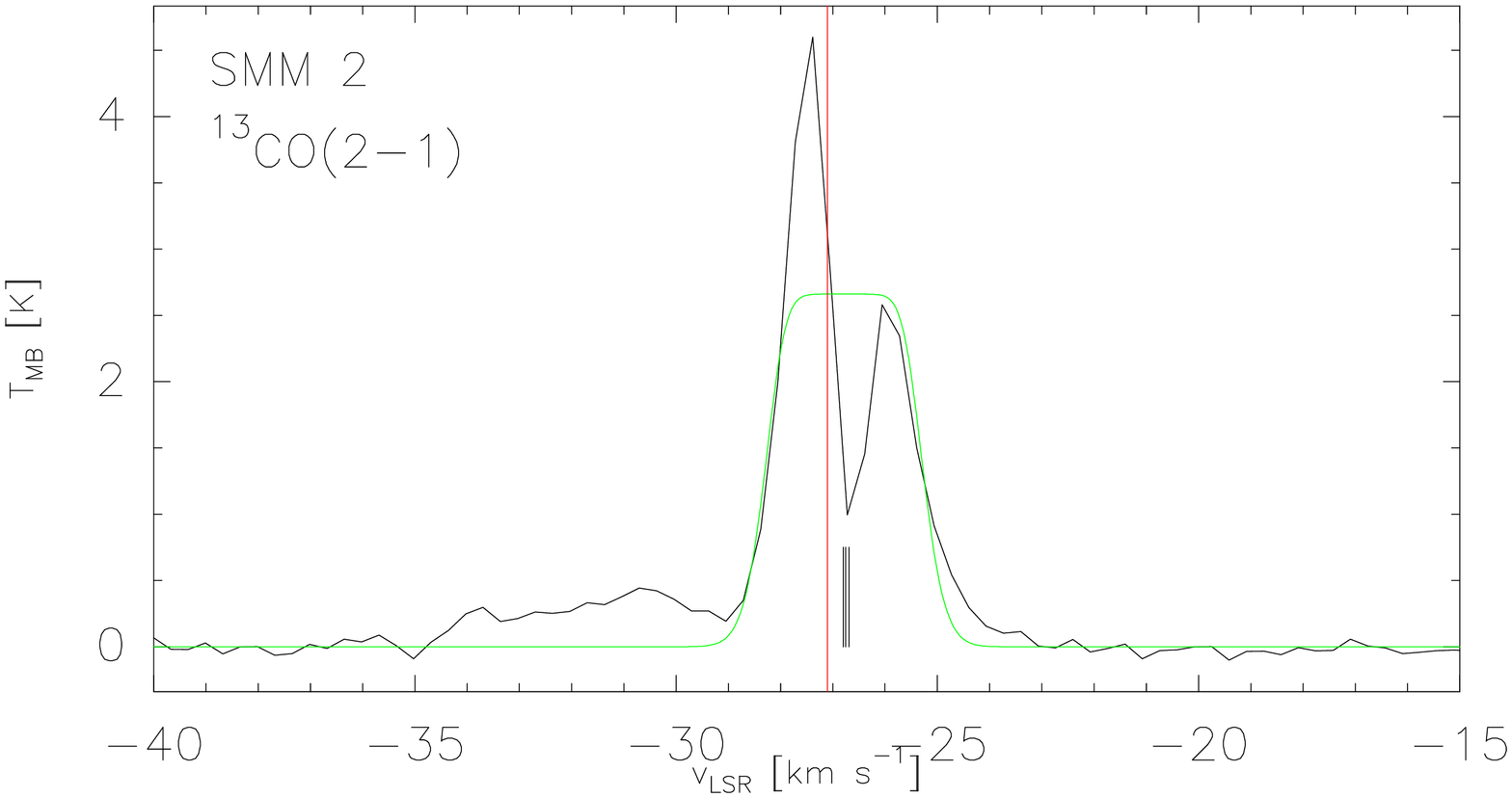}
\includegraphics[width=3.1cm, angle=-90]{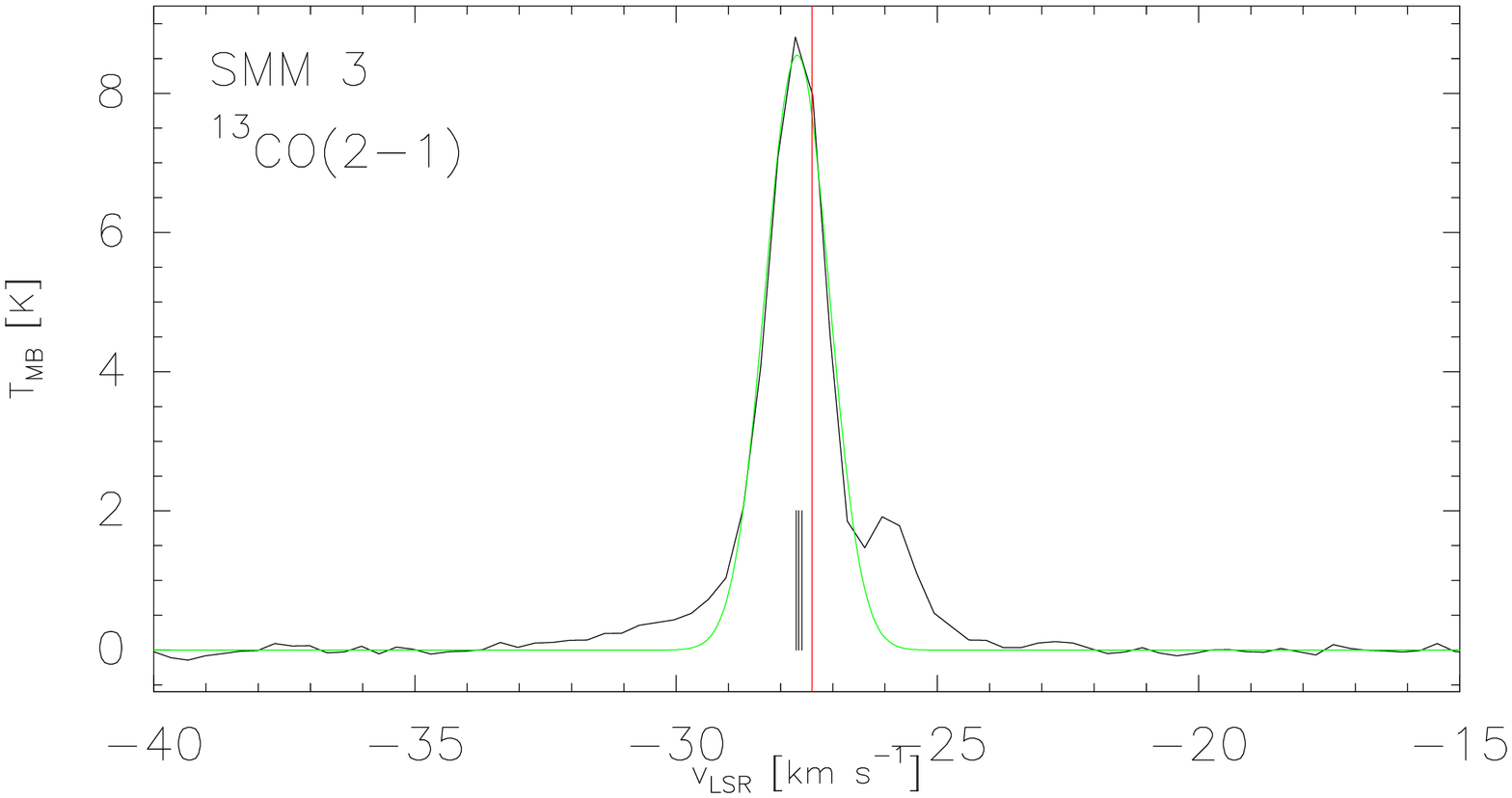}
\includegraphics[width=3.1cm, angle=-90]{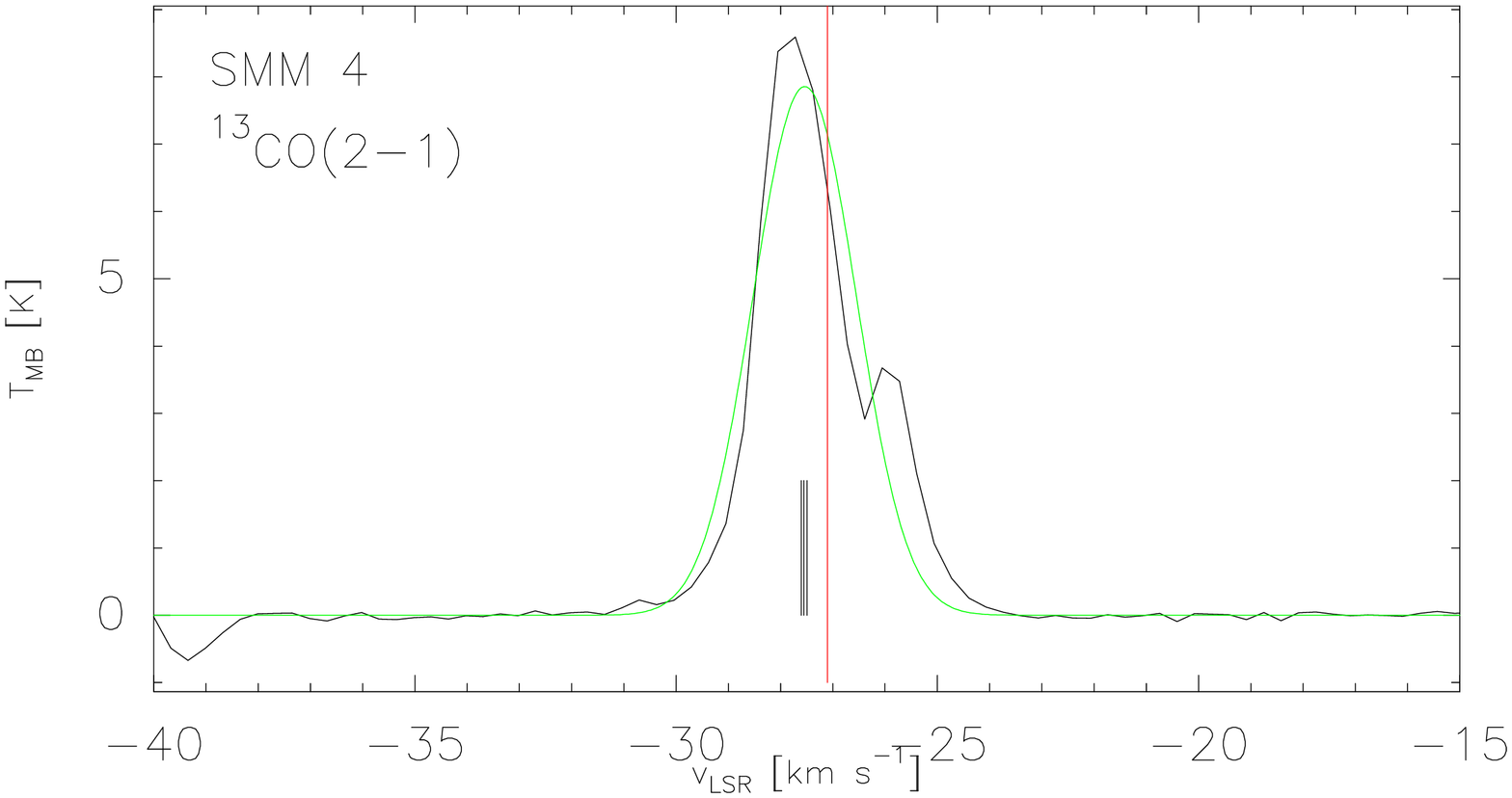}
\includegraphics[width=3.1cm, angle=-90]{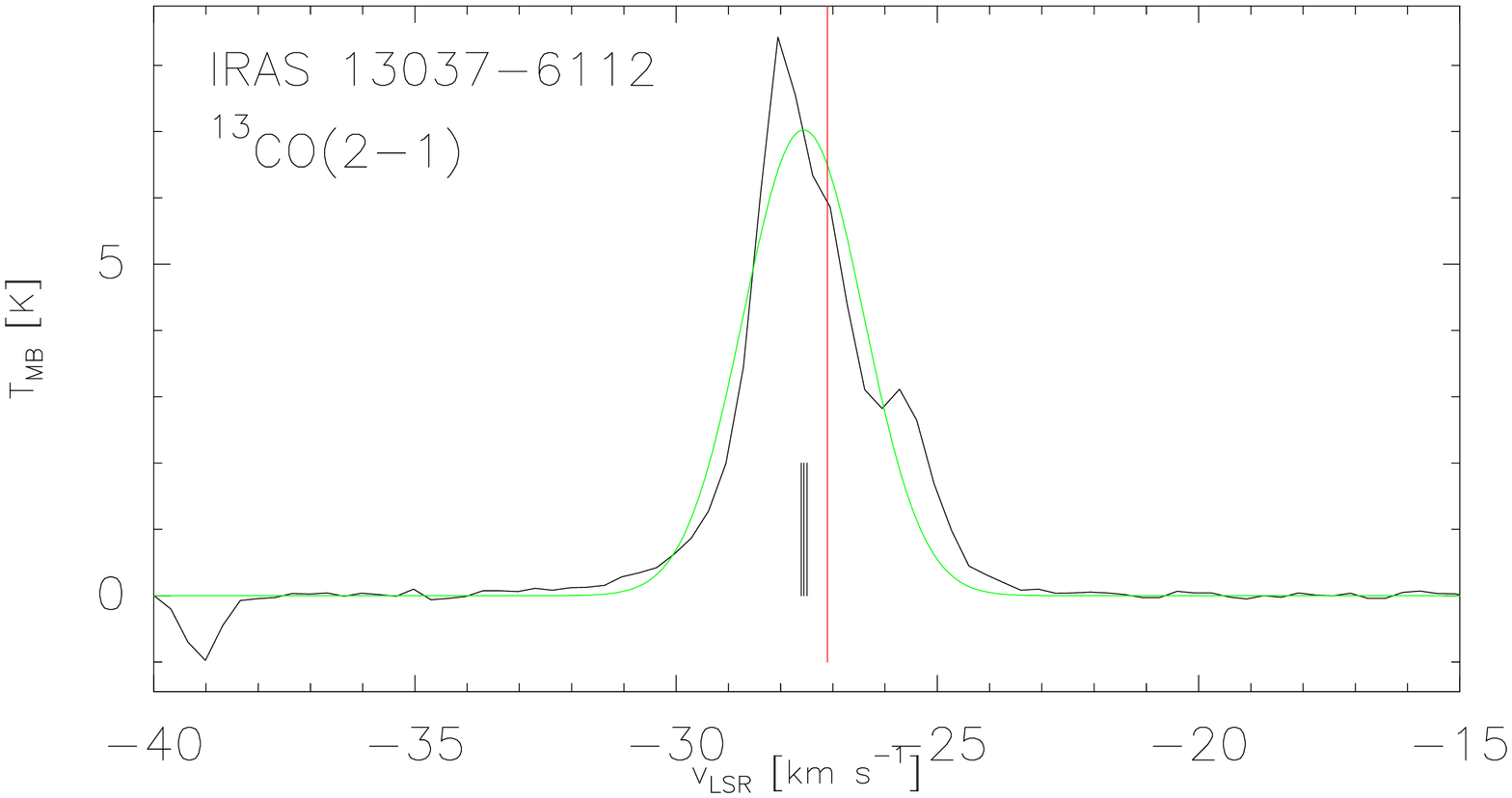}
\includegraphics[width=3.1cm, angle=-90]{blank.eps}
\caption{Smoothed $^{13}$CO$(2-1)$ spectra overlaid with hf-structure 
fits. The relative velocities of individual hf components 
are indicated with vertical lines under the spectra. The red vertical line 
shows the systemic velocity as derived from C$^{17}$O$(2-1)$.}
\label{figure:13CO}
\end{center}
\end{figure*}

\begin{figure*}
\begin{center}
\includegraphics[width=3.1cm, angle=-90]{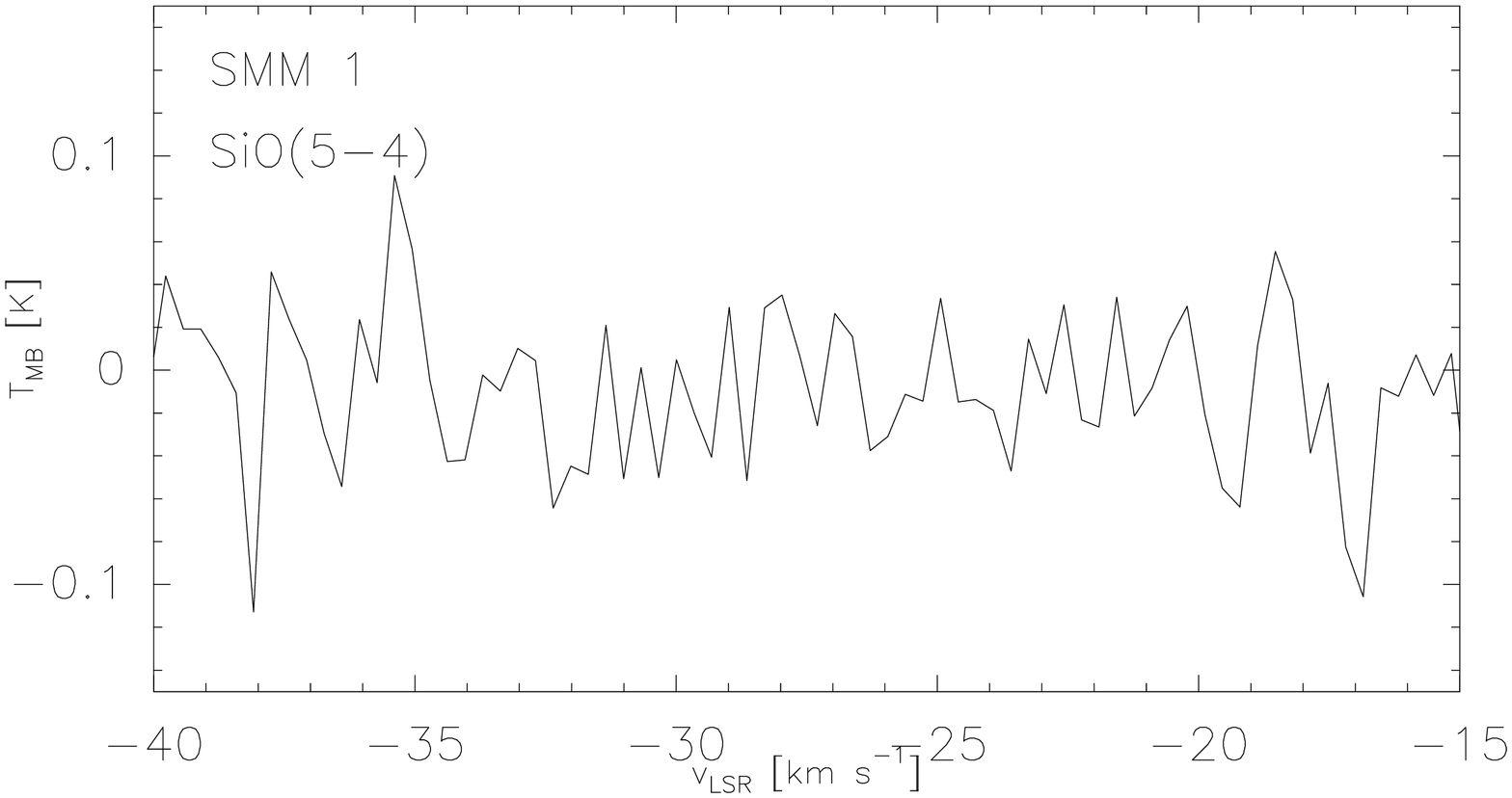}
\includegraphics[width=3.1cm, angle=-90]{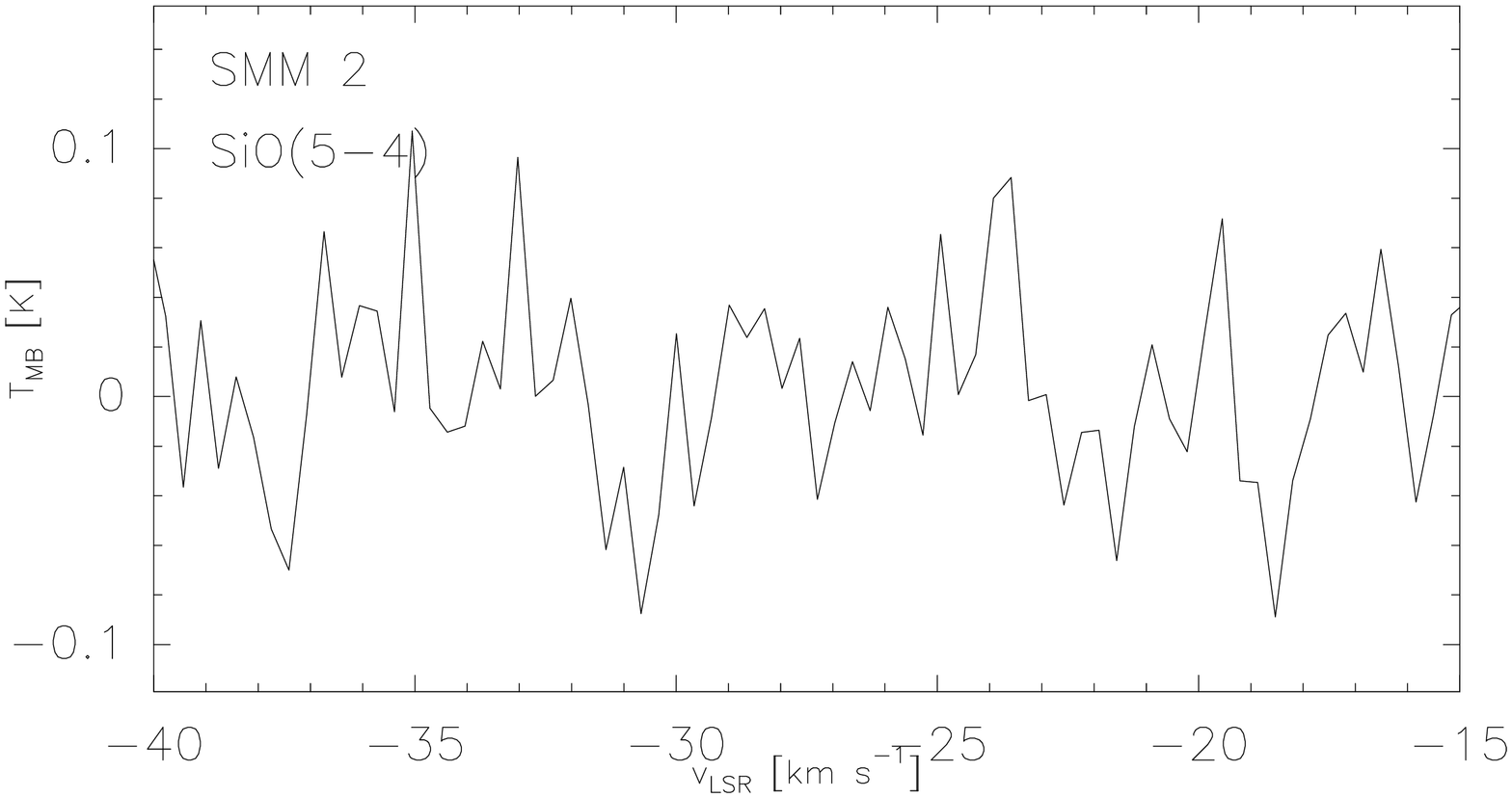}
\includegraphics[width=3.1cm, angle=-90]{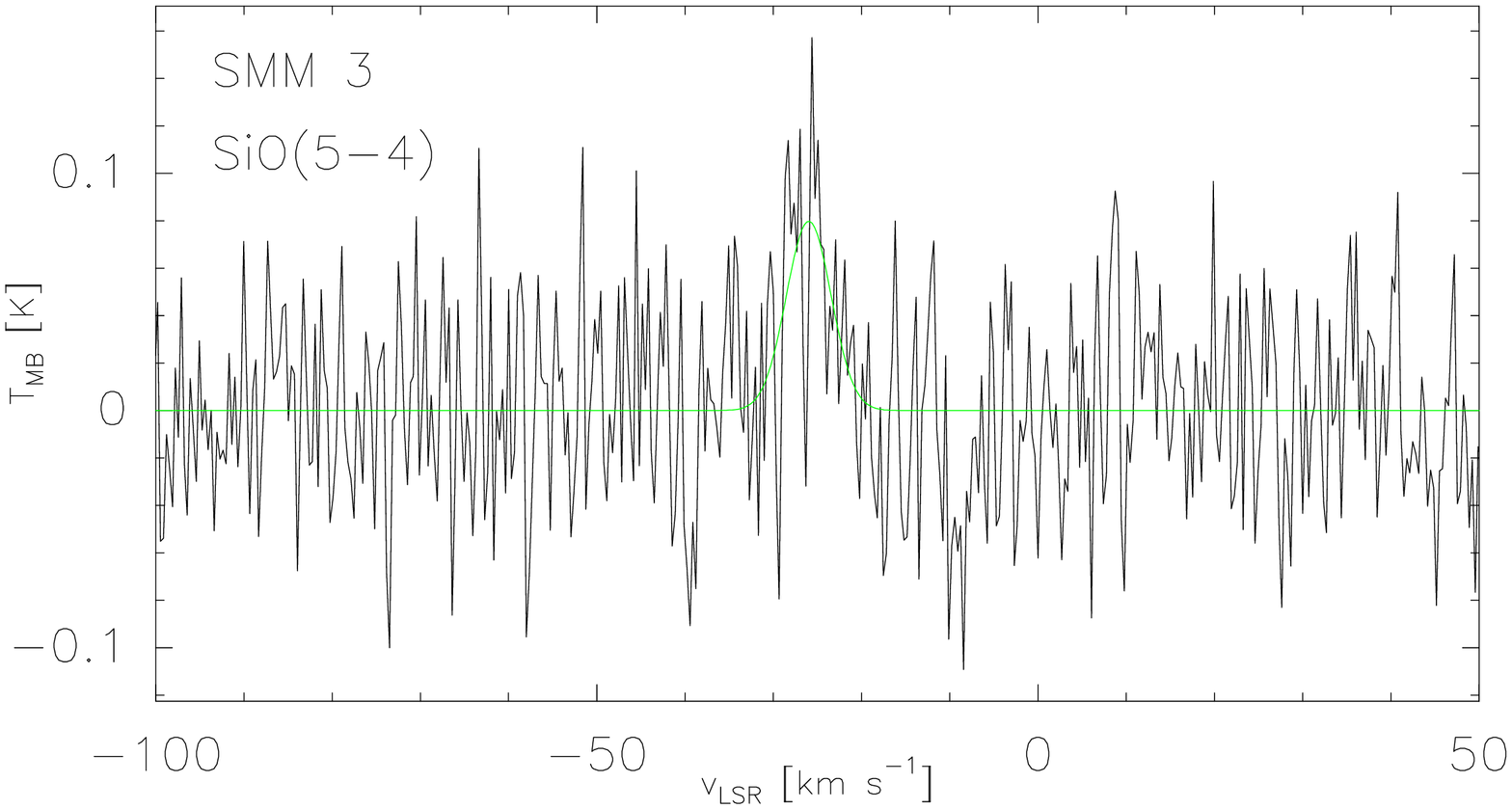}
\includegraphics[width=3.1cm, angle=-90]{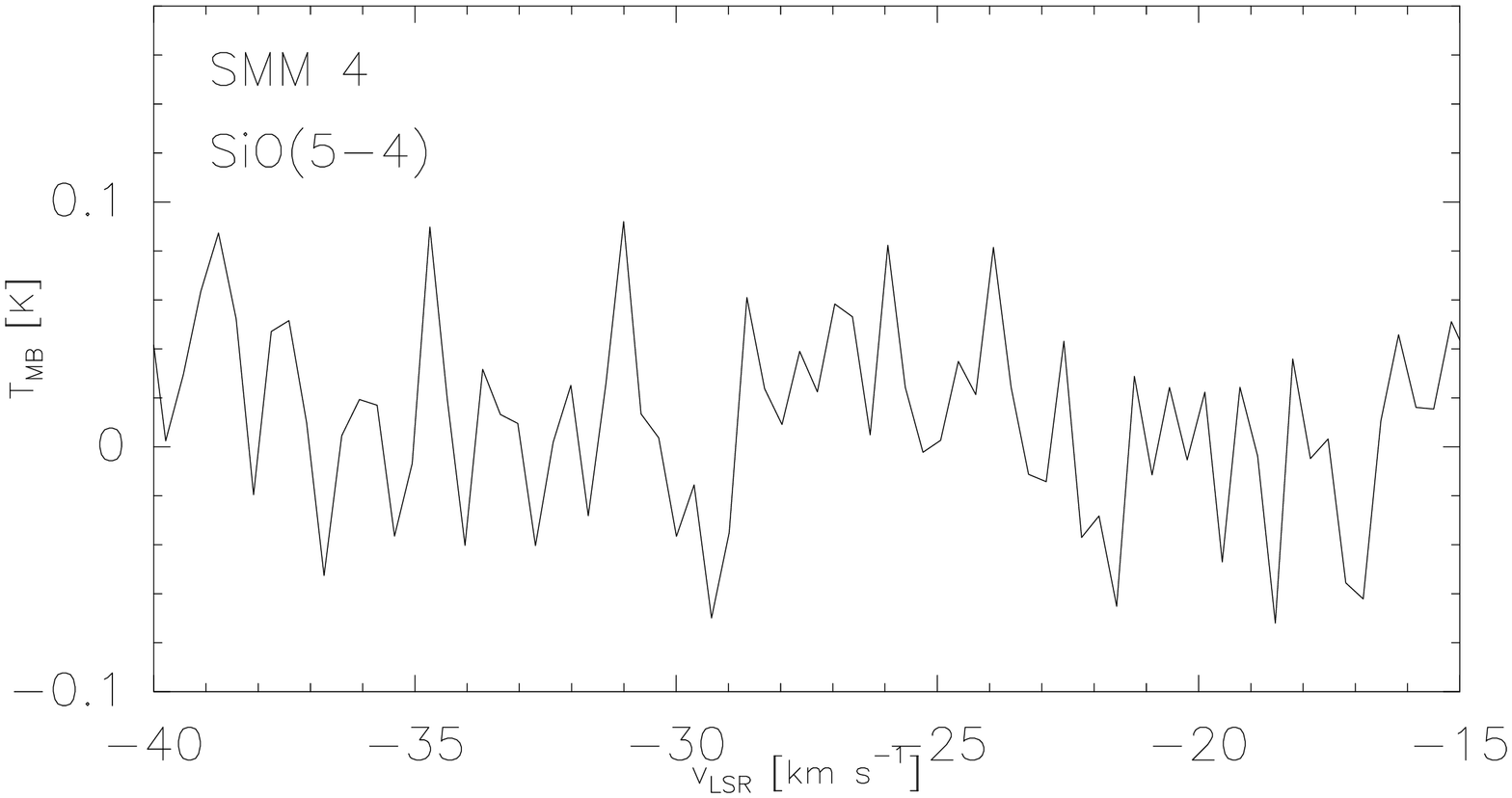}
\includegraphics[width=3.1cm, angle=-90]{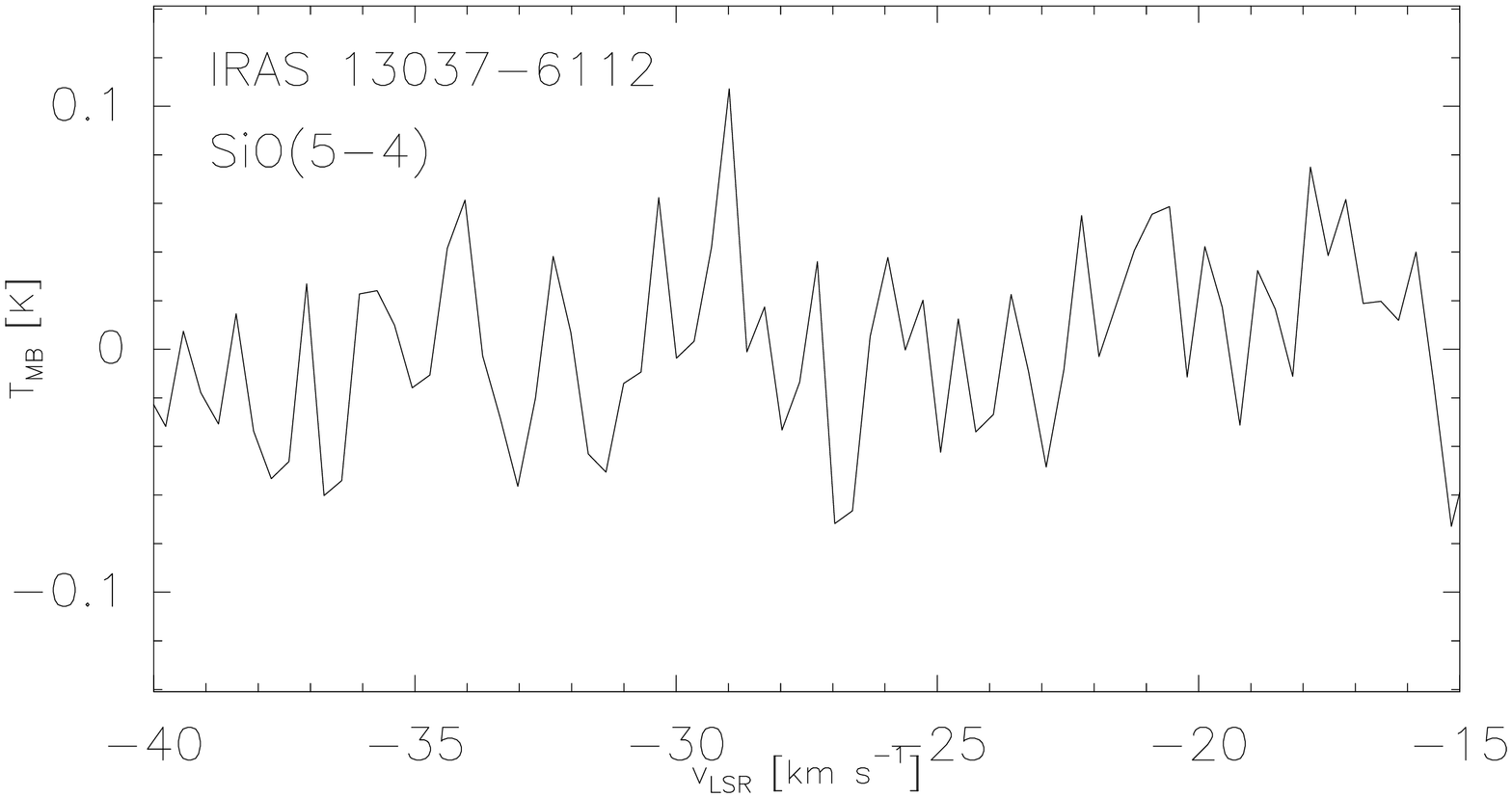}
\includegraphics[width=3.1cm, angle=-90]{blank.eps}
\caption{Smoothed SiO$(5-4)$ spectra. The line was detected only towards SMM 3 
(overlaid with a Gaussian fit). Note that the spectrum towards SMM 3 is shown 
with a wider velocity range for a better illustration.}
\label{figure:SiO}
\end{center}
\end{figure*}

\begin{figure*}
\begin{center}
\includegraphics[width=3.1cm, angle=-90]{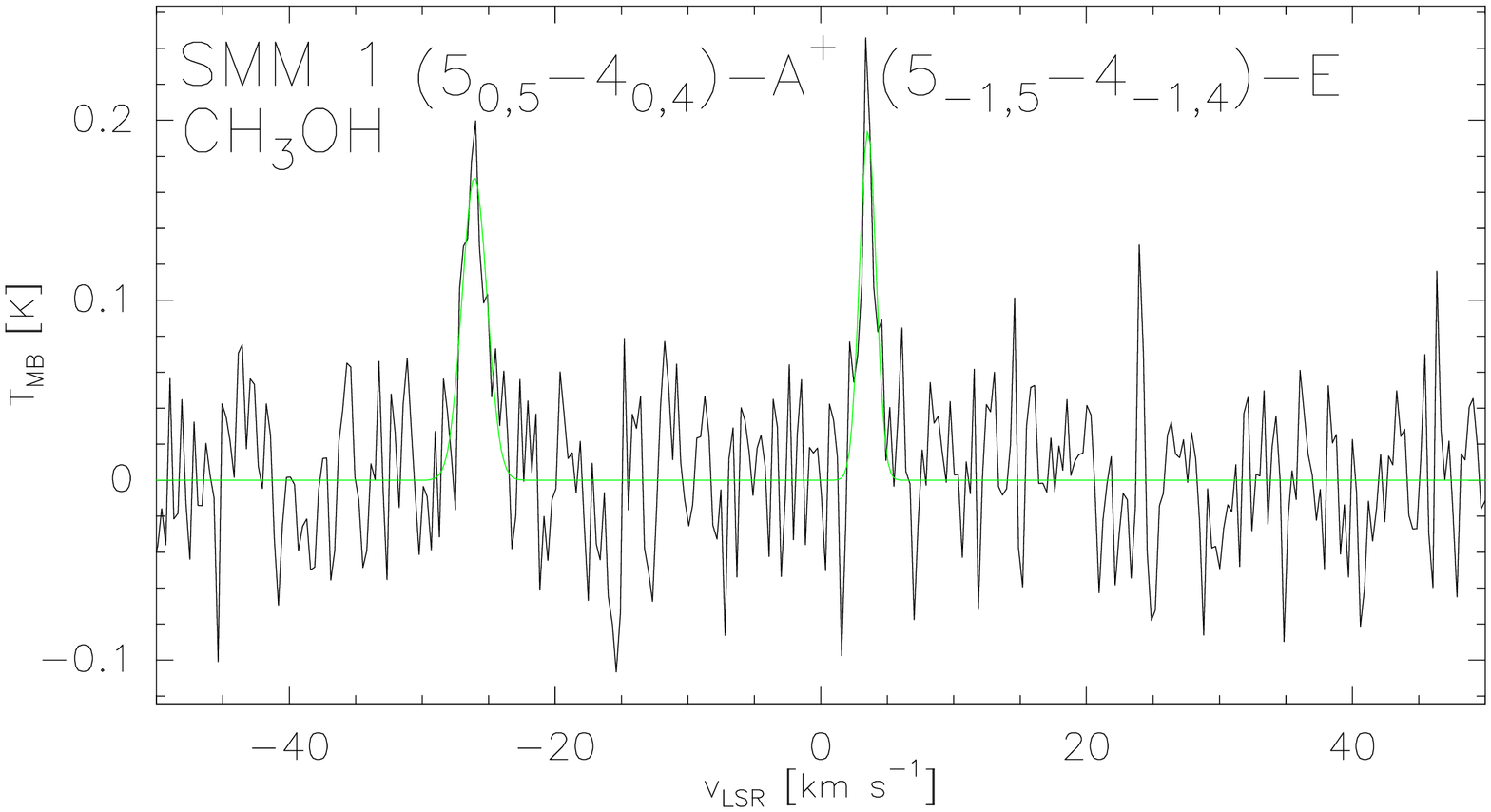}
\includegraphics[width=3.1cm, angle=-90]{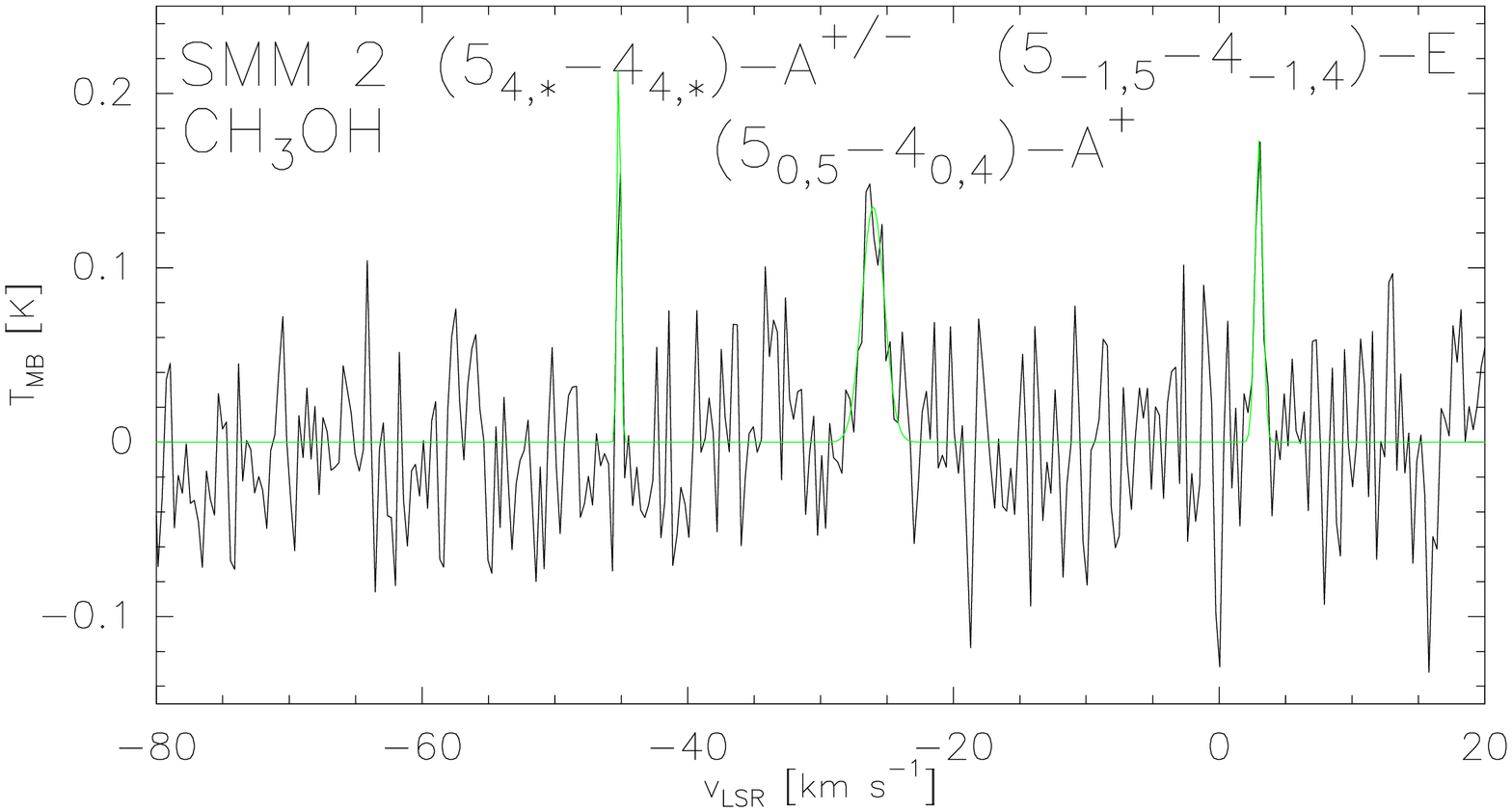}
\includegraphics[width=3.1cm, angle=-90]{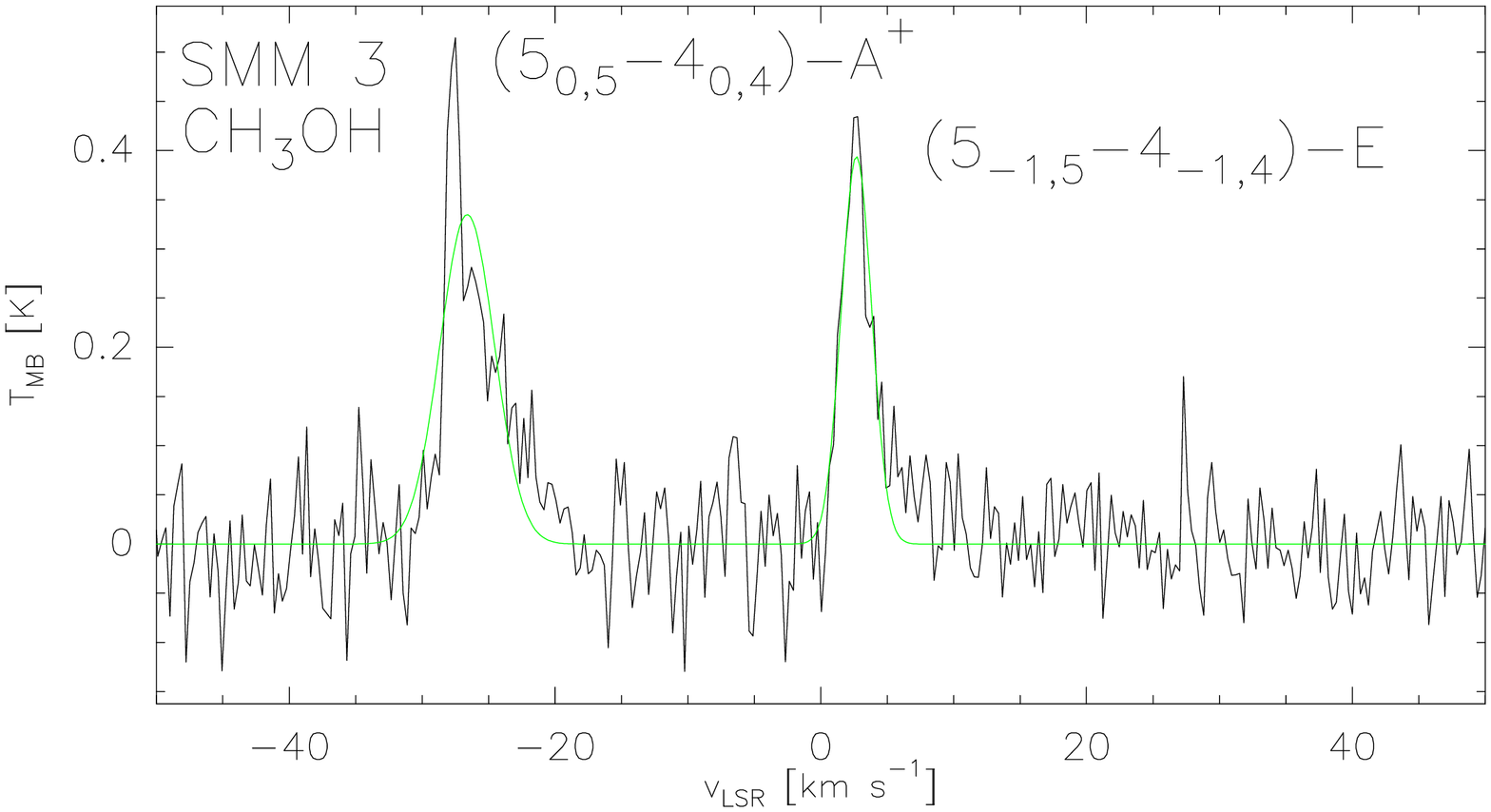}
\includegraphics[width=3.1cm, angle=-90]{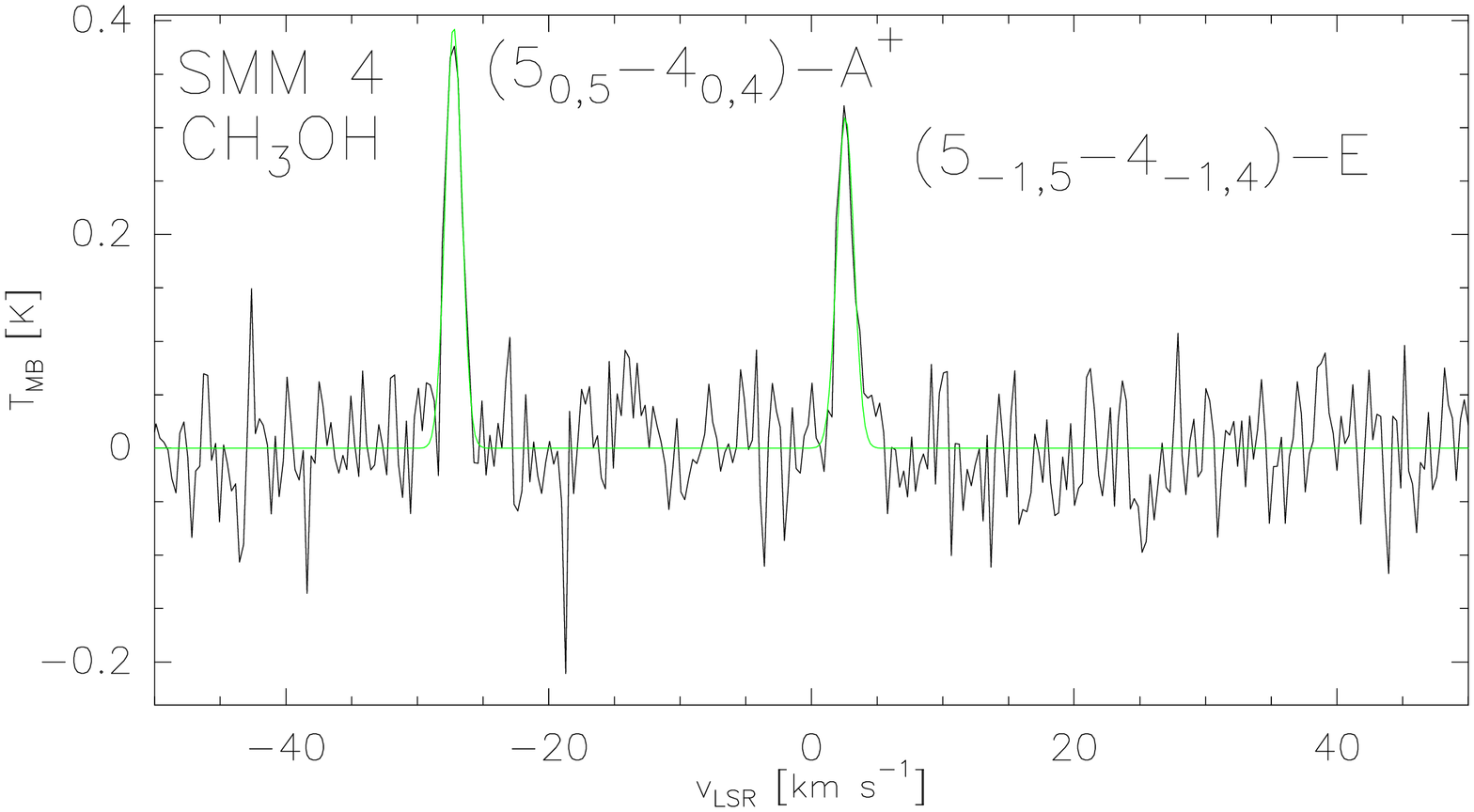}
\includegraphics[width=3.1cm, angle=-90]{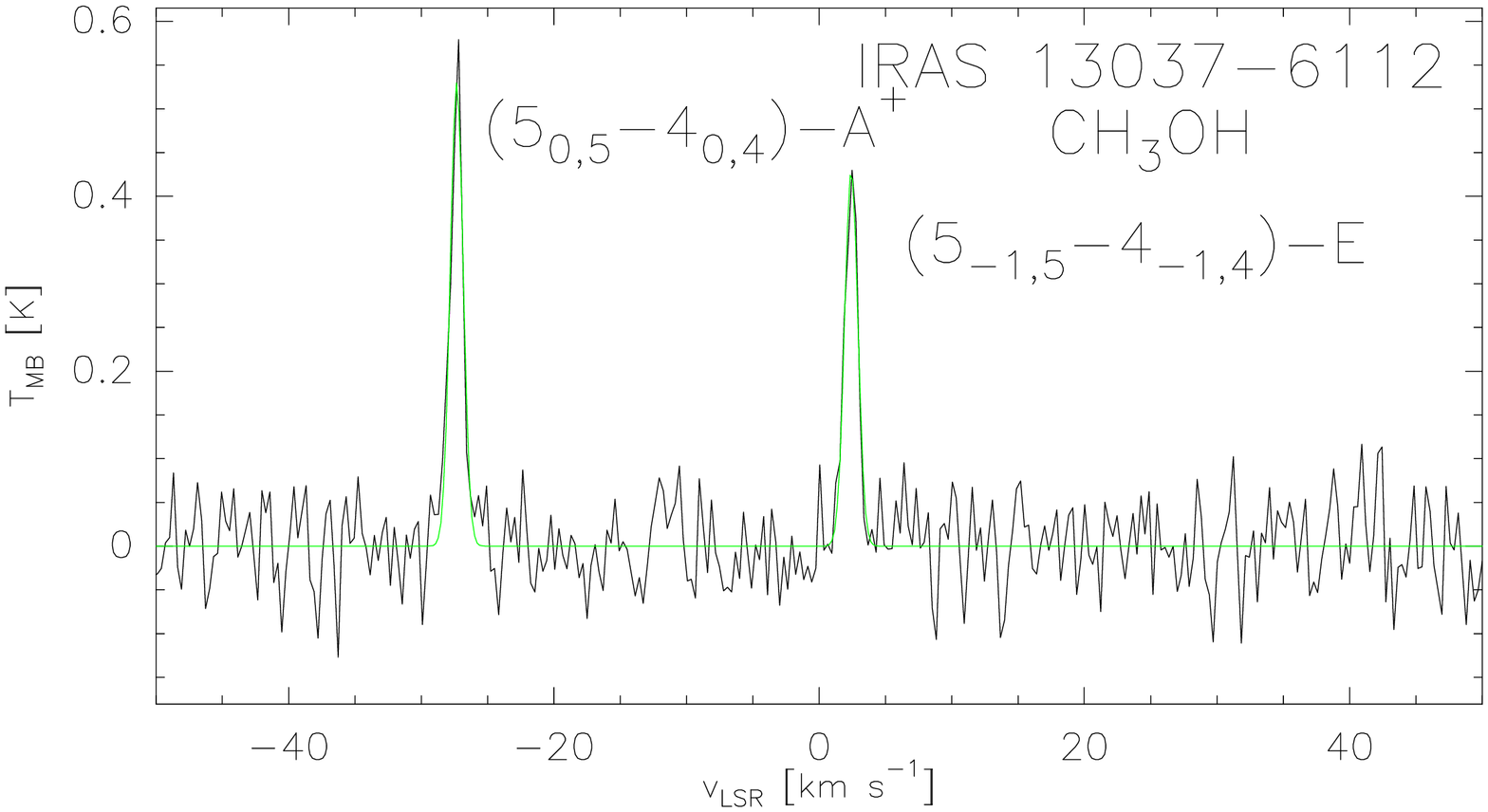}
\includegraphics[width=3.1cm, angle=-90]{blank.eps}
\caption{Smoothed CH$_3$OH$(5_k-4_k)$ spectra overlaid with Gaussian fits. 
In the spectrum towards SMM 2, the CH$_3$OH$(5_{4,1}-4_{4,0})$-A$^+$ and 
CH$_3$OH$(5_{4,2}-4_{4,1})$-A$^-$ transitions are blended. Note that the velocity 
range shown is different for SMM 2 for illustrative purposes.}
\label{figure:CH3OH}
\end{center}
\end{figure*}

\begin{figure}[!h]
\centering
\resizebox{0.75\hsize}{!}{\includegraphics[angle=-90]{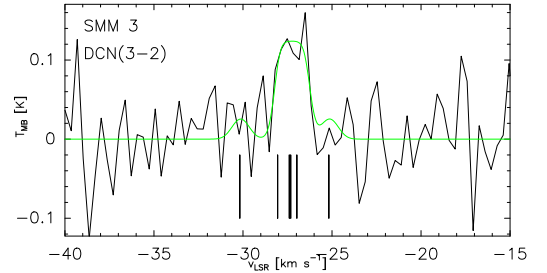}}
\caption{The smoothed DCN$(3-2)$ spectrum towards SMM 3. The line is overlaid 
with a hf-structure fit, and the vertical lines show the relative velocities of 
individual hf components.}
\label{figure:DCN}
\end{figure}

The smoothed C$^{17}$O$(2-1)$ spectra towards all target positions are 
shown in Fig.~\ref{figure:spectra}. The line was clearly detected towards 
all sources, but its hf components are mostly blended in all cases. Towards 
a few sources, such as SMM 2, the hf structure is just partially resolved, but 
not sufficiently well to derive the line optical thickness. 
Figures~\ref{figure:13CO}--\ref{figure:CH3OH} show the $^{13}$CO$(2-1)$, 
SiO$(5-4)$, and CH$_3$OH$(5_k-4_k)$ spectra towards the observed five clumps 
in the southern part of the filament. All the $^{13}$CO$(2-1)$ lines 
show a blue asymmetric profile, i.e., the blueshifted peak is stronger than the 
redshifted peak, and there is a self-absorption dip near the systemic 
velocity. The self-absorbed profiles of the $^{13}$CO lines indicate 
that the lines are optically thick. Moreover, the lines show non-Gaussian 
line-wing emission. In particular, SMM 2 shows high-velocity blueshifted wing 
emission up to $-35$ km~s$^{-1}$. The small ``absorption''-like features seen 
in the $^{13}$CO spectra of SMM 4 and IRAS13037 are caused by weak emission 
in the off-source reference position (off-beam). SiO$(5-4)$ was detected only 
towards SMM 3, though the line is also very weak towards this source. 
We additionally detected the DCN$(3-2)$ line towards SMM 3 in the 
frequency band covering SiO$(5-4)$ (see Fig.~\ref{figure:DCN}). The detected 
$J=3-2$ rotational line of DCN is split into six hf components. In fitting 
this hf structure, we used the rest frequencies from the CDMS catalogue.  
The E-type methanol transition CH$_3$OH$(5_{-1,5}-4_{-1,4})$-E, and the A-type 
transition CH$_3$OH$(5_{0,5}-4_{0,4})$-A$^+$ at $\sim241.8$ GHz were detected 
towards all the observed clumps. Moreover, towards SMM 2, a blend of 
CH$_3$OH$(5_{4,1}-4_{4,0})$-A$^+$ and CH$_3$OH$(5_{4,2}-4_{4,1})$-A$^-$ was 
detected.

The spectral-line parameters derived from hf- and single Gaussian fits are 
given in Col.~(3)--(5) of Table~\ref{table:lineparameters}. For SiO$(5-4)$ 
non-detections, the $3\sigma$ upper limit on the line intensity is given. 
The integrated line intensities listed in Col.~(6) of 
Table~\ref{table:lineparameters} were computed over the velocity range given 
in square brackets in the corresponding column.  
The quoted uncertainties in ${\rm v}_{\rm LSR}$ and $\Delta {\rm v}$ are 
formal $1\sigma$ fitting errors, whereas those in $T_{\rm MB}$ and 
$\int T_{\rm MB} {\rm dv}$ also include the 10\% calibration uncertainty. In 
the last column we give the estimated peak line optical thickness, which will 
be derived in Sect.~4.3/Appendix A.3.

\begin{table*}
\caption{Spectral-line parameters.}
{\scriptsize
\begin{minipage}{2\columnwidth}
\centering
\renewcommand{\footnoterule}{}
\label{table:lineparameters}
\begin{tabular}{c c c c c c c}
\hline\hline 
Source & Transition & ${\rm v}_{\rm LSR}$ & $\Delta {\rm v}$ & $T_{\rm MB}$\tablefootmark{a} & $\int T_{\rm MB} {\rm dv}$\tablefootmark{b} & $\tau_0$\\
     & & [km~s$^{-1}$] & [km~s$^{-1}$] & [K] & [K~km~s$^{-1}$] &\\
\hline
SMM 1 & SiO$(5-4)$ & \ldots & \ldots & $<0.13$ & \ldots & \ldots \\
      & $^{13}$CO$(2-1)$ & $-26.8\pm0.01$ & $3.18\pm0.01$ & $4.92\pm0.63$(B)/$4.16\pm0.57$(R) & $15.78\pm1.59$ [-30.13, -22.65] &  \ldots \\
      & C$^{17}$O$(2-1)$ & $-26.3\pm0.1$ & $0.68\pm0.13$ & $0.40\pm0.08$ & $0.87\pm0.14$ [-28.88, -25.37] & $0.04\pm0.02$\\
      & CH$_3$OH$(5_{-1,5}-4_{-1,4})$-E & $-26.5\pm0.1$ & $1.47\pm0.27$ & $0.19\pm0.04$ & $0.35\pm0.05$ [-28.24, -24.57] & $0.01\pm0.001-0.04\pm0.01$\\
      & CH$_3$OH$(5_{0,5}-4_{0,4})$-A$^+$ & $-26.1\pm0.1$ & $2.22\pm0.37$ & $0.17\pm0.03$ & $0.39\pm0.06$ [-27.57, -23.23] & $0.01\pm0.001-0.03\pm0.01$\\
SMM 2 & SiO$(5-4)$ & \ldots & \ldots & $<0.12$ & \ldots & \ldots\\
      & $^{13}$CO$(2-1)$ & $-26.8\pm0.03$ & $1.53\pm0.08$ & $4.57\pm0.51$(B)/$2.51\pm0.4$(R) & $8.47\pm0.87$ [-28.96, -23.11] &  \ldots \\
      & C$^{17}$O$(2-1)$ & $-27.1\pm0.1$ & $0.59\pm0.19$ & $0.53\pm0.10$ & $0.63\pm0.10$ [-29.05, -25.66] & $0.05\pm0.03$\\
      & CH$_3$OH$(5_{-1,5}-4_{-1,4})$-E & $-27.0\pm0.1$ & $0.67\pm0.23$ & $0.17\pm0.03$ & $0.14\pm0.03$ [-28.31, -26.10] & $0.01\pm0.001-0.03\pm0.01$\\
      & CH$_3$OH$(5_{0,5}-4_{0,4})$-A$^+$ & $-26.0\pm0.2$ & $1.98\pm0.46$ & $0.14\pm0.02$ & $0.30\pm0.06$ [-28.13, -23.53] & $0.01\pm0.01-0.03\pm0.01$  \\
      & CH$_3$OH$(5_{4,*}-4_{4,*})$-A$^{+/-}$\tablefootmark{c} & $-26.5\pm0.04$ & $0.30\pm0.80$ & $0.13\pm0.02$ & $0.17\pm0.03$ [-27.02, -25.55] & \ldots \\
SMM 3 & SiO$(5-4)$ & $-26.0\pm0.5$ & $5.99\pm1.06$ & $0.08\pm0.05$ & $0.49\pm0.09$ [-32.02, -19.99] & $0.003\pm0.001-0.07\pm0.01$ \\
      & $^{13}$CO$(2-1)$ & $-27.7\pm0.01$ & $1.41\pm0.03$ & $8.81\pm0.93$(B)/$1.91\pm0.35$(R) & $14.69\pm1.48$ [-29.26, -24.91] & \ldots \\
      & C$^{17}$O$(2-1)$ & $-27.4\pm0.1$ & $0.84\pm0.35$ & $0.52\pm0.11$ & $0.91\pm0.15$ [-29.55, -25.75] & $0.05\pm0.03$\\
      & CH$_3$OH$(5_{-1,5}-4_{-1,4})$-E & $-27.3\pm0.1$ & $2.70\pm0.21$ & $0.39\pm0.09$ & $1.66\pm0.18$ [-29.75, -21.73] & $0.02\pm0.01-0.08\pm0.01$\\
      & CH$_3$OH$(5_{0,5}-4_{0,4})$-A$^+$ & $-26.6\pm0.2$ & $4.82\pm0.60$ & $0.51\pm0.09$ & $1.85\pm0.24$ [-30.75, -18.38] & $0.03\pm0.01-0.10\pm0.01$\\
      & DCN$(3-2)$ & $-27.4\pm0.1$ & $1.09\pm0.27$ & $0.13\pm0.04$ & $0.19\pm0.05$ [-28.76, -26.00; 92.6\%] & $0.01\pm0.001-0.11\pm0.01$\\
SMM 4 & SiO$(5-4)$ & \ldots & \ldots & $<0.11$ & \ldots & \ldots \\
      & $^{13}$CO$(2-1)$ & $-27.6\pm0.01$ & $2.27\pm0.03$ & $8.94\pm0.93$(B)/$3.32\pm0.42$(R) & $20.04\pm2.01$ [-30.13, -23.91] & \ldots \\
      & C$^{17}$O$(2-1)$ & $-27.1\pm0.03$ & $1.20\pm0.15$ & $1.33\pm0.18$ & $2.80\pm0.30$ [-30.55, -25.54] & $0.14\pm0.07$\\
      & CH$_3$OH$(5_{-1,5}-4_{-1,4})$-E & $-27.4\pm0.1$ & $1.53\pm0.18$ & $0.31\pm0.04$ & $0.61\pm0.08$ [-28.91, -24.40] & $0.02\pm0.001-0.06\pm0.01$\\ 
      & CH$_3$OH$(5_{0,5}-4_{0,4})$-A$^+$ & $-27.2\pm0.1$ & $1.47\pm0.13$ & $0.40\pm0.05$ & $0.60\pm0.08$ [-28.74, -25.07] & $0.02\pm0.01-0.08\pm0.01$\\
IRAS 13037-6112 & SiO$(5-4)$ & \ldots & \ldots & $<0.12$ & \ldots & \ldots \\
                & $^{13}$CO$(2-1)$ & $-27.6\pm0.01$ & $2.64\pm0.04$ & $7.86\pm0.87$(B)/$2.73\pm0.46$(R) & $20.66\pm2.08$ [-30.34, -23.78] & \ldots \\
                & C$^{17}$O$(2-1)$ & $-27.1\pm0.02$ & $1.10\pm0.08$ & $1.91\pm0.27$ & $3.99\pm0.41$ [-30.30, -25.29] & $0.12\pm0.01$ \\
                & CH$_3$OH$(5_{-1,5}-4_{-1,4})$-E & $-27.3\pm0.03$ & $1.15\pm0.09$ & $0.53\pm0.07$ & $0.73\pm0.08$ [-29.41, -24.73] & $0.03\pm0.01-0.11\pm0.01$\\ 
                & CH$_3$OH$(5_{0,5}-4_{0,4})$-A$^+$ & $-27.6\pm0.04$ & $1.17\pm0.10$ & $0.43\pm0.06$ & $0.70\pm0.08$ [-29.41, -25.74] & $0.02\pm0.01-0.09\pm0.01$\\
SMM 5 & C$^{17}$O$(2-1)$ & $-26.4\pm0.04$ & $0.77\pm0.11$ & $0.81\pm0.16$ & $1.40\pm0.17$ [-28.59, -25.33] & $0.08\pm0.04$ \\
SMM 6 & C$^{17}$O$(2-1)$ & $-27.0\pm0.03$ & $1.66\pm0.08$ & $1.56\pm0.20$ & $3.64\pm0.38$ [-30.01, -24.66] & $0.17\pm0.09$ \\
SMM 7 & C$^{17}$O$(2-1)$ & $-26.3\pm0.04$ & $1.36\pm0.10$ & $1.21\pm0.16$ & $2.46\pm0.27$ [-28.96, -24.45] & $0.13\pm0.07$ \\
IRAS 13039-6108 & C$^{17}$O$(2-1)$ & $-26.1\pm0.04$ & $1.14\pm0.07$ & $1.12\pm0.14$ & $1.92\pm0.21$ [-28.63, -24.66] & $0.07\pm0.01$ \\
SMM 8 & C$^{17}$O$(2-1)$ & $-26.2\pm0.1$ & $1.42\pm0.33$ & $0.47\pm0.08$ & $1.28\pm0.18$ [-29.17, -24.33] & $0.05\pm0.02$ \\
SMM 9 & C$^{17}$O$(2-1)$ & $-26.5\pm0.1$ & $2.01\pm0.46$ & $0.46\pm0.10$ & $1.22\pm0.19$ [-29.26, -25.24] & $0.05\pm0.02$ \\
IRAS 13042-6105 & C$^{17}$O$(2-1)$ & $-26.5\pm0.1$ & $1.00\pm0.24$ & $0.40\pm0.07$ & $0.65\pm0.11$ [-28.21, -24.66] & $0.04\pm0.02$\\
\hline 
\end{tabular} 
\tablefoot{Columns~(3)--(8) of this table are as follows: (3) LSR velocity; 
(4) FWHM linewidth; (5) peak line intensity; (6) integrated intensity; (7) 
peak optical thickness (see Sect.~4.3/Appendix A.3).\tablefoottext{a}{For 
the $^{13}$CO lines the intensities of the blue (B) and red (R) peak are 
given.}\tablefoottext{b}{Integrated intensity 
is derived by integrating over the velocity range indicated in square 
brackets. Note that the high-velocity blueshifted wing in the $^{13}$CO 
spectrum towards SMM 2 was excluded from the integrated intensity.
For DCN$(3-2)$, the percentage 92.6\% in brackets indicates the 
contribution of hf component's intensity lying within the detected 
line.}\tablefoottext{c}{A blend of CH$_3$OH$(5_{4,1}-4_{4,0})$-A$^+$ and 
CH$_3$OH$(5_{4,2}-4_{4,1})$-A$^-$.}}
\end{minipage} }
\end{table*}

\section{Analysis and results}

\subsection{Kinematic distance of G304.74+01.32}

The LSR velocities derived from C$^{17}$O$(2-1)$ lie in the range 
${\rm v}_{\rm LSR}\in [-27.4,\, -26.1]$ km~s$^{-1}$, i.e., within 1.3 km~s$^{-1}$. 
Such velocity coherence shows that the filamentary IRDC G304.74 is a 
coherent structure, and not just a group of physically independent objects 
seen in projection in the plane of the sky. 

The mean and median radial velocities derived from C$^{17}$O$(2-1)$ are 
$-26.7$ and $-26.5$ km~s$^{-1}$, respectively. To revise the cloud kinematic 
distance, we adopt the mean velocity of $-26.7$ km~s$^{-1}$, and 
employ the recent rotation curve of Reid et al. (2009). The resulting near 
kinematic distance, as expected to be appropriate for an IRDC seen in 
absorption, is $d=2.54\pm0.66$ kpc. The corresponding galactocentric distance 
is about $R_{\rm GC}\simeq 7.26$ kpc (see Appendix A.1 for further details).  

\subsection{Revision of clump properties presented in Paper I}

We used the revised cloud distance $2.54\pm0.66$ kpc to 
recalculate the distance-dependent clump parameters presented in Paper I, 
where the value $d=2.4$ kpc was adopted. 
These include the clump radius ($R \propto d$), mass ($M \propto d^2$), and 
volume-averaged H$_2$ number density 
[$\langle n({\rm H_2}) \rangle \propto M/R^3 \propto d^{-1}$]; see 
Eqs.~(6)--(8) in Paper I. The results are presented in Cols.~(4)--(7) of 
Table~\ref{table:sources}. The revised values are in the range 
$R\sim0.3-0.5$ pc, $M\sim40-250$ M$_{\sun}$, $N({\rm H_2})\sim1-3\times10^{22}$ 
cm$^{-2}$, and $\langle n({\rm H_2}) \rangle \sim 0.3-1\times10^4$ cm$^{-3}$. 
For more details, see Appendix A.2.

Figure~\ref{figure:massradius} shows the clump masses as a function of 
effective radius. A least-squares fit to the data gives 
$\log(M)=(3.08\pm0.23)+(2.82\pm0.55)\log(R)$, with the linear correlation 
coefficient $r=0.85$. For comparison, we also show the mass-radius thres\-hold
for massive star formation, $M(R)=870$ M$_{\sun}\times(R/{\rm pc})^{1.33}$, 
recently inferred by Kauffmann \& Pillai (2010).

\begin{figure}[!h]
\centering
\resizebox{0.8\hsize}{!}{\includegraphics[angle=0]{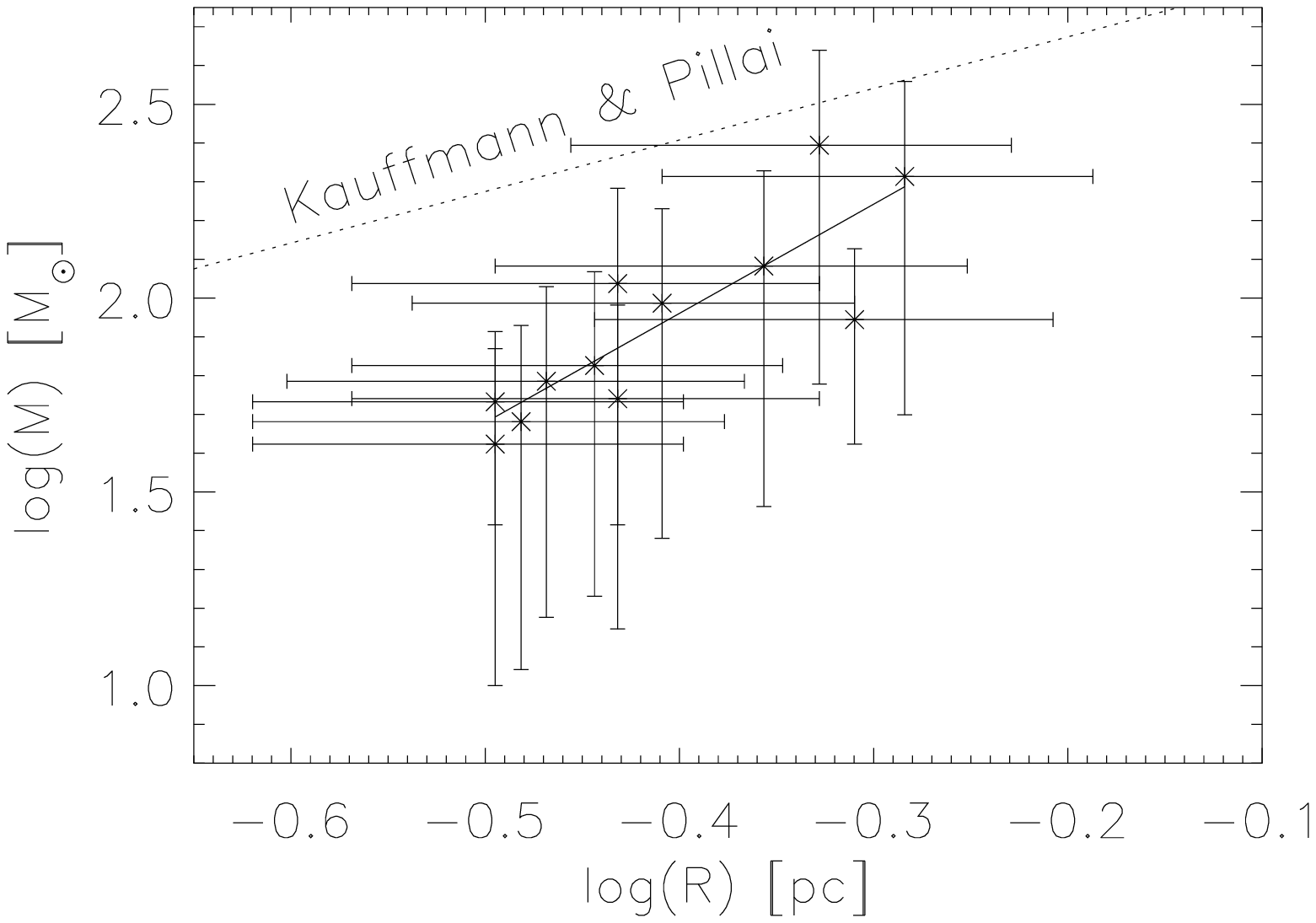}}
\caption{Relation between mass and effective radius for the clumps in G304.74. 
The solid line shows the least-squares fit to the data, i.e., 
$\log(M)\approx3.1+2.8\log(R)$. The dotted line represents the mass-radius 
threshold for massive star formation recently proposed by Kauffmann \& Pillai 
(2010), i.e., $M(R)=870$ M$_{\sun}\times(R/{\rm pc})^{1.33}$.}
\label{figure:massradius}
\end{figure}

\subsection{Molecular column densities and fractional abundances}

Details on how the molecular column densities and fractional 
abundances were calculated are given in Appendix A.3. In brief, we used 
Eq.~(\ref{eq02}) to estimate the peak optical thickness
for those lines which were used in the column-density calculations 
[see Col.~(7) of Table~\ref{table:lineparameters}]. In all cases the lines 
appear to be optically thin ($\tau_0 \ll 1$). Therefore, the beam-averaged 
column densities of C$^{17}$O, SiO, CH$_3$OH, and DCN were calculated using 
Eq.~(\ref{eq03}). The fractional abundances of the molecules were calculated 
by dividing the molecular column density by the H$_2$ column density. For 
this purpose, the values of $N({\rm H_2})$ were derived from the LABOCA dust 
continuum map smoothed to the corresponding resolution of the line 
observations. The obtained column densities and fractional abundances are 
listed in Table~\ref{table:column}.

\begin{table*}
\caption{Molecular column densities, fractional abundances with respect to 
H$_2$, and CO depletion factors ($f_{\rm D}$).}
\begin{minipage}{2\columnwidth}
\centering
\renewcommand{\footnoterule}{}
\label{table:column}
\begin{tabular}{c c c c c c}
\hline\hline 
Source & $N({\rm C^{17}O})$ & $x({\rm C^{17}O})$ & $f_{\rm D}$ & $N({\rm CH_3OH})$ & $x({\rm CH_3OH})$\\
        & [$10^{14}$ cm$^{-2}$] & [$10^{-8}$] & & [$10^{13}$ cm$^{-2}$] & [$10^{-9}$]\\
\hline
SMM 1 & $4.7\pm1.3$ & $3.1\pm1.7$ & $2.2\pm1.2$ & $2.1\pm0.3$--$4.4\pm0.7$ & $1.3\pm0.8$--$2.7\pm1.6$ \\
SMM 2 & $3.4\pm1.0$ & $4.1\pm2.3$ & $1.7\pm0.9$ & $1.6\pm0.3$--$3.4\pm0.7$ & $0.2\pm0.1$--$0.4\pm0.1$ \\
SMM 3\tablefootmark{a} & $4.9\pm1.4$ & $3.0\pm1.7$ & $2.3\pm1.3$ & $9.9\pm1.3$--$20.7\pm2.7$ & $5.8\pm3.3$--$12.2\pm6.9$ \\
SMM 4 & $15.1\pm3.9$ & $6.3\pm3.4$ & $1.1\pm0.6$ & $3.2\pm0.4$--$6.7\pm0.9$ & $1.3\pm0.7$--$2.8\pm1.5$\\
IRAS 13037-6112 & $21.8\pm2.2$ & $21.9\pm12.1$ & $0.3\pm0.2$ & $3.8\pm0.4$--$7.9\pm0.9$ & $3.8\pm2.3$--$7.9\pm4.4$\\
SMM 5 & $7.6\pm2.0$ & $13.0\pm7.2$ & $0.5\pm0.3$ & \ldots & \ldots\\
SMM 6 & $19.7\pm5.1$ & $12.2\pm6.7$ & $0.6\pm0.3$ & \ldots & \ldots\\
SMM 7 & $13.3\pm3.5$ & $16.5\pm9.1$ & $0.4\pm0.2$ & \ldots & \ldots\\
IRAS 13039-6108 & $10.5\pm1.1$ & $12.0\pm6.7$ & $0.6\pm0.3$ & \ldots & \ldots\\
SMM 8 & $6.9\pm1.9$ & $6.8\pm3.7$ & $1.0\pm0.5$ & \ldots & \ldots\\
SMM 9 & $6.6\pm1.9$ & $5.9\pm3.2$ & $1.2\pm0.6$ & \ldots & \ldots\\
IRAS 13042-6105 & $3.5\pm1.0$ & $5.1\pm2.8$ & $1.3\pm0.7$ & \ldots & \ldots\\
\hline 
\end{tabular} 
\tablefoot{\tablefoottext{a}{Towards SMM 3, we also determined the SiO and DCN 
column densities and abundances of 
$N({\rm SiO})=6.5\pm1.2\times10^{11}-2.5\pm0.5\times10^{13}$ cm$^{-2}$, 
$x({\rm SiO})=4.0\pm2.4\times10^{-11}-1.5\pm0.9\times10^{-9}$, and 
$N({\rm DCN})=6.7\pm1.8-48.7\pm12.8\times10^{10}$ cm$^{-2}$, 
$x({\rm DCN})=4.1\pm2.5-30.1\pm18.3\times10^{-12}$.}}
\end{minipage} 
\end{table*}

\subsection{CO depletion factors}

To estimate the amount of CO depletion in the clumps, we calculated 
the CO depletion factors, $f_{\rm D}$, following the analysis outlined in 
Appendix A.4. The derived CO depletion factors, $0.3-2.3$, are listed in 
Col.~(4) of Table~\ref{table:column}.

\subsection{Gas kinematics and internal pressure}

\begin{figure*}
\begin{center}
\includegraphics[scale=0.4]{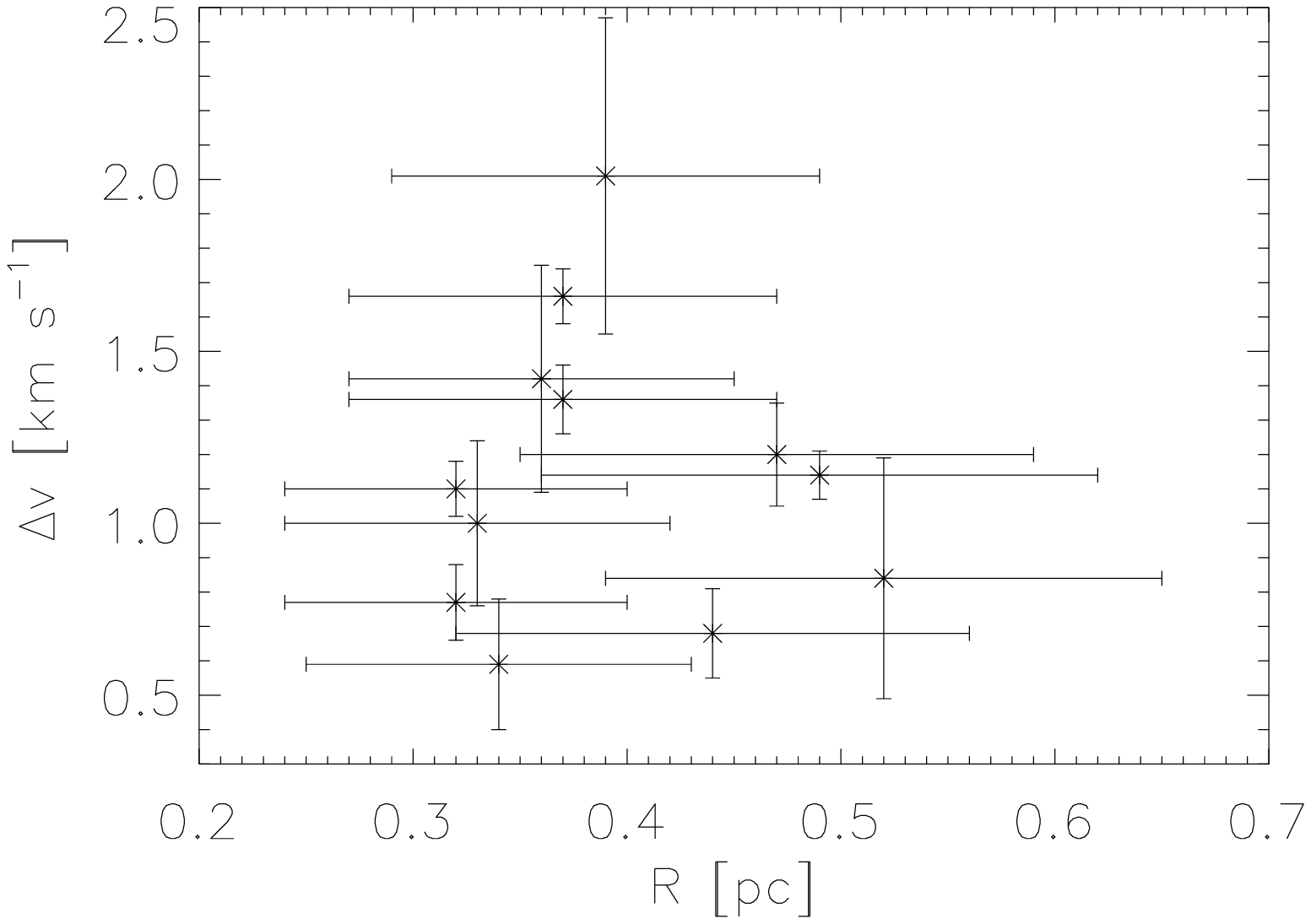}
\includegraphics[scale=0.4]{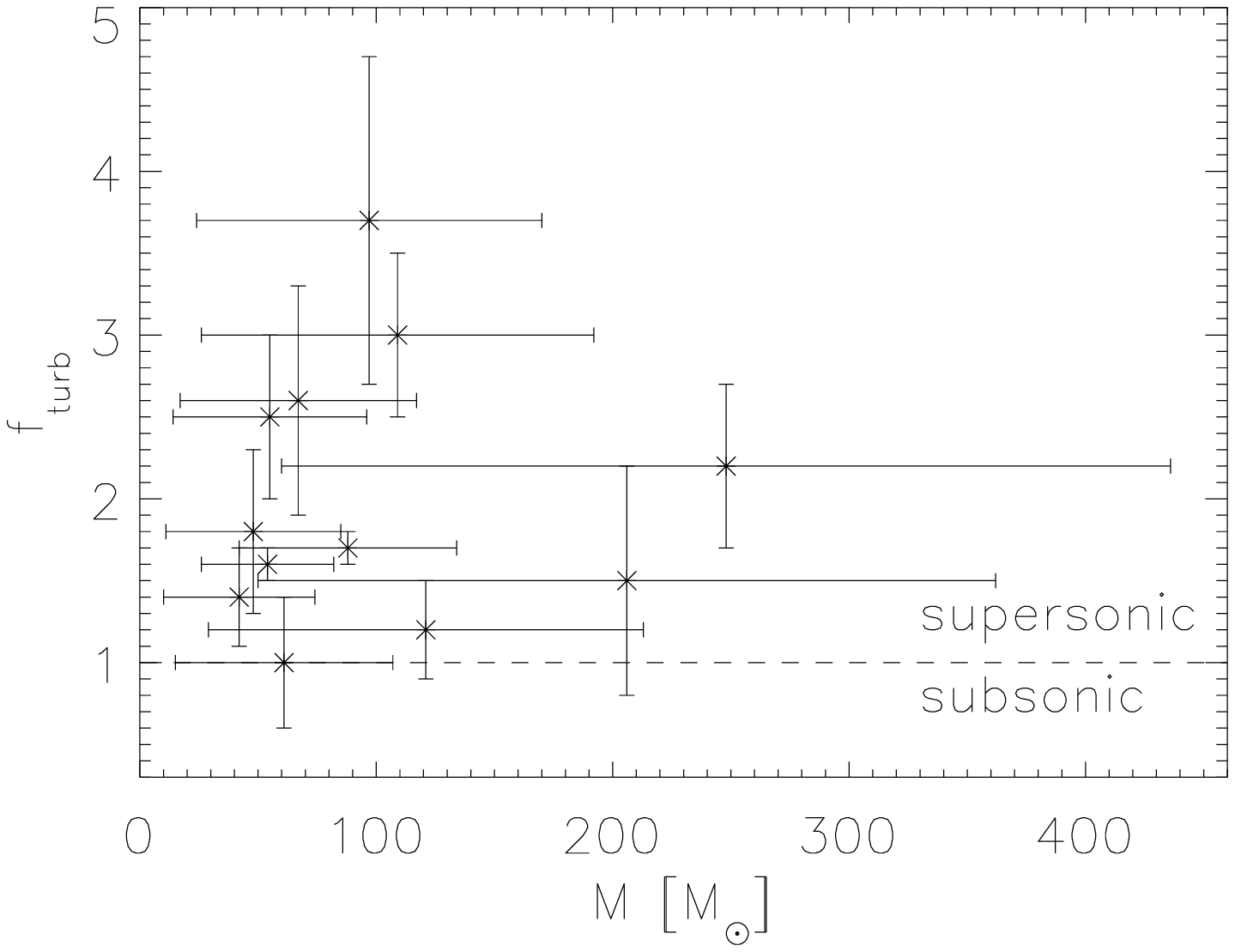}
\caption{\textbf{Left:} The FWHM linewidth of C$^{17}$O$(2-1)$ versus clump 
effective radius. \textbf{Right:} The $\sigma_{\rm NT}/c_{\rm s}$ ratio versus 
clump mass. The dashed line represents the limit between subsonic and 
supersonic non-thermal motions, i.e.,  $\sigma_{\rm NT}/c_{\rm s}=1$.}
\label{figure:kinematics}
\end{center}
\end{figure*}

We used the measured C$^{17}$O$(2-1)$ linewidths to 
calculate \textit{i)} the non-thermal portion of the line-of-sight velocity 
dispersion (averaged over a $27\farcs8$ beam), and \textit{ii)} the level of 
internal 'turbulence' (see below). The observed velocity dispersion is 
related to the FWHM linewidth as 
$\sigma_{\rm obs}=\Delta {\rm v}/\sqrt{8\ln 2}\simeq\Delta {\rm v}/2.355$. 
The non-thermal velocity dispersion can then be calculated as (e.g., 
\cite{myers1991})

\begin{equation}
\label{eq:sigmaNT}
\sigma_{\rm NT}=\sqrt{\sigma_{\rm obs}^2-\frac{k_{\rm B}T_{\rm kin}}{m_{\rm C^{17}O}}} \, ,
\end{equation}
where $m_{\rm C^{17}O}=29$ amu is the mass of the C$^{17}$O molecule. 
Furthermore, the level of internal turbulence is given by 
$f_{\rm turb}=\sigma_{\rm NT}/c_{\rm s}$, where $c_{\rm s}$ is the 
one-dimensional isothermal sound speed (0.23 km~s$^{-1}$ in a 15 K H$_2$ gas 
with 10\% He, where the mean molecular weight per free particle is 2.33). 
The values of $\sigma_{\rm NT}$ and $f_{\rm turb}$ 
are given in Cols.~(2) and (3) of Table~\ref{table:dispersion}, respectively. 
The errors in these parameters were derived by propagating the errors 
in $\Delta {\rm v}$ and $T_{\rm kin}$. In Fig.~\ref{figure:kinematics} we 
plot the FWHM linewidths as a function of clump effective radius (left panel), 
and the derived $f_{\rm turb}$ va\-lues as a function of clump mass 
(right panel). 
The linewidths do not show any clear correlation with the clump size. Instead, 
they are scattered around a horizontal regression line (the fit is not shown),
which suggests that the linewidths are more or less ``constant'' as a function 
of clump radius. As can be seen from the right panel of 
Fig.~\ref{figure:kinematics}, the clumps are characterised by trans- to 
supersonic ($\sigma_{\rm NT}\gtrsim c_{\rm s}$) non-thermal motions.

We also calculated the total (thermal$+$non-thermal) kinetic pressure within 
the clumps from

\begin{equation}
P_{\rm kin}=P_{\rm T}+P_{\rm NT}=nk_{\rm B}T_{\rm kin}+\rho \sigma_{\rm NT}^2=
\rho(c_{\rm s}^2+\sigma_{\rm NT}^2) \, ,
\end{equation}
where $\rho=2.33 m_{\rm H}\langle n({\rm H_2}) \rangle$ is the average gas mass 
density of the clump, and $m_{\rm H}$ is the mass of a hydrogen atom. The 
results, $\sim2-18\times10^5$ K~cm$^{-3}$ ($6.7\times10^5$ K~cm$^{-3}$ on 
average), are listed in Col.~(4) of Table~\ref{table:dispersion}. Note that 
the pressures are associated with considerable errors because of the assumed 
$\pm5$ K uncertainty in temperature. 

\begin{table}
\renewcommand{\footnoterule}{}
\caption{Non-thermal velocity dispersion ($\sigma_{\rm NT}$), the level of 
internal turbulence ($f_{\rm turb}$), and the total internal kinetic pressure 
($P_{\rm kin}$)\tablefootmark{a}.}
\begin{minipage}{1\columnwidth}
\centering
\label{table:dispersion}
\begin{tabular}{c c c c c}
\hline\hline 
Source & $\sigma_{\rm NT}$ & $f_{\rm turb}$ & $P_{\rm kin}/k_{\rm B}$\\
        & [km~s$^{-1}$] & & [$10^5$ K~cm$^{-3}$]\\
\hline
SMM 1 & $0.28\pm0.06$ & $1.2\pm0.3$ & $2.6\pm2.0$\\
SMM 2 & $0.24\pm0.08$ & $1.0\pm0.4$ & $2.2\pm1.7$\\
SMM 3 & $0.35\pm0.15$ & $1.5\pm0.7$ & $3.5\pm3.2$\\
SMM 4 & $0.51\pm0.06$ & $2.2\pm0.5$ & $9.7\pm7.3$\\
IRAS 13037-6112 & $0.46\pm0.03$ & $1.6\pm0.1$ & $6.6\pm3.3$\\ 
SMM 5 & $0.32\pm0.05$ & $1.4\pm0.3$ & $2.7\pm1.8$\\
SMM 6 & $0.70\pm0.03$ & $3.0\pm0.5$ & $15.3\pm12.3$\\
SMM 7 & $0.57\pm0.04$ & $2.5\pm0.5$ & $5.3\pm4.3$\\
IRAS 13039-6108 & $0.47\pm0.03$ & $1.7\pm0.1$ & $2.5\pm1.7$\\ 
SMM 8 & $0.60\pm0.14$ & $2.6\pm0.7$ & $8.2\pm6.7$\\
SMM 9 & $0.85\pm0.20$ & $3.7\pm1.0$ & $17.6\pm15.2$\\
IRAS 13042-6105 & $0.42\pm0.10$ & $1.8\pm0.5$ & $3.9\pm3.6$\\
\hline 
\end{tabular} 
\tablefoot{\tablefoottext{a}{All the parameters were calculated assuming 
$T_{\rm kin}=15\pm5$ K for all the other sources except the \textit{IRAS} 
sources 13037 and 13039, for which we used the value $T_{\rm kin}=22$ K.}}
\end{minipage} 
\end{table}

\subsection{Virial analysis}

In order to examine the dynamical state of the clumps, we calculated their 
virial masses using the following formula, where the effects of external 
pressure, rotation, and magnetic field are ignored:

\begin{equation}
\label{eq:Mvir}
M_{\rm vir}=\frac{5}{8\ln 2}\frac{R\Delta {\rm v_{\rm ave}^2}}{aG} \, ,
\end{equation}
where $\Delta {\rm v_{\rm ave}}$ is the 
width of the spectral line emitted by the molecule of mean mass $\mu=2.33$, 
and $G$ is the gravitational constant. The parameter $a=(1-p/3)/(1-2p/5)$, 
where $p$ is the power-law index of the density profile 
[$n(r)\propto r^{-p}$], is a correction for deviations from uniform density 
(\cite{bertoldi1992}; hereafter, BM92; their Appendix A). 
In the clump's gravitational potential energy, the $a$ parameter appears as 
follows:

\begin{equation}
\mathcal{W}=-\frac{3}{5}a\frac{GM^2}{R} \, .
\end{equation}
We note that the effect of the clump's ellipticity is neglected here 
(see BM92). We adopted the value $p=1.6$ ($a=1.3$) as found by Beuther
et al. (2002) for a sample of high-mass star-forming clumps. 
As a function of the observed linewidth, $\Delta {\rm v}_{\rm obs}$, 
$\Delta {\rm v_{\rm ave}}$ is given by (\cite{fuller1992})

\begin{equation}
\label{eq:Dv}
\Delta {\rm v_{\rm ave}^2}=\Delta {\rm v}_{\rm T}^2+\Delta {\rm v}_{\rm NT}^2=\Delta {\rm v_{\rm obs}^2}+8\ln 2 \times \frac{k_{\rm B}T_{\rm kin}}{m_{\rm H}}\left(\frac{1}{\mu}-\frac{1}{\mu_{\rm mol}} \right) \, ,
\end{equation}
where $\Delta {\rm v}_{\rm T}$ is the thermal linewidth, $\mu=2.33$, and 
$\mu_{\rm mol}=29$ for C$^{17}$O, the molecule that we have employed in the 
analysis. The virial masses are listed in Col.~(2) of 
Table~\ref{table:virial}. The associated error was propagated from 
those of $R$, $\Delta {\rm v}_{\rm obs}$, and $T_{\rm kin}$. 

The virial parameters of the clumps were calculated following the definition 
by BM92, i.e.,

\begin{equation}
\label{eq:virial}
\alpha_{\rm vir}=\frac{2\mathcal{T}}{\vert \mathcal{W} \vert}=\frac{M_{\rm vir}}{M}\, ,
\end{equation}
where $\mathcal{T}$ is the internal kinetic energy of the clump. The value 
$\alpha_{\rm vir}=1$ corresponds to the virial equilibrium, i.e., 
$2\langle \mathcal{T} \rangle+\langle \mathcal{W} \rangle=0$. The value 
$\alpha_{\rm vir}\leq 2$ corresponds to the self-gravitating limit defined by 
$\langle \mathcal{T} \rangle+\langle \mathcal{W} \rangle \leq 0$ 
(e.g., \cite{pound1993})\footnote{Actually, in this case, it could be more 
appropriate to use the term ``equipartition'' parameter instead of ``virial'' 
parameter (\cite{ballesteros2006}; \cite{heitsch2008}).}. The derived 
$\alpha_{\rm vir}$ values are given in Col.~(3) of Table~\ref{table:virial}. 
The uncertainty was derived by propagating the errors in both mass estimates.
 
The derived $\alpha_{\rm vir}$ values, $\sim0.4-2.8$, are plotted as a 
function of clump mass in Fig.~\ref{figure:virial}. Although the 
$\alpha_{\rm vir}$ values are associated with considerable uncertainties, 
most of the clumps appear to be gra\-vitationally bound, and some of them 
are near virial equilibrium. Moreover, there is a hint of negative 
correlation bet\-ween $\alpha_{\rm vir}$ and $M$. A least-squares fit to the 
data points yields 
$\log(\alpha_{\rm vir})=(1.38\pm0.63)-(0.69\pm0.32)\log(M)$, with the 
correlation coefficient $r=-0.56$. 

\subsection{The confining effect of external pressure}

\begin{figure}[!h]
\centering
\resizebox{0.8\hsize}{!}{\includegraphics[angle=0]{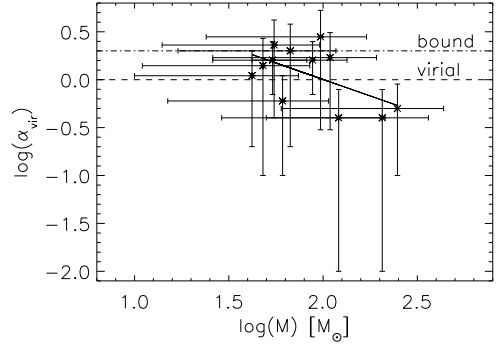}}
\caption{Relation between virial parameter and mass for the clumps in G304.74. 
The solid line is the least-squares fit to the data, i.e., 
$\log(\alpha_{\rm vir})\simeq1.4-0.7\log(M)$. The dashed line indicates the 
virial-equilibrium limit of $\alpha_{\rm vir}=1$, and the dash-dotted line 
shows the limit of gravitational boundedness or $\alpha_{\rm vir}=2$.}
\label{figure:virial}
\end{figure}

The slope of the  $\alpha_{\rm vir}-M$ relation found above, -0.69, is quite 
close to the value $-2/3 \simeq-0.67$ characteristic of pressure-confined 
clumps (BM92). Therefore, the external pressure may play 
an important role in the dynamics of the clumps. Taking the kinetic energy 
resulting from the surface pressure on the clump, $\mathcal{T}_{\rm ext}$, 
into account, we can write the virial theorem in the following form 
(e.g., \cite{mckee1992})\footnote{If the magnetic energy, $\mathcal{M}$, is 
also taken into account, the virial theorem is $2(\langle \mathcal{T}\rangle - \langle \mathcal{T}_{\rm ext}\rangle)+ \langle \mathcal{M}\rangle + \langle \mathcal{W} \rangle=0$. The only expansionary terms here are $\mathcal{T}$ and $\mathcal{M}$ (the angle brackets to denote time averages have been dropped for simplicity). Therefore, from the condition $\mathcal{M}\ll 2\mathcal{T}$, we can estimate the upper limit to the magnetic field strength, $B$, when Eq.~(\ref{eq:virial2}) is still expected to be valid. If we neglect any contribution from the magnetic field outside the clump, and assume that the field inside the clump is 
uniform, we can write $\mathcal{M}=B^2V_{\rm cl}/2\mu_0$, where $\mu_0$ is 
the vacuum permeability. This leads to an upper limit of $B \ll \sqrt{4\mu_0 \mathcal{T}/V_{\rm cl}}=\sqrt{6\mu_0P_{\rm kin}}$.}:

\begin{equation}
\label{eq:virial2}
2(\langle \mathcal{T}\rangle - \langle \mathcal{T}_{\rm ext}\rangle)+ \langle \mathcal{W} \rangle=0 \,.
\end{equation}
The above energies can be written as a function of pressure as follows:

\begin{equation}
\label{eq:kin}
\mathcal{T}=\frac{3}{2}P_{\rm kin}V_{\rm cl} \, ,
\end{equation}

\begin{equation}
\label{eq:ext}
\mathcal{T}_{\rm ext}=\frac{3}{2}P_{\rm ext}V_{\rm cl} \, ,
\end{equation}
and
\begin{equation}
\label{eq:grav}
\mathcal{W}=-3P_{\rm grav}V_{\rm cl} \, .
\end{equation}
In the above formulas, $V_{\rm cl}=\frac{4}{3}\pi R^3$ is the clump volume, 
$P_{\rm ext}$ is the external pressure on the clump, and $P_{\rm grav}$ is the 
gravitational pressure. We can write $P_{\rm grav}$ as follows:

\begin{equation}
\label{eq:gravpressure}
P_{\rm grav}=-\frac{1}{3}\frac{\mathcal{W}}{V_{\rm cl}}=-\frac{1}{3}\left(-\frac{3}{5}a\frac{GM^2}{R}\right)\frac{1}{\frac{4}{3}\pi R^3}=\frac{3a}{20\pi}\frac{GM^2}{R^4} \, .
\end{equation}
Substituting Eqs.~(\ref{eq:kin})--(\ref{eq:grav}) into Eq.~(\ref{eq:virial2}), 
we get

\begin{equation}
\label{eq:virial3}
P_{\rm kin}-P_{\rm ext}-P_{\rm grav}=0 \, .
\end{equation}
From Eq.~(\ref{eq:virial3}) we can estimate the required $P_{\rm ext}$ to bring 
the clumps into virial equilibrium, which is roughly suggested by 
Fig.~\ref{figure:virial}.
The estimated $P_{\rm ext}^{\rm eq}$ values are listed in Col.~(4) of 
Table~\ref{table:virial}. We note that the values are associated with very 
large errors, and we only give the central values in the table. The clumps 
SMM 1--4 in the southern filament are formally characterised by ne\-gative 
external pressures. Although there are large uncertainties in the analysis, 
these clumps should be either warmer or have some additional internal pressure 
acting outward, such as magnetic pressure, in order to achieve virial 
equlibrium. The positive values of $P_{\rm ext}^{\rm eq}$ are found for the 
clumps north of SMM 4, and they lie in the range $\sim0.5-12\times10^5$ 
K~cm$^{-3}$ ($\sim3.3\times10^5$ K~cm$^{-3}$ on average).

\begin{table}
\caption{The clump virial masses and virial parameters, and the estimated 
external pressure required to bring the clump into virial equili\-brium.}
\begin{minipage}{1\columnwidth}
\centering
\renewcommand{\footnoterule}{}
\label{table:virial}
\begin{tabular}{c c c c}
\hline\hline 
Source & $M_{\rm vir}$ & $\alpha_{\rm vir}$ & $P_{\rm ext}^{\rm eq}/k_{\rm B}$\\
        & [M$_{\sun}$] & & [$10^5$ K~cm$^{-3}$] \\
\hline
SMM 1 & $54\pm20$ & $0.4\pm0.4$ & -2.5\tablefootmark{a}\\
SMM 2 & $35\pm16$ & $0.6\pm0.5$ & -1.4\tablefootmark{a}\\
SMM 3 & $84\pm54$ & $0.4\pm0.4$ & -4.1\tablefootmark{a}\\
SMM 4 & $132\pm44$ & $0.5\pm0.4$ & -6.8\tablefootmark{a}\\
IRAS 13037-6112 & $85\pm23$ & $1.6\pm0.9$ & 3.0\\
SMM 5 & $46\pm15$ & $1.1\pm0.9$ & 0.5\\
SMM 6 & $183\pm52$ & $1.7\pm1.4$ & 7.0\\
SMM 7 & $128\pm39$ & $2.3\pm1.9$ & 3.2\\
IRAS 13039-6108 & $137\pm39$ & $1.6\pm0.9$ & 0.7\\
SMM 8 & $135\pm64$ & $2.0\pm1.8$ & 4.7\\
SMM 9 & $273\pm136$ & $2.8\pm2.5$ & 12.3\\
IRAS 13042-6105 & $69\pm32$ & $1.4\pm1.3$ & 1.4\\
\hline 
\end{tabular} 
\tablefoot{\tablefoottext{a}{The negative values suggest that 
stronger/additional internal pressure is needed to bring the clump into 
virial balance (see text).}}
\end{minipage} 
\end{table}

\subsection{Mass infall}

The detected $^{13}$CO$(2-1)$ lines, most notably towards SMM 2, show the 
'classic' signature of the infalling gas motions, i.e, redshifted absorption 
dips, or blue asymmetric profiles (e.g., \cite{lucas1976}; \cite{myers1996}).  
The optically thin C$^{17}$O lines peak close to the radial velocity of the 
self-absorption dip. Therefore, we can presume that the double-peak profiles 
are not caused by two distinct velocity components along the line of sight.
Note, however, that many of the absorption features are slightly redshifted 
with respect to the systemic velocity (Fig.~\ref{figure:13CO}). The blue 
asymmetry is weakest towards SMM 1, which indicates that in this clump there 
is the slowest infalling gas. We estimated the infall velocity, $V_{\rm inf}$, 
using Eq.~(9) of Myers et al. (1996):

\begin{equation}
V_{\rm inf}\approx \frac{\sigma^2}{{\rm v}_{\rm red}-{\rm v}_{\rm blue}}\ln \left(\frac{1+e\times T_{\rm BD}/T_{\rm dip}}{1+e\times T_{\rm RD}/T_{\rm dip}} \right) \, ,
\end{equation}
where $\sigma$ is the velocity dispersion derived from appropriate optically 
thin line (we used the C$^{17}$O lines), ${\rm v}_{\rm red}-{\rm v}_{\rm blue}$ is 
the difference between the radial velocities of the red and blue peaks, 
$T_{\rm BD}$ and $T_{\rm RD}$ are the heights of the blue and red peaks above 
the absorption dip, and $T_{\rm dip}$ is the intensity of the absorption dip 
[see also Fig.~2 of Myers et al. (1996) for the definitions]. The obtained 
values, $\sim0.03-0.20$ km~s$^{-1}$, are listed in Col.~(2) of 
Table~\ref{table:infall}. The uncertainty in $V_{\rm inf}$ was estimated from 
the error in the FWHM linewidth of C$^{17}$O$(2-1)$.

Assuming spherically symmetric infall, and that the mass contained in the 
spherical shell is ${\rm d}M(r)=4\pi r^2 \rho {\rm d}r$, the infall velocities 
can be used to estimate the mass infall rate as 

\begin{equation}
\dot{M}_{\rm inf}=\frac{{\rm d}M}{{\rm d}t}=\frac{{\rm d}M}{{\rm d}r}\frac{{\rm d}r}{{\rm d}t}=4\pi r_{\rm inf}^2\rho V_{\rm inf} \, ,
\end{equation}
where $r_{\rm inf}$ is the size of the infalling region (taken to be the clump 
effective radius), and $\rho=2.33m_{\rm H}\langle n({\rm H_2})\rangle$ is again 
the average gas mass density\footnote{We note, however, that the 
density of the absorbing and infalling gas may be somewhat lower than the 
volume-average clump density.}. The derived values, $\sim2-36\times10^{-5}$ 
M$_{\sun}$~yr$^{-1}$, with uncertainties estimated from the errors in 
$V_{\rm inf}$, are given in Col.~(3) of Table~\ref{table:infall}.

Assuming that $\dot{M}_{\rm inf}$ remains constant, the infall timescale can 
be calculated from 

\begin{equation}
\tau_{\rm inf}=\frac{M}{\dot{M}_{\rm inf}} \,.
\end{equation}
The obtained timescales, $\sim6-46\times10^5$ yr, are listed in Col.~(4) of
Table~\ref{table:infall} (the uncertainty is based on that in 
$\dot{M}_{\rm inf}$). 

We also calculated the free-fall velocities of the clumps

\begin{equation}
V_{\rm ff}=\sqrt{\frac{2GM}{R}} \, ,
\end{equation}
which were then compared with the derived infall velocities. In the last 
column of Table~\ref{table:infall} we list the $V_{\rm inf}/V_{\rm ff}$ 
ratios. The associated uncertainty was propagated from those of 
$V_{\rm inf}$, $M$, and $R$. In all cases, the infall velocity is much lower 
than the corresponding free-fall velocity.

\begin{table}
\caption{Mass-infall parameters.}
\tiny
\begin{minipage}{1\columnwidth}
\centering
\renewcommand{\footnoterule}{}
\label{table:infall}
\begin{tabular}{c c c c c}
\hline\hline 
Source & $V_{\rm inf}$ & $\dot{M}_{\rm inf}$ & $\tau_{\rm inf}$ & $V_{\rm inf}/V_{\rm ff}$\\
        & [km~s$^{-1}$] & [$10^{-5}$ M$_{\sun}$~yr$^{-1}$] & [$10^5$ yr] &\\
\hline
SMM 1 & $0.03\pm0.01$ & $2.6\pm1.0$ & $45.8\pm17.4$ & $0.02\pm0.01$\\
SMM 2 & $0.03\pm0.02$ & $1.8\pm1.2$ & $33.1\pm21.4$ & $0.02\pm0.02$\\
SMM 3 & $0.17\pm0.14$ & $23.7\pm19.6$ & $8.7\pm7.2$ & $0.09\pm0.08$\\
SMM 4 & $0.20\pm0.05$ & $36.1\pm9.0$ & $6.9\pm1.7$ & $0.09\pm0.04$\\
IRAS13037 & $0.16\pm0.02$ & $9.8\pm1.2$ & $5.5\pm0.7$ & $0.13\pm0.04$\\  
\hline 
\end{tabular} 
\end{minipage} 
\end{table}

\subsection{Fragmentation and stability analysis of the filament}

The IRDC G304.74 is highly filamentary in shape, its total mass is about 
$\sim1200$ M$_{\sun}$ (the sum of all clump masses), and it is resolved into 
twelve submm clumps at $\sim20\arcsec$ resolution. The projected 
linear extent of filament's long axis is about $12\arcmin$ or 8.9 pc, and the 
filament's mass per unit length, or line mass, is $M_{\rm line}\simeq135$ 
M$_{\sun}$~pc$^{-1}$. We note that this line mass may only re\-present a lower 
limit, because there is likely to be a lot of mass on larger scales 
(\cite{kainulainen2011b}). The projected separations between the clump submm 
peak positions lie in the range $22\arcsec-115\arcsec$, or 0.27--1.42 pc at 
the cloud distance ($61\arcsec=0.75$ pc on average). The shortest separations 
are found near SMM 6 and IRAS13037, whereas the separations are largest 
around SMM 7.

To examine the fragmentation of G304.74 in more detail, we first calculate its
thermal Jeans length from

\begin{equation}
\label{eq:jeans}
\lambda_{\rm J}=\frac{c_{\rm s}^2}{G \Sigma_0} \,, 
\end{equation}
where $c_{\rm s}$ was adopted to be 0.23 km~s$^{-1}$ (Sect.~4.5), 
$\Sigma_0=2.33m_{\rm H}N({\rm H_2})$ is the surface density, and $N({\rm H_2})$ 
refers to the central column density for which we use the median value 
$\sim1.4\times10^{22}$ cm$^{-2}$ [see Col.~(6) of Table~\ref{table:sources}]. 
The resulting thermal Jeans length is only $\sim0.05$ pc, which is about five 
times less than the spatial resolution of 
our LABOCA map, and therefore much shorter than the observed minimum clump 
separation. Thermal Jeans instability does not appear to be 
responsible for the cloud fragmentation \textit{at the scale of clumps}. 

On the other hand, we found that the filament is dominated by supersonic 
non-thermal motions. Therefore, it may be appropriate to replace $c_{\rm s}$ in 
Eq.~(\ref{eq:jeans}) with the effective sound speed, which takes both thermal 
and non-thermal motions into account, i.e., 
$c_{\rm eff}=(c_{\rm s}^2+1/3 \times \sigma_{\rm NT,\,3D}^2)^{1/2}$, where 
$\sigma_{\rm NT,\,3D}$ is the three dimensional non-thermal velocity 
dispersion (e.g., \cite{chandrasekhar1951}; \cite{bonazzola1987}). Here we have 
assumed isotropic non-thermal motions, i.e., 
$\sigma_{\rm NT,\,3D}=\sqrt{3}\sigma_{\rm NT,\,1D}$. Using the median value of 
$\sigma_{\rm NT,\,1D}=0.47$ km~s$^{-1}$ [see Col.~(2) of 
Table~\ref{table:dispersion}], the effective, or ``turbulent'' Jeans length 
becomes $\sim0.24$ pc. This is about five times larger than the thermal Jeans 
length, but still much smaller than the average clump separation of 
$\sim0.75$ pc. Therefore, the observed fragmentation of the fi\-lament cannot 
be attributed to non-thermal Jeans-type instabi\-lity either. 

According to the magnetohydrodynamic ``sausage''-instability theory, 
fragmentation of a self-gravitating fluid cylinder will lead to the 
formation of distinct condensations with almost periodic separations. This
fragmentation length-scale cor\-responds to the wavelength, at which the 
instability grows the fastest (e.g., \cite{chandrasekhar1953}; 
\cite{nagasawa1987}; see also \cite{jackson2010}). If the cylinder's radius 
is $R_{\rm cyl}$, then, for an incompressible fluid, perturbation analysis 
shows that the above critical wavelength is
$\lambda_{\rm max}^{\rm incomp.}\simeq 11R_{\rm cyl}$. If 
we use as $R_{\rm cyl}$ the mean clump effective radius of $\sim0.4$ pc, we 
obtain $\lambda_{\rm max}^{\rm incomp.}\simeq4.4$ pc. This is clearly longer 
length scale than observed in G304.74. On the other hand, for an isothermal, 
infinitely-long gas cylinder the fastest growing mode appears at 
$\lambda_{\rm max}^{\rm isotherm.}\simeq 22H$, where 
$H=c_{\rm s}/\sqrt{4\pi G \rho_0}$ is the isothermal scale height with 
$\rho_0$ the central gas mass density along the axis of the cylinder (e.g., 
\cite{nagasawa1987}). If we compute $\rho_0$ assuming that the 
central number density is $10^5$ cm$^{-3}$, we obtain the scale height 
$H=0.013$ pc, and thus $\lambda_{\rm max}^{\rm isotherm.}\simeq0.3$ pc. This 
is compar\-able to the effective Jeans-length calculated above. 
If the value of $H$ is calculated using the effective sound speed, 
$c_{\rm eff}$, i.e., we calculate the effective scale height, we get 
$H_{\rm eff}=0.03$ pc, and $\lambda_{\rm max}^{\rm isotherm.}\simeq0.7$ pc. 
This wavelength is close to the observed average clump separation. We 
also note that it is comparable to the average width of the filament, 
$\sim0.8$ pc, in accordance with the theory (\cite{nakamura1993}).

Next, we examine whether the G304.74 filament could be unstable to 
axisymmetric perturbations or radial collapse. For an unmagnetised, infinite, 
isothermal filament the instability is reached if its $M_{\rm line}$ value 
exceeds the critical equilibrium value of 
$M_{\rm line}^{\rm crit}=2c_{\rm s}^2/G$ (e.g., \cite{ostriker1964}; 
\cite{inutsuka1992}). At $T_{\rm kin}=15$ K, the critical 
value is $M_{\rm line}^{\rm crit}\simeq25$ M$_{\sun}$~pc$^{-1}$. Using this 
value we estimate that the ratio 
$M_{\rm line}/M_{\rm line}^{\rm crit}\simeq5$ for G304.74. 
Thus, G304.74 appears to be a thermally supercritical filament susceptible 
to fragmentation, in agreement with the detected clumpy structure. 
If the external pressure is not negligible, as may be the case for G304.74, 
then $M_{\rm line}^{\rm crit}$ is actually smaller than the above value 
(e.g, \cite{fiege2000a}), making the filament even more supercritical. 
However, again due to the presence of non-thermal 
motions, it may be appropriate to replace $M_{\rm line}^{\rm crit}$ with the 
virial mass per unit length, 
$M_{\rm line}^{\rm vir}=2 \langle \sigma^2\rangle /G$, 
where $\langle \sigma^2\rangle$ is the square of the total velocity dispersion 
obtained from Eq.~(\ref{eq:Dv}) as 
$\langle \sigma^2\rangle=\Delta {\rm v_{\rm ave}^2}/8 \ln 2$ 
(\cite{fiege2000a}). Using again the value $T_{\rm kin}=15$ K and the 
C$^{17}$O linewidths, we obtain the mean velocity dispersion of 
$\langle \sigma^2 \rangle^{1/2} =0.54$ 
km~s$^{-1}$. This leads to the va\-lues $M_{\rm line}^{\rm vir}\simeq136$ 
M$_{\sun}$~pc$^{-1}$ and $M_{\rm line}/M_{\rm line}^{\rm vir}\simeq1$. 
In this case, G304.74 as a whole would be near virial equilibrium. 
The fragmentation timescale for a filament of radius $R_{\rm cyl}$ is 
expected to be comparable to its radial signal crossing time, 
$\tau_{\rm cross}=R_{\rm cyl}/\langle \sigma^2 \rangle^{1/2}$ [see Eq.~(26) in 
\cite{fiege2000b}], which for G304.74 is estimated to be 
$\sim7.2\times10^5$ yr. 
We note that the value of $\langle \sigma^2\rangle^{1/2}$, 
and thus $M_{\rm line}^{\rm vir}$, may increase if the cloud's gravitational 
potential energy is converted into gas kinetic e\-nergy during the 
contraction of the cloud. We also note that in the case of a magnetised 
molecular-cloud filament, the critical mass per unit length is unlikely to 
differ from that of an unmagnetised filament by more than a factor of order 
unity (\cite{fiege2000a},b). 

\section{Discussion}

\subsection{Fragmentation and dynamical state of G304.74+01.32}

The average projected clump separation in G304.74 is about 0.75 pc.
As shown above, this value can be explained if the fragmentation of the 
parent filament is caused by ``sausage''-type (or pinch) instability, 
where the dominant unstable mode appears at the wavelength depending on the 
effective sound speed. Such instability could have been triggered by, e.g., 
increased radial inward force or radial contraction leading to the increase of 
the tension force of the azimuthal magnetic field. This conforms to the results 
of Jackson et al. (2010), who speculated that the ``sausage''-type 
instability is the main physical driver of fragmentation of filamentary IRDCs. 
Henning et al. (2010) and Kang et al. (2011) recently found that the mean 
clump separation is 0.9 pc in both the filamentary IRDC G011.11-0.12 or the 
Snake, and IRDC G049.40-00.01, respectively. This resembles the situation in 
G304.74, and could therefore suggest a similar fragmentation process. 
In Paper I we discussed the possibility that the origin of the G304.74 
fi\-lament, and its fragmentation into clumps is caused by supersonic 
turbulent flows. Even the formation of the cloud 
could have been caused by converging turbulent flows, the role of random 
turbulence in fragmenting the filament appears unclear in the light of the 
present results. Nevertheless, the new results support the importance of 
non-thermal motions in G304.74. 

The slope of the $\alpha_{\rm vir}$--$M$ relation we found is roughly 
consistent with the prediction of BM92 for clumps confined by ambient 
pressure from their surrounding medium, i.e., 
$\alpha_{\rm vir}\propto M^{-2/3}$. This slope can be understood as follows. 
On one hand, the clump mass was found to be roughly proportional to $R^3$ 
(Fig.~\ref{figure:massradius}). On the other hand, the linewidth--radius 
'relation' shown in Fig.~\ref{figure:kinematics} is roughly flat, i.e., 
$\Delta {\rm v}\approx$constant. Therefore, from Eqs.~(\ref{eq:Mvir}) and 
(\ref{eq:virial}) we see that $\alpha_{\rm vir} \propto R \Delta {\rm v}^2/M \propto M^{1/3}M^{-1}\propto M^{-2/3}$. Although our clumps, or most of them, 
appear to be gravitationally bound ($\alpha_{\rm vir} \leq 2$), it is 
possible that the ambient pressure (still) plays a non-negligible role in the 
clump dyna\-mics. Note that for real pressure confined clumps, 
$\alpha_{\rm vir} \gg 1$, and they are \textit{not} gravitationally bound 
(BM92).

Could the external pressure be caused by turbulent flow motions outside the 
cloud ? To estimate the turbulent ram pressure, 
$P_{\rm ram}=\rho \sigma_{\rm NT}^2$, we use the average $^{13}$CO$(2-1)$ 
linewidth of $\sim2.2$ km~s$^{-1}$ derived for five of our clumps, and 
assume the density of the $^{13}$CO emitting gas to be $10^3$ cm$^{-3}$ 
(following \cite{lada2008}). If $T_{\rm kin}$ is in the range 10--20 K, the 
cor\-responding non-thermal velocity dispersion is $\sigma_{\rm NT}\simeq0.93$ 
km~s$^{-1}$, which yields the value $P_{\rm ram}/k_{\rm B}\sim3\times10^5$ 
K~cm$^{-3}$. This is only about two times lower than the average internal 
kinetic pressure within the clumps, and very close to the average 
external pressure required for virial equilibrium 
($P_{\rm ram}\simeq P_{\rm ext}^{\rm eq}$; Sect.~4.7). Therefore, this 
supports the scenario of the importance of turbulence presented in Paper I. 
The obtained $P_{\rm ram}/k_{\rm B}$ value is also consistent with intercloud 
pressures of $10^4-10^6$ K~cm$^{-3}$ deduced from theore\-tical estimates 
and different observations (see \cite{field2011} and references therein). 
However, we note that the large-scale turbulent motions are anisotropic, 
and therefore the associated pressure may not be able to confine the cloud, 
but could rather distort and/or even disrupt it 
(e.g., \cite{ballesteros2006}). The finding 
that $M_{\rm line}/M_{\rm line}^{\rm vir}\simeq1$ for the G304.74 filament 
suggests that it is close to virial equili\-brium. The virial balance may be 
difficult to explain in the context of turbulent fragmentation discussed in 
Paper I, because turbulent motions are expected to 
induce a flux of mass, momentum, and energy between the cloud and its 
surrounding medium (\cite{ballesteros2006}). Also, the detection of 
large-scale infall motions towards G304.74 suggests it to be out-of-equilibrium.
For comparison, the recent \textit{Herschel} study of the Aquila rift 
and Polaris Flare regions by Andr\'e et al. (2010) showed that practically 
all filaments with $M_{\rm line}/M_{\rm line}^{\rm crit}>1$ were associated with 
prestellar cores and/or embedded protostars. On the other hand, filaments 
with $M_{\rm line}/M_{\rm line}^{\rm crit}<1$ were generally found to lack such 
star-formation signatures. More observations with higher resolution would be 
needed to better characterise the star-formation activity in G304.74, and 
to compare it with filamentary structures revealed by \textit{Herschel}.

If the $M_{\rm line}/M_{\rm line}^{\rm vir}$ ratio is indeed $\simeq1$, then from 
Eq.~(11) of Fiege \& Pudritz (2000a), and the average ratio of the external 
to internal pressure of 
$\langle P_{\rm ext}^{\rm eq} \rangle / \langle P_{\rm kin} \rangle \simeq0.49$, 
we estimate that the ratio between the total magnetic and kinetic energies per 
unit length is $\mathcal{M}/\vert \mathcal{W} \vert \simeq0.49$. This value is 
$>0$, and therefore suggests that \textit{i)} the overall effect of the 
magnetic field on the cloud is to provide support, and \textit{ii)} the 
magnetic field is poloidally do\-minated. This is in contrast with the finding 
of Fiege \& Pudritz (2000a), i.e., that the magnetic fields in the 
filamentary clouds they analysed are likely to be helical and toroidally 
dominated. However, Fiege et al. (2004) analysed the Snake IRDC, which is 
more similar to G304.74, and concluded that it is likely do\-minated by the 
poloidal component of the magnetic field (or is magnetically neutral), 
and is therefore an excellent candidate for a 
magnetically supported filament. This may also be the case for the G304.74 
filament. Interestingly, Fiege \& Pudritz (2000b) found that models 
with purely toroidal fields are stable against ``sausage'' modes, whereas 
models with a poloidal field component may be unstable.
We note, however, that the observed fragmentation of the filament strongly 
suggests that the magnetic field is not \textit{purely} 
poloidal (or toroidal), because otherwise the fragmentation would have 
presumably been stabilised (\cite{fiege2000b}). If the filament fragmentation 
is caused by sausage-type instabi\-lity, some toroidal magnetic field is 
probably needed, because magnetic pressure associated with the 
poloidal field would resist the ``squeezing'' of the cloud. 

On the other hand, the ambient pressure can also arise from the 
cloud's self-gravity. This was found to be the case in the Pipe Nebula 
by Lada et al. (2008), where all the cores within the cloud where found to have 
$\alpha_{\rm vir}>1$, and $\alpha_{\rm vir}\propto M^{-0.66}$. The pressure 
due to the weight of the cloud is 

\begin{equation}
\label{eq:cloudpressure}
P_{\rm cl}=\frac{3\pi}{20}G\Sigma^2 \phi_{\rm G} \, ,
\end{equation}
where $\Sigma=M/ \pi R^2$ is the mean mass surface density of the cloud, and 
$\phi_{\rm G}$ is a dimensionless correction factor to account for the 
non-spherical geometry of a cloud [BM92; cf. our 
Eq.~(\ref{eq:gravpressure})]. Following BM92, we use as the mean effective
cloud radius the value $R\simeq \sqrt{0.8 \times 8.9 \,{\rm pc}^2}\simeq2.7$ 
pc, where 0.8 pc is the average width of the filament. The corresponding 
surface density is $\Sigma \simeq 1$ g~cm$^{-2}$. The aspect ratio of the 
filament is about $y \simeq 8.9\,{\rm pc}/0.8 \,{\rm pc}\sim11$ (prolate), and 
therefore we adopt the value $\phi_{\rm G} \simeq 2$ (see Appendix B of BM92).
Thus, we obtain the value $P_{\rm cl}\sim5.5\times10^4$ K~cm$^{-3}$. This is 
about an order of magnitude lower than the average internal kinetic pressure 
within the clumps or the average $P_{\rm ext}^{\rm eq}$ value, suggesting 
that the cloud's weight is not contributing significantly to the overall 
confining pressure. We note, however, that Kainulainen et al. (2011b) 
recently found that the masses of IRDCs traced by 
the thermal dust emission are gene\-rally only $\lesssim10-20\%$ of the masses 
traced by near-infrared and CO data. Therefore, the pressure due to the weight 
of the cloud itself, which is mostly caused by the material at low column 
densities, is likely to be higher than estimated above. 

Some of the clumps in the southern end of the filament were found to be 
characterised by negative external pressures in the case of virial 
equilibrium (or 'energy equipartition'; Sect.~4.7), 
suggesting that they should be either warmer or have some additional 
internal pressure. Interestingly, this conforms to the results 
obtained in Paper I, where we used the dust opacities at 8 $\mu$m and 870 
$\mu$m to estimate the dust temperatures towards the MIR-dark submm peaks. 
The clumps SMM 1 and 2 in the southern tip of the filament were estimated to 
have temperatures of $\sim30$ K, but the va\-lues were associated with very 
large relative errors of $\sim60-70\%$ (therefore we have adopted 
$T_{\rm dust}=15$ K). The masses (mean densities) of SMM 1 and 2 would be 
$\sim44$  M$_{\sun}$ ($0.2\times10^4$ cm$^{-3}$) and 22 M$_{\sun}$ 
($0.3\times10^4$ cm$^{-3}$), respectively, if $T_{\rm dust}=30$ K.
The non-thermal velocity dispersions would be 0.27 and 0.23 km~s$^{-1}$, and 
the total kinetic pressures about $1.2\times10^5$ K~cm$^{-3}$ and 
$1.6\times10^5$ K~cm$^{-3}$, respectively. Consequently, the external 
pressures required to virial balance would become positive, i.e., 
$P_{\rm ext}^{\rm eq}/k_{\rm B}\sim5.2\times10^4$ and $\sim1.1\times10^5$ 
K~cm$^{-3}$, for SMM 1 and 2, respectively. These are in reasonable agreement 
with the $P_{\rm ext}^{\rm eq}$ values estimated towards other clumps 
[see Col.~(4) of Table~\ref{table:virial}]. The present results seem to 
support the possibility of north-to-south temperature gradient of 
$\sim15\rightarrow30$ K in G304.74.

\subsection{Clump kinematics, infall motions, and prospects for 
massive star formation}

The clumps in G304.74 appear to be characterised by trans- to supersonic 
internal non-thermal motions. According to the Turbulent Core Model by 
McKee \& Tan (2003), the clump (or core) has to be supersonically turbulent 
in order to achieve a high mass accretion rate and be able to form massive 
stars. The presence of such turbulence is also expected to lead to the 
formation of substructure within the clump. Higher-resolution observations 
would be needed to find out whether the clumps are fragmented into smaller 
cores. For example, interferometric observations of a sample of 
clumps associated with IRDCs have revealed the presence of embedded cores 
within them (e.g., \cite{rathborne2008}; \cite{hennemann2009}; 
\cite{wang2011}).

The clump SMM 6, which has the second largest $f_{\rm turb}$ value 
($\sim3$), is associated with two \textit{MSX} point sources, suggesting that 
it indeed hosts subfragments, and that they have already been collapsed. 
On the other hand, as shown in Fig.~\ref{figure:massradius}, none of the 
clumps appear to clearly fullfil the mass-radius threshold for massive star 
formation proposed by Kauffmann \& Pillai (2010). Only within the 
uncertainties, the two most massive clumps, SMM 3 and 4, could exceed this 
limit.

The supersonic ``turbulent'' motions in the clumps may actually be due to
gravitational infall motions (see \cite{heitsch2009} and references therein).
Evidence of large-scale infall motions in $^{13}$CO lines was indeed found 
towards the southern clumps (see further discussion below).
If the whole fi\-lament is in a process of hierarchical gravitational collapse 
with clumps collapsing locally, a complex velocity pattern is expected to be 
created. Such ``turbulence'' is not able to resist gravitational collapse; 
instead, it is generated from the collapse itself (\cite{ballesteros2011}). 

Radiative transfer models have shown that if the clump centre is collapsing 
faster than the outer layers, the observed spectral line should have two 
peaks with the blueshifted peak being stronger than the redshifted one 
(\cite{zhou1993}; \cite{myers1996}). The blue peak originates from the rear  
side of the clump, whereas the red peak comes from the front part of the 
clump. The central dip between the two peaks is caused by the envelope layer 
(see also \cite{evans1999}). Such line profiles were seen towards SMM 1--4 and 
IRAS13037, and therefore they are likely undergoing large-scale inward motions.
We derived the infall velocites and mass infall rates of $\sim0.03-0.20$ 
km~s$^{-1}$ and $\sim2-36\times10^{-5}$ M$_{\sun}$~yr$^{-1}$, respectively.
These are comparable to the values 0.14 km~s$^{-1}$ and 
$3\times10^{-5}$ M$_{\sun}$~yr$^{-1}$ found by Hennemann et al. (2009) towards 
J18364 SMM 1 South, which is a massive clump within an IRDC. 
Chen et al. (2010) measured higher infall velocities and rates of $\sim0.7-4.6$ 
km~s$^{-1}$ and $\sim1-28\times10^{-3}$ M$_{\sun}$~yr$^{-1}$, respectively, 
for their sample of MYSOs associated with extended green objects (EGOs) and 
IRDCs. Typically, the mass infall rates in MYSOs are 
$\sim10^{-4}-10^{-2}$ M$_{\sun}$~yr$^{-1}$ (e.g., \cite{klaassen2007}; 
\cite{zapata2008}; \cite{beltran2011}). We note that the infall timescales 
we derived, $\sim6-46\times10^5$ yr, are quite long compared to the 
characteristic timescale of $\sim10^5$ yr for high-mass star formation (e.g., 
\cite{davies2011} and references therein). The infall rate, however, 
may not stay constant during the star-formation process, which could also 
explain some of the large observed variations in values quoted above. 

We found that the infall velocities are one to two orders of magnitude lower 
than what would be expected from a free-fall collapse. This suggests that 
the large-scale collapse is not gravitationally dominated, but it could be 
retarded by, e.g., magnetic fields. The predominantly poloidal magnetic field 
could resist motions perpendicular to the poloidal field lines, such as 
appears to be the case in the Snake IRDC (\cite{fiege2000b}; 
\cite{fiege2004}). We also emphasise that the derived infall parameters refer 
to the infall of the surrounding gas onto a (pre-)protocluster, not onto 
individual stars. For example, the timescale for (disk) accretion from the 
infalling envelope may be substantially shorter than the infall timescale. 
The current data do not allow us to decide whether the observed large-scale 
mass infall could feed competitive accretion in the clumps/protoclusters 
(\cite{bonnell2006}), or be related to monolithic collapse leading to massive 
star formation (\cite{mckee2003}). On one hand, the formation of high-mass 
stars in G304.74 seems uncertain on the basis of the mass-radius threshold 
proposed by Kauffmann \& Pillai (2010). On the other hand, the high 
luminosites of the \textit{IRAS} sources 13037 and 13039, 
$\sim1.5-2\times10^3$ L$_{\sun}$ (Paper I), shows that at least 
intermediate-mass star- and/or cluster formation is taking place in this IRDC.

\subsection{The observed chemical properties of the clumps}

\subsubsection{CO depletion and  the SiO detection towards SMM 3}

The CO depletion factors we derived are in the range $f_{\rm D}\sim0.3-2.3$. 
Therefore, CO molecules do not appear to be signi\-ficantly depleted at the
scale of clumps in G304.74, which conforms to the low average H$_2$ densities. 
Higher-resolution observations would be needed to investigate whether the 
clumps contain smaller and more CO-depleted \textit{cores}. The values 
$f_{\rm D}<1$ could be due to too small a canonical CO abundance used, or due 
to conta\-mination of non-depleted gas along the line of sight (see also 
\cite{miettinen2011} and references therein). The latter possibi\-lity, 
however, may be less important because the cloud is located above the galactic 
plane, and thus relatively free of contamina\-ting emission along the line of 
sight. Of course, uncertainties in $N({\rm H_2})$, particularly due to the 
dust opacity, pro\-pagate into $x({\rm C^{17}O})$ and $f_{\rm D}$.

It is not surprising that we found no evidence of CO freeze-out towards the  
\textit{IRAS} sources and SMM 6, which are all MIR bright and very likely 
associated with ongoing star formation. The physical processes associated 
with star-formation acti\-vity, such as outflows, heating, etc., are expected 
to release CO from the icy grain mantles back into the gas phase. 
Consistently, the largest $f_{\rm D}$ values are seen towards the MIR-dark 
clumps SMM 1 and 3. However, the detection of SiO emission in SMM 3 suggests 
the presence of outflows, because gas-phase SiO is expected to be produced 
when shocks disrupt the dust grains (e.g., \cite{hartquist1980}; 
\cite{schilke1997}). This is in agreement with the finding of Sakai 
et al. (2010; henceforth, SSH10), i.e., that the fraction of shocked gas 
(as traced by SiO) is higher in the \textit{MSX}-dark sources than in the 
\textit{MSX}-bright sources. This could indicate that the MIR-dark clumps are 
associated with newly formed shocks giving rise to stronger SiO emission, 
whereas the weaker emission from the MIR-bright clumps would be due to shocks 
being older, and thus milder. We note that the highest estimated SiO column 
density value in SMM 3, $\sim2.5\times10^{13}$ cm$^{-2}$, is comparable to 
the values $\sim1.1-2.6\times10^{13}$ cm$^{-2}$ found by SSH10 towards their 
\textit{MSX}-dark clumps. 

For comparison, Hernandez et al. (2011) found depletion factors up to 
$f_{\rm D}\sim5$ in the highly filamentary IRDC G035.39-00.33. 
Miettinen et al. (2011) derived the values $f_{\rm D}\sim0.6-2.7$ towards a 
sample of clumps within different IRDCs, which are very similar to those 
obtained in the present work with the same telescope/angular resolution. 
Higher values of $f_{\rm D}\sim6-19$ were recently found by Chen et al. (2011) 
for their sample of high-mass star-forming clumps, including sources within 
IRDCs, u\-sing the 10-m SMT observations of C$^{18}$O at $34\arcsec$ 
resolution. This shows that strong CO depletion is possible even at the 
scale of clumps. We note that Hernandez et al. (2011) used the CO depletion 
timescale to constrain the lifetime of G035.39-00.33 (or the part of the 
cloud showing the strongest depletion). Such investigations are useful 
considering the cloud formation timescales and mechanisms. Unfortunately, 
the fact that we found no clear evidence of CO depletion in G304.74 does 
not allow us to perform similar analysis.

\subsubsection{CH$_3$OH}

The fractional CH$_3$OH abundances in the observed clumps were estimated to 
lie in the range $\sim0.2-12\times10^{-9}$. The lowest values are 
found towards the MIR-dark clump SMM 2, and the highest ones towards SMM 3 and 
IRAS13037. 

The pure gas-phase chemical models suggest that CH$_3$OH is formed 
from methylium cation through the radiative association reaction 
${\rm CH_3^+}+{\rm H_2O}\rightarrow {\rm CH_3OH_2^+}+h \nu$, followed by 
the electron recombination reaction
${\rm CH_3OH_2^+}+{\rm e^-}\rightarrow {\rm CH_3OH}+{\rm H}$. The latter 
reaction is, however, too inefficient to explain the observed 
CH$_3$OH abundances in star-forming regions alone 
(see, e.g., \cite{wirstrom2011} and references therein). 
For example, the CH$_3$OH observations by van der Tak et al. (2000) 
towards a sample of 13 massive young stellar objects (MYSOs) suggested three 
types of abundance profiles: the coldest sources have CH$_3$OH abundances 
of $\sim10^{-9}$, warmer sources are cha\-racterised by a range of 
abundances from $\sim10^{-9}$ to $\sim10^{-7}$, whereas the hot cores have 
$x({\rm CH_3OH})\sim10^{-7}$. These high abundances indicate that solid 
CH$_3$OH must have been eva\-porated from the icy grain mantles into the gas 
phase. In the grain-surface chemistry scheme, CH$_3$OH is formed through 
successive hydrogenation of CO on dust grains 
(${\rm CO}\rightarrow {\rm HCO}\rightarrow{\rm H_2CO}\rightarrow{\rm CH_2OH}\rightarrow{\rm CH_3OH} $), and subsequent grain mantle evaporation is then 
predicted to cause the gas-phase abundance of CH$_3$OH to be $>10^{-8}$ 
(e.g., \cite{vandertak2000}; \cite{charnley2004}; \cite{garrod2008}).
Besides grain heating, methanol can be released from grain mantles 
by mo\-derate shocks (e.g., \cite{bachiller1997} and refe\-rences 
therein). 

What comes to the observed methanol abundances in the present work, 
the lowest values, which are found in SMM 2, do not appear to necessarily 
require mantle eva\-poration (see below). The detection of high-velocity 
$^{13}$CO wing emission in SMM 2, however, suggests the presence of outflow 
activity. The rest of the $x({\rm CH_3OH})$ values, particularly in SMM 3 and 
IRAS13037, appear to require that CH$_3$OH has been injected from dust grains 
into the gas. In SMM 3 the high CH$_3$OH abundance together with the SiO 
detection and $^{13}$CO wing emission, are likely related to outflow activity. 
The latter wing-emission signature was also found towards 
IRAS13037, but the SED of this source suggests the presence of hot dust 
(Paper I). Therefore, in addition to grain processing in 
shocks, the central YSO(s) may have raised the dust tempe\-rature above 100 K 
in IRAS13037, which is the approximate sublimation temperature of CH$_3$OH 
(e.g., \cite{garrod2006}). Interestingly, besides the highest 
CH$_3$OH abundance, the strongest CO depletion was found towards SMM 3. 
This could be a sign of enhanced methanol production in SMM 3 due to CO 
freeze-out (e.g., \cite{whittet2011} and references therein); we still see 
a remnant of stronger CO depletion when CO hydrogenation on dust grains was 
arised, but also the product of this process, i.e., the gas-phase CH$_3$OH.

The CH$_3$OH observations towards IRDCs have also been done by other 
authors. Beuther \& Sridharan (2007) studied a sample of massive clumps within 
IRDCs, and found CH$_3$OH column densities and abundances of 
$1.7-37.2\times 10^{13}$ cm$^{-2}$ and $0.4-10.7\times10^{-10}$, respectively. 
Although their column densities are very similar to our values, our 
abundances are generally higher. As discussed by Beuther \& Sridharan 
(2007), their a\-verage CH$_3$OH abundance of $4.3\times10^{-10}$ is 
similar to the CH$_3$OH abundances of $6\times10^{-10}$ found in the low-mass 
starless cores L1498 and L1517B by Tafalla et al. (2006). There are no internal
heating/outflows in starless cores, so grain processing is not expected in 
such sources. Sakai et al. (2010) derived CH$_3$OH column densities in 
the range $<2.9-24.8\times10^{14}$ cm$^{-2}$ towards a sample of 20 massive 
clumps within IRDCs, which are much higher than those found in 
the present work. Sakai et al. (2010) pointed out that their much higher 
$N({\rm CH_3OH})$ values compared to those of Beuther \& Sridharan (2007) 
could be due to the source selection criteria; SSH10 studied \textit{MSX}-dark 
clumps with signatures of star-formation activity. 
Based on the observed correlation between the CH$_3$OH and SiO abundances, 
SSH10 concluded that the origin of CH$_3$OH in the 
\textit{MSX}-dark clumps is related to shocks. G{\'o}mez et al. (2011) 
found CH$_3$OH abundances of $\sim4\times10^{-9}-3\times10^{-8}$ in the IRDC 
core G11.11-0.12 P1, and also suggested that outflow desorption is 
responsible in releasing CH$_3$OH from ice mantles. With comparable CH$_3$OH 
abundances, this could also be the case in the MIR-dark clump SMM 3.

\subsubsection{DCN}

We detected the $J=3-2$ transition of DCN towards SMM 3. To our 
knowledge, this is the first reported detection of DCN in an IRDC [Sakai et 
al. (2012) recently detected the DNC isotopomer towards different IRDC 
clumps]. The first detection of DCN in interstellar medium was made by 
Jefferts et al. (1973) towards the 
Orion Nebula. Mangum et al. (1991) mapped the Orion-KL region in DCN, and 
they suggested that most of the gas-phase DCN could have been released from 
icy grain mantles. Both the DCN and DNC molecules are the major products of 
the association reaction ${\rm D}+{\rm CN}$ on grain surfaces 
(\cite{hiroaka2006}), and DCN can be thermally desorbed at $\sim100$ K (e.g., 
\cite{albertson2011}). In SMM 3 we may see shock-originated DCN emission, 
which conforms to the results discussed above.

\section{Summary and conclusions}

We have carried out a molecular-line study of the IRDC G304.74 using 
the APEX telescope. All the clumps along the filamentary cloud were observed 
in C$^{17}$O$(2-1)$, and selected positions in the souther part of the 
filament were observed in the rotational lines of SiO, $^{13}$CO, and CH$_3$OH. 
The main results are summarised as follows:

\begin{enumerate}
\item The $\sim9$ pc long filamentary IRDC G304.74 is spatially coherent 
structure as indicated by the uniform C$^{17}$O$(2-1)$ radial velocities along 
the cloud.
\item The fragmentation of the filament into clumps appears to be caused by 
``sausage''-type fluid instability, in agreement with results from other IRDCs, 
such as the ``Nessie'' Nebula (\cite{jackson2010}). At smaller scales, 
however, clumps may have fragmented into smaller cores because of, e.g., 
Jeans-type instability. Studying this possibility would require 
higher-resolution observations.
\item Most of the clumps appear to be gravitationally bound 
($\alpha_{\rm vir} \lesssim 2$), although we found evidence that the external 
pressure may play an important role in their overall dynamics. The ambient 
pressure could be due to the turbulent ram pressure, in agreement with the 
hypothesis that the formation of the whole fi\-lament is caused by converging
supersonic turbulent flows (Paper I).
\item The whole filament might be close to virial balance, as suggested 
by the analysis of the virial mass per unit length. 
Radial collapse could be retarded by the poloidal magnetic field component. 
Clump-by-clump virial analysis suggests that some of the clumps in the southern 
part of the filament may be warmer than the rest of the clumps. This 
is consistent with our analysis of dust opacity ratios at 8 and 870 $\mu$m 
in Paper I, which suggested dust temperatures of $\sim30$ K in SMM 1 and 2, 
although with considerable uncertainties.
\item The clumps show trans- to supersonic non-thermal internal motions. 
Moreover, the observed clumps in the southern part of the filament show 
large-scale infall motion signatures in the $^{13}$CO$(2-1)$ line. The infall 
velocities and mass infall rates were estimated to be $\sim0.03-0.20$ 
km~s$^{-1}$ and $\sim2-36\times10^{-5}$ M$_{\sun}$~yr$^{-1}$, respectively.
\item None of the clumps clearly fulfill the mass-radius limit of massive star 
formation recently proposed by Kauffmann \& Pillai (2010). The clumps may 
``only'' form stellar clusters and/or intermediate-mass stars. 
Two \textit{IRAS} sources in the cloud have luminosities 
$\sim1.5-2\times10^3$ L$_{\sun}$ (Paper I), which supports the latter scenario.
\item The CO molecules do not appear to be significantly depleted in the 
clumps ($f_{\rm D}\lesssim 2$). The star-formation activity, such as outflows, 
may have released CO from the icy grain mantles into the gas phase. Some of the 
$^{13}$CO$(2-1)$ lines also show broad wings of emission, suggesting the 
presence of outflowing gas. 
\item We estimate that the fractional CH$_3$OH abundances in the clumps are 
around $2\times10^{-10}-1\times10^{-8}$. No evidence of hot-core type methanol 
abundance of $\sim10^{-7}$ was found among the observed five clumps. Outflow 
activity appears to be a likely explanation for the observed CH$_3$OH 
abundance in the MIR-dark clump SMM 3. The presence of outflows in 
SMM 3 is also indicated by the detection of SiO and DCN emission towards this 
source.
\end{enumerate}

\begin{acknowledgements}

I express my gratitude to the anonymous referee for his/her constructive 
comments and suggestions on the original manuscript. 
I wish to thank the staff at the APEX telescope for performing the 
service-mode observations presented in this paper. I would also like to 
thank the people who maintain the CDMS and JPL molecular spectroscopy 
databases, and the \textit{Splatalogue} Database for Astronomical Spectroscopy. 
The Academy of Finland is acknow\-ledged for the financial support through 
grant 132291. This research has made use of NASA's Astrophysics Data System 
and the NASA/IPAC Infrared Science Archive, which is operated by the JPL, 
California Institute of Technology, under contract with the NASA.

\end{acknowledgements}

\begin{appendix}
\section{Derivation of the physical and chemical properties of 
the cloud/clumps}

This appendix presents details on the analysis and formulas discussed 
in Sects.~4.1 -- 4.4.

\subsection{Kinematic distance}

To calculate the cloud kinematic distance, we adopted
the mean C$^{17}$O$(2-1)$ radial velocity of $-26.7$ km~s$^{-1}$. 
We also employed the recent rotation curve of Reid et al. (2009), 
which is based on measurements of trigonometric parallaxes and proper motions 
of masers in high-mass star-forming regions. The best-fit rotation parameters 
of Reid et al. (2009) are ($\Theta_0$, $R_0$)$=$(254 km~s$^{-1}$, 8.4 kpc), 
where $R_0$ is the solar galactocentric distance, and $\Theta_0$ is the 
rotation velocity at $R_0$. The resulting near kinematic distance, as expected 
to be appropriate for an IRDC seen in absorption, is $d=2.54\pm0.66$ kpc. 
The corresponding galactocentric distance is about $R_{\rm GC}\simeq 7.26$ kpc. 
Actually, besides being an IRDC, the galactic latitude of G304.74, 
$b=+1\fdg32$, supports the idea that it lies at the near distance, because 
then the cloud is $\sim59$ pc above the galactic plane. In the case of the 
far kinematic distance, which is $\sim7.05$ kpc, it would be suspiciously 
high above the galactic plane, i.e., $\sim162$ pc. This would be more than 
twice the scale height of the molecular disk, $z_{1/2}\sim70$ pc 
(\cite{bronfman1988}; see also \cite{fontani2005}).

\subsection{Revision of clump properties presented in Paper I}

The revised cloud distance was used to recalculate the 
distance-dependent clump parameters presented in Paper I, i.e., the clump 
radius ($R \propto d$), mass ($M \propto d^2$), and volume-averaged H$_2$ 
number density [$\langle n({\rm H_2}) \rangle \propto M/R^3 \propto d^{-1}$]; 
see Eqs.~(6)--(8) in Paper I. Previously, we assumed a uniform dust 
temperature of $T_{\rm dust}=15$ K for all the other sources except IRAS13037 
and IRAS13039, for which we used the value $\simeq 22$ K derived from their 
spectral energy distributions (SEDs; see Fig.~5 of Paper I). Here, we adopt 
these same temperature values but the assumed 15 K temperature is now 
assumed to have a $\pm5$ K uncertainty. This is expected to be a reasonable 
choice, because previous molecular-line observations of clumps within IRDCs 
have shown the typical gas kinetic temperature to lie in the range 
$T_{\rm kin}\approx10-20$ K (\cite{carey1998}; \cite{teyssier2002}; 
\cite{sridharan2005}; \cite{pillai2006}; \cite{sakai2008}, \cite{zhang2011}; 
\cite{devine2011}; \cite{ragan2011}). Moreover, at high densities of
$n({\rm H_2})\gtrsim3\times10^4$ cm$^{-3}$, where collisional coupling between 
the gas and dust becomes efficient, the gas and dust temperatures are expected
to be similar (e.g., \cite{galli2002}). For comparison, Nguy{\^e}n Luong et 
al. (2011) recently derived a mean dust temperature of $\sim14.5$ K for the 
clumps/cores in the filamentary IRDC G035.39-00.33. To estimate the errors in 
$R$, $M$, and $\langle n({\rm H_2}) \rangle$, we took the temperature 
uncertainty, and the distance error of $\pm0.66$ kpc into account. The 
$\pm5$-K temperature error was also used to estimate the uncertainty in the 
H$_2$ column density. 

\subsection{Molecular column densities and fractional abundances}

The spectral line's intensity can be expressed as the main-beam brightness 
temperature, which is given by

\begin{equation}
T_{\rm MB}=\frac{h\nu}{k_{\rm B}}\left[F(T_{\rm ex})-F(T_{\rm bg})\right]\left(1-e^{-\tau_{\nu}} \right) \, ,
\label{eq01}
\end{equation}
where $h$ is the Planck constant, $k_{\rm B}$ is the Boltzmann constant, 
$T_{\rm ex}$ is the line excitation temperature, $T_{\rm bg}$ is the 
background tempe\-rature (taken to be the cosmic background radiation 
temperature of 2.725 K), $\tau_{\nu}$ is the line optical thickness, and 
the function $F(T)$ is identical to the (average) photon occupation number:

\begin{equation}
F(T) \equiv \frac{1}{e^{h\nu/k_{\rm B}T}-1} \, .
\end{equation}
We note that the often used Rayleigh-Jeans equivalent temperature is given 
by $J(T)=h \nu/k_{\rm B} \times F(T)$. From Eq.~(\ref{eq01}) we can solve for 
the peak optical thickness:

\begin{equation}
\tau_0=-\ln \left \{ 1-\frac{T_{\rm MB}}{\frac{h\nu}{k_{\rm B}}\left[F(T_{\rm ex})-F(T_{\rm bg})\right]} \right \} \, .
\label{eq02}
\end{equation}
Equation~(\ref{eq02}) was used to estimate the value of $\tau_0$ for those 
lines which were used in the column-density calculations [see Col.~(7) of 
Table~\ref{table:lineparameters}]. The assumptions made concerning the value 
of $T_{\rm ex}$ are described below. 

The beam-averaged column densities, $N({\rm mol})$, of C$^{17}$O, SiO, 
CH$_3$OH, and DCN were calculated under the assumption of local thermodynamic 
equilibrium (LTE), and assuming optically thin emission. The latter assumption 
seems to be valid for our lines. In this case, $N({\rm mol})$ is directly 
proportional to the integrated line intensity as 

\begin{equation}
N({\rm mol}) = \frac{3k_{\rm B}\epsilon_0}{2\pi^2} \frac{1}{\nu \mu_{\rm el}^2S} \frac{Z_{\rm rot}(T_{\rm ex})}{g_{K}g_{I}} \frac{e^{E_{\rm u}/k_{\rm B}T_{\rm ex}}}{1 - \frac{F(T_{\rm bg})}{F(T_{\rm ex})}} \int \, T_{\rm MB} \, {\rm dv} \; ,
\label{eq03}
\end{equation}
where $\epsilon_0$ is the vacuum permittivity, $\mu_{\rm el}$ is the permanent 
electric dipole moment, $S$ is the line strength, $Z_{\rm rot}$ is the 
rotational partition function, $g_{K}$ is the $K$-level 
degeneracy, $g_{I}$ is the reduced nuclear spin degeneracy (see, e.g., 
\cite{turner1991}), and $E_{\rm u}$ is the e\-nergy of the upper-transition 
state. 
We note that in the derivation of Eq.~(\ref{eq03}), the electric dipole moment 
matrix element is defined to be 
$\left\vert \mu_{\rm ul} \right\vert^2=\mu_{\rm el}^2S/g_{\rm u}$, where 
$g_{\rm u}=2J+1$ is the rotational degene\-racy of the upper state 
(\cite{townes1975}). Note that for linear molecules, $g_K = g_I = 1$ 
for all levels, whereas for $A$-type CH$_3$OH, $g_K=1$ and $g_I=2$, 
and for $E$-type CH$_3$OH, $g_K=2$ and $g_I=1$ (\cite{turner1991}).
We derived the CH$_3$OH column densities from the integrated intensity of both
the CH$_3$OH$(5_{0,5}-4_{0,4})$-A$^+$ and CH$_3$OH$(5_{-1,5}-4_{-1,4})$-E 
lines, and the results were found to be comparable to each other. 
In this paper we only give the $N({\rm CH_3OH})$ values derived from the above 
A$^+$-transition. Note that we have detected too few lines to 
employ the population diagram analysis to derive the rotational 
temperature and column density of CH$_3$OH (\cite{goldsmith1999}).

For the linear molecules C$^{17}$O, SiO, and DCN, we approximated the 
partition function as

\begin{equation}
Z_{\rm rot}(T_{\rm ex})\simeq \frac{k_{\rm B}T_{\rm ex}}{hB}+\frac{1}{3} \, ,
\end{equation}
where $B$ is the rotational constant. The above expression is appropriate for 
heteropolar molecules at the high temperature limit 
($hB/k_{\rm B}T_{\rm ex} \ll 1$). The partition function 
of methanol, which is a slightly asymmetric rotor with one internal rotor and 
$A$- and $E$-symmetry states, was calculated as 

\begin{equation}
Z_{\rm rot}(T_{\rm ex})=Z_{\rm rot}(A)+Z_{\rm rot}(E)=2\sqrt{\frac{\pi(k_{\rm B}T_{\rm ex})^3}{h^3ABC}}\, ,
\end{equation}
where $A$, $B$, and $C$ are the rotational constants (\cite{turner1991}).
The spectroscopic properties of the observed molecules/transitions are given 
in Table~\ref{table:spec}

The unknown parameter in the above optical-thickness and column-density 
calculations is the excitation temperature, $T_{\rm ex}$. In the case of 
C$^{17}$O we assumed that the line is thermalised at the clump temperature, 
and adopted the value $T_{\rm ex}=15\pm5$ K for all the clumps except the 
\textit{IRAS} sources 13037 and 13039, for which the value $T_{\rm ex}=22$ K 
was used (Appendix A.2). This is likely to be true for our clumps, 
because even their average H$_2$ densities are comparable to the 
critical density of C$^{17}$O$(2-1)$ [$9.5\times10^3$ cm$^{-3}$; see Col.~(2) 
of Table~\ref{table:spec}]. Fontani et al. (2005) actually derived the 
C$^{17}$O rotational excitation temperature of 18 K towards IRAS13039, very 
close to the value we adopted. The approximate-equality 
$T_{\rm kin}\simeq T_{\rm ex}[{\rm C^{17}O}(2-1)]$ was also recently found by 
Miettinen et al. (2011) for a sample of clumps associated with IRDCs. For the 
linear rotors SiO and DCN, we assumed that $T_{\rm ex}$ is in the range 
$[5\,{\rm K},\, E_{\rm u}/k_{\rm B}]$, whereas for CH$_3$OH it was assumed that 
$T_{\rm ex} \in [10\,{\rm K},\, 2E_{\rm u}/3k_{\rm B}]$. The adopted 
upper $T_{\rm ex}$ limits result from the turning point of the 
temperature-dependent part of $N({\rm mol})$, i.e., from the derivative 

\begin{equation}
\frac{{\rm d}N({\rm mol})}{{\rm d}T_{\rm ex}}=\frac{{\rm d}}{{\rm d}T_{\rm ex}}Z_{\rm rot}(T_{\rm ex})e^{E_{\rm u}/k_{\rm B}T_{\rm ex}}=0\, ,
\end{equation}
where it is assumed that $T_{\rm ex}\gg T_{\rm bg}$ 
(e.g., \cite{hatchell1998}). The upper $T_{\rm ex}$ limits thus derived lead 
to the lower limit to $N({\rm mol})$. We note that the 
$T_{\rm ex}({\rm CH_3OH})$ range we have used, 10--23.2 K, is comparable to 
the CH$_3$OH$(2_k-1_k)$ rotational tempe\-ratures of $\sim13.4-24.5$ K 
derived by SSH10 towards a sample of massive clumps within IRDCs.

We calculated the fractional abundances of the molecules by dividing the 
molecular column density by the H$_2$ column density: 
$x({\rm mol})=N({\rm mol})/N({\rm H_2})$. For this purpose, the values of 
$N({\rm H_2})$ were derived from the LABOCA dust continuum map smoothed to 
the corresponding resolution of the line observations. 

\begin{table*}
\caption{Spectroscopic properties of the observed transitions.}
\begin{minipage}{2\columnwidth}
\centering
\renewcommand{\footnoterule}{}
\label{table:spec}
\begin{tabular}{c c c c c c c}
\hline\hline 
Transition & $n_{\rm crit}$ & $E_{\rm u}/k_{\rm B}$ & $\mu_{\rm el}^2S$ & $A$ & $B$ & $C$\\
           & [cm$^{-3}$] & [K] & [D$^2$] & [MHz] & [MHz] & [MHz] \\
\hline
SiO$(5-4)$ & $2.3\times10^6$ & 31.3 & 48.14651 & \ldots & 21\,711.96 & \ldots\\
DCN$(3-2)$ & $\sim10^7$ & 20.9 & 80.50709 & \ldots & 36\,207.46 & \ldots\\
$^{13}$CO$(2-1)$ & $8.9\times10^3$ & 15.9 & 0.02437 & \ldots & 55\,101.01 & \ldots\\
C$^{17}$O$(2-1)$ & $9.5\times10^3$ & 16.2 & 0.02432 & \ldots & 56\,179.99  & \ldots\\
CH$_3$OH$(5_{-1,5}-4_{-1,4})$-E & $9.5\times10^5$ & 40.4 & 3.88240 & 127\,484.0 & 24\,679.98 & 23\,769.70\\
CH$_3$OH$(5_{0,5}-4_{0,4})$-A$^+$ & $1.1\times10^6$ & 34.8 & 4.04297 & $- \mid \mid -$ & $- \mid \mid -$ & $- \mid \mid -$ \\
CH$_3$OH$(5_{4,*}-4_{4,*})$-A$^{+/-}$\tablefootmark{a} & $3.5\times10^5$ & 115.2 & 1.45278 & $- \mid \mid -$ & $- \mid \mid -$ & $- \mid \mid -$ \\
\hline 
\end{tabular} 
\tablefoot{Column~(2) gives the critical density of the transition, 
$n_{\rm crit}=A_{\rm ul}/C_{\rm ul}$, where $A_{\rm ul}$ is the spontaneous 
decay rate and $C_{\rm ul}$ is the collisional de-excitation rate. It was 
calculated at $T=15$ K using the collisional-rate data 
available in the Leiden Atomic and Molecular Database (LAMDA; 
{\tt http://www.strw.leidenuniv.nl/$\sim$moldata/}) (\cite{schoier2005}). 
For DCN, there is no molecular datafile available, so we used the collisional 
rate coefficient of HCN to estimate $n_{\rm crit}$. Columns~(3) and (4) give 
the upper-state energy and the product $\mu_{\rm el}^2S$, where 
$\mu_{\rm el}$ is the permanent electric dipole moment, and $S$ is the line 
strength. Columns~(5)--(7) list the rotational constants of the molecules. 
The data were compiled from the Jet Propulsion Laboratory (JPL; 
\cite{pickett1998}) and CDMS (\cite{muller2005}) 
databases.\tablefoottext{a}{A blend of CH$_3$OH$(5_{4,1}-4_{4,0})$-A$^+$ and 
CH$_3$OH$(5_{4,2}-4_{4,1})$-A$^-$, which have the same spectroscopic 
parameters.}}
\end{minipage} 
\end{table*}

\subsection{CO depletion factors}

Because the clumps under investigation (or at least most of them) are 
presumably cold, the CO molecules should efficiently freeze out onto dust 
grain surfaces. To estimate the amount of this CO depletion, we 
calculated the CO depletion factors, $f_{\rm D}$, following the analysis 
presented in the paper by Fontani et al. (2006). If $x({\rm CO})_{\rm can}$ is 
the ``canonical'' (undepleted) abundance of CO, and $x({\rm CO})_{\rm obs}$ 
is the observed CO abundance, $f_{\rm D}$ is given by 

\begin{equation}
\label{eq:depletion}
f_{\rm D}=\frac{x({\rm CO})_{\rm can}}{x({\rm CO})_{\rm obs}}\, .
\end{equation}
The ``canonical'' CO abundance at the galactocentric distance $R_{\rm GC}$ 
was calculated using the relationship [Eq.~(7) in \cite{fontani2006}]

\begin{equation}
x({\rm CO})_{\rm can}=9.5\times10^{-5}{\rm e}^{1.105-0.13R_{\rm GC}[{\rm kpc}]}\, .
\end{equation}
Note that this relationship results from using the value $R_0=8.5$ kpc, 
whereas our $R_{\rm GC}$ value was computed using $R_0=8.4$ kpc. Such a small 
discrepancy is however negligible. At $R_{\rm GC}=8.5$ kpc, the above 
relationship gives the standard value $9.5\times10^{-5}$ for the 
abundance of the main CO isotopologue in the solar neighbourhood 
(\cite{frerking1982}). At the galactocentric distance of G304.74, 
$x({\rm CO})_{\rm can}\simeq1.1\times10^{-4}$.
To calculate the ``canonical'' C$^{17}$O abundance we 
employ the oxygen-isotopic ratio (\cite{wilson1994})

\begin{equation}
\label{eq:O-ratio}
\frac{[^{16}{\rm O}]}{[^{18}{\rm O}]}=58.8\times R_{\rm GC}[{\rm kpc}]+37.1 \,,
\end{equation}
and the ratio $[^{18}{\rm O}]/[^{17}{\rm O}]=3.52$ (\cite{frerking1982}):

\begin{eqnarray}
x({\rm C^{17}O})_{\rm can} &=& \frac{x({\rm CO})_{\rm can}}{[^{18}{\rm O}]/[^{17}{\rm O}]\times [^{16}{\rm O}]/[^{18}{\rm O}]}\nonumber \\
   &=& \frac{x({\rm CO})_{\rm can}}{3.52\times(58.8\times R_{\rm GC}[{\rm kpc}]+37.1)} \, .
\end{eqnarray}
For G304.74, we get 
$x({\rm C^{17}O})_{\rm can}\simeq x({\rm CO})_{\rm can}/1633\simeq 6.8\times10^{-8}$.
The depletion factor $f_{\rm D}$ is then calculated from
$f_{\rm D}=x({\rm C^{17}O})_{\rm can}/x({\rm C^{17}O})_{\rm obs}$. 

\end{appendix}

\end{document}